\documentclass[]{JHEP3}

\pdfoutput=1

\usepackage{amsmath}
\usepackage{epsfig,multicol,bbm}
\usepackage{graphicx}
\usepackage{bm}
\usepackage{epsfig}
\usepackage{url}

\renewcommand{\thesection}{\arabic{section}}
\renewcommand{\emph}[1]{{\it #1}}

\newcommand{\be}{\begin{equation}}
\newcommand{\ee}{\end{equation}}
\newcommand{\nn}{\nonumber} 

\newcommand{\MeV}{\text{ MeV}}
\newcommand{\GeV}{\text{ GeV}}
\newcommand{\ab}{\text{ ab}}

\newcommand{\eq}[1]{Eq.~\eqref{#1}}
\newcommand{\eqs}[2]{Eqs.~\eqref{#1} and \eqref{#2}}

\renewcommand{\sec}[1]{Sec.~\ref{#1}}
\newcommand{\appx}[1]{App.~\ref{#1}}
\newcommand{\fig}[1]{Fig.~\ref{#1}}
\newcommand{\figs}[2]{Figs.~\ref{#1} and \ref{#2}}

\newcommand{\df}{\mathrm{d}}

\title{Dark Force Detection in Low Energy e-p Collisions}

\author{Marat Freytsis, Grigory Ovanesyan, and Jesse Thaler \\
Berkeley Center for Theoretical Physics, University of California, Berkeley, CA 94720 and
Theoretical Physics Group, Lawrence Berkeley National Laboratory, Berkeley, CA 94720 \\
E-mail: \email{freytsis@berkeley.edu}, \email{ovanesyan@berkeley.edu}, \email{jthaler@jthaler.net} }
 
\preprint{}

\abstract{We study the prospects for detecting a light boson $X$ with mass $m_X \lesssim 100 \MeV$ at a low energy electron-proton collider.  We focus on the case where $X$ dominantly decays to $e^+ e^-$ as motivated by recent ``dark force'' models.  In order to evade direct and indirect constraints, $X$ must have small couplings to the standard model ($\alpha_X \lesssim 10^{-8}$) and a sufficiently large mass ($m_X \gtrsim 10 \MeV$).  By comparing the signal and background cross sections for the $e^{-} p \,  e^{+} e^{-}$ final state, we conclude that dark force detection requires an integrated luminosity of around 1 ab$^{-1}$, achievable with a forthcoming JLab proposal.}

\begin{document}

\section{Introduction}
\label{sec:intro}

While the gravitational evidence for dark matter is overwhelming~\cite{Rubin:1970zz,AdelmanMcCarthy:2005se,Komatsu:2008hk,Briel:1997hz}, direct measurements of the spectrum and properties of dark matter have so far been elusive.  However, recent astrophysical anomalies---including the WMAP Haze~\cite{Finkbeiner:2004us,Hooper:2007kb}, the PAMELA, FERMI, and H.E.S.S. $e^+/e^-$ excesses~\cite{Adriani:2008zr, Abdo:2009zk,Collaboration:2008aaa,Aharonian:2009ah}, and the INTEGRAL 511 keV excess~\cite{Weidenspointner:2006nua,Knodlseder:2003sv}---could be evidence for dark matter annihilation, decay, or up-scattering in our galactic halo.  With these observations, an intriguing paradigm for dark matter has emerged, where TeV-scale dark matter interacts with a GeV-scale boson~\cite{Finkbeiner:2007kk, Pospelov:2007mp,ArkaniHamed:2008qn,Nomura:2008ru}.  This new light boson $X$ typically has a mass in the range
\be
2 m_e < m_X \lesssim \mbox{few} \GeV,
\ee
with an $\mathcal{O}(1)$ branching fraction $X \rightarrow e^+ e^-$.\footnote{For a recent study of models with even lighter bosons, see Ref.~\cite{Goodsell:2009xc}.}

What is the best way to look for light bosons with small couplings to the standard model?  Indirect constraints from lepton anomalous magnetic moments require the coupling of $X$ to leptons to be $\alpha_X \lesssim 10^{-8}$ \cite{Fayet:2007ua,Pospelov:2008zw}, much smaller than the electromagnetic coupling $\alpha_{\rm EM} \simeq 1/137$.\footnote{There are additional direct constraints on $X$ from rare meson decays \cite{Fayet:2007ua,Kahn:2007ru,Pospelov:2008zw,Reece:2009un}.}  Therefore, any direct production mode for $X$ faces a large irreducible background from an equivalent process where $X$ is replaced by an off-shell photon $\gamma^*$.  A number of studies at lepton colliders have concluded that around 1 ab$^{-1}$ of data is needed to see the process $e^+ e^- \rightarrow \gamma + X$~\cite{Borodatchenkova:2005ct,Fayet:2007ua,Batell:2009yf,Essig:2009nc,Reece:2009un,Morrissey:2009ur,Yin:2009mc}.  While such large integrated luminosities have been achieved at the $B$-factories, it is worthwhile to consider alternative experimental setups that might be more easily scaled to multi-ab$^{-1}$ data sets.  

One standard method to find new particles with small couplings is fixed-target experiments, either with a high intensity beam on a thin target or a ``beam dump'' experiment on a thick target.  As we will see, such experiments already constrain the $X$ parameter space \cite{Riordan:1987aw,Bross:1989mp}.  Recent studies in Refs.~\cite{Reece:2009un,Bjorken:2009mm} have concluded that improved fixed-target experiments can cover a wide range of masses and coupling for $X$, especially if $X$ has a sufficiently long lifetime to yield a displaced vertex or if $X$ has a decay mode to penetrating muons.  Even in the case of prompt $X$ boson decay, the luminosity achievable in traditional fixed-target experiments approaches 1 ab$^{-1}$/day, so with good energy resolution and control over systematics, the irreducible $\gamma^*$ background could be beaten by statistics, and one can simply search for electron pairs that reconstruct a narrow $X$ resonance.  However, full event reconstruction is impossible in this context, since one cannot measure the spectrum of the recoiling nucleus, so traditional fixed-target experiments lack a crucial kinematic cross-check that is available in lepton colliders.

In this paper, we propose searching for an $X$ boson in low-energy electron-proton collisions through the process
\be
e^- p \rightarrow e^- p + X, \qquad X \rightarrow e^+e^-.
\ee
With a high intensity electron beam on a diffuse hydrogen gas target, one combines the high statistics of a traditional fixed-target experiment with the full event reconstruction potential of a lepton collider.  To our knowledge, this experimental setup was first suggested in Ref.~\cite{Heinemeyer:2007sq}, motivated by a different dark matter scenario with an invisibly decaying $X$ boson \cite{Boehm:2003hm}.  In that context, the recoiling proton spectrum was crucial for discovery.  Here, we focus on $X$ bosons that decay visibly to $e^+ e^-$.  Like the fixed-target proposals in Refs.~\cite{Reece:2009un,Bjorken:2009mm}, one is still looking for a narrow $X$ resonance on top of a huge radiative QED background, but here the recoiling proton and electron spectrum can be used to over-constrain the kinematics.

As in Ref.~\cite{Heinemeyer:2007sq}, we consider an electron beam with energy $E_e \simeq 100 \MeV$, where the scattering is dominated by elastic scattering and associated radiative processes.   In particular, pion production is kinematically forbidden as are nuclear excitations.  Such a setup is being actively considered for installation at the Free Electron Laser (FEL) at the Thomas Jefferson Lab National Accelerator Facility (JLab), replacing the laser cavity with a hydrogen gas target.\footnote{For an alternative low-energy search using a positron beam incident on a hydrogen target, see Ref.~\cite{Wojtsekhowski:2009vz}.}  Where the reach in $X$ parameter space overlaps, this proposal of a high intensity beam on a diffuse target is complementary to the proposal in Ref.~\cite{Bjorken:2009mm} of a diffuse beam on a high density target.  We will argue that for the same integrated luminosity, the $X$ reach in $ep$ collisions is comparable to $e^+ e^-$ and $e^- e^-$ collisions.  Since 1 ab$^{-1}$/month is achievable with the FEL beam on a hydrogen gas target, low-energy $ep$ collisions are in principle a cost-effective way to search for the $X$ boson.

In our study, we will focus on irreducible physics backgrounds and assume idealized detectors.  While there are important experimental backgrounds, we will assume that these can be controlled using, for example, information about recoiling proton and electron.  The JLab FEL setup is in principle sensitive to:
\begin{align}
m_X < 2 m_e &:  X \rightarrow \gamma \gamma, \mbox{invisible}\nn \\
2 m_e < m_X \lesssim 100 \MeV  &:  X \rightarrow e^+ e^-, \gamma \gamma, \mbox{invisible}
\end{align}
In models with rich dark sectors, one can even imagine multi-body $X$ decay modes or more than one $X$ field \cite{Baumgart:2009tn,Essig:2009nc}.  For simplicity, we will only look at $X \rightarrow e^+ e^-$, and focus on the case that $X$ couples only to electrons and not to protons.  To capture a wide range of possible ``dark boson'' scenarios, we allow $X$ to be scalar, pseudoscalar, vector, or axial-vector.\footnote{In the case of a light vector boson, $X$ is often referred to as a $U$-boson.}  

In the next section, we summarize the conclusion of our study, that with 1 month to 1 year of running at the JLab FEL, one can probe an interesting parameter space for the $X$ boson.  In \sec{sec:theory}, we outline our theoretical setup, and review indirect and direct constraints on the $X$ boson properties.  We study the reach for $X$ in $ep$ collisions in \sec{sec:reach} and show how a matrix element method can be used to extend the $X$ boson reach.  Comparisons to other $X$ boson collider searches appear in \sec{sec:compexp}.  Two benchmark scenarios appear in \sec{sec:kinedist}, showing an example analysis strategy as well as a variety of kinematic distributions.  We consider an alternative displaced vertex search in \sec{sec:displacedpossibility} and conclude in \sec{sec:conclude}.

\FIGURE[!p]{

~\\
~\\
~\\

\centerline{\includegraphics[scale=0.7]{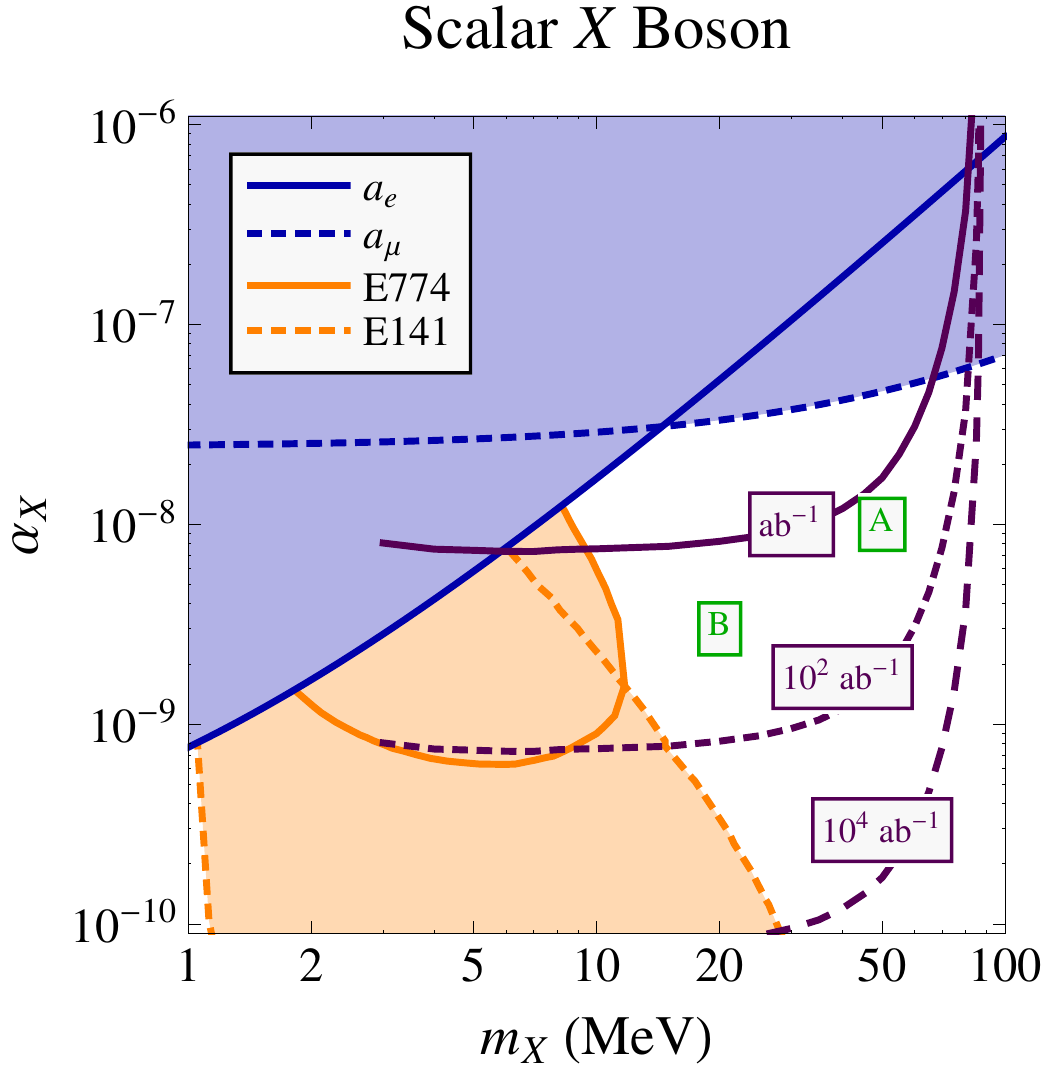} \includegraphics[scale=0.7]{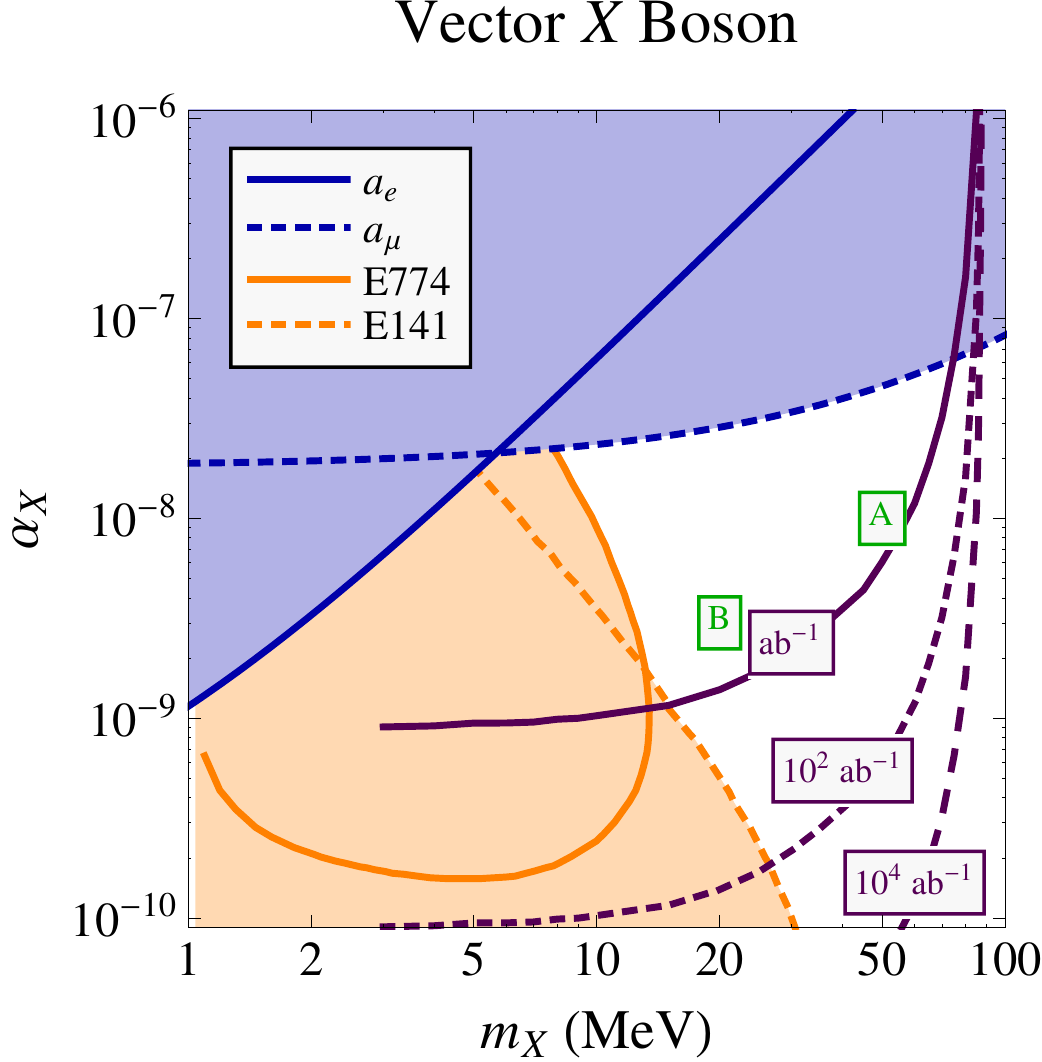}}

~\\

\centerline{\includegraphics[scale=0.7]{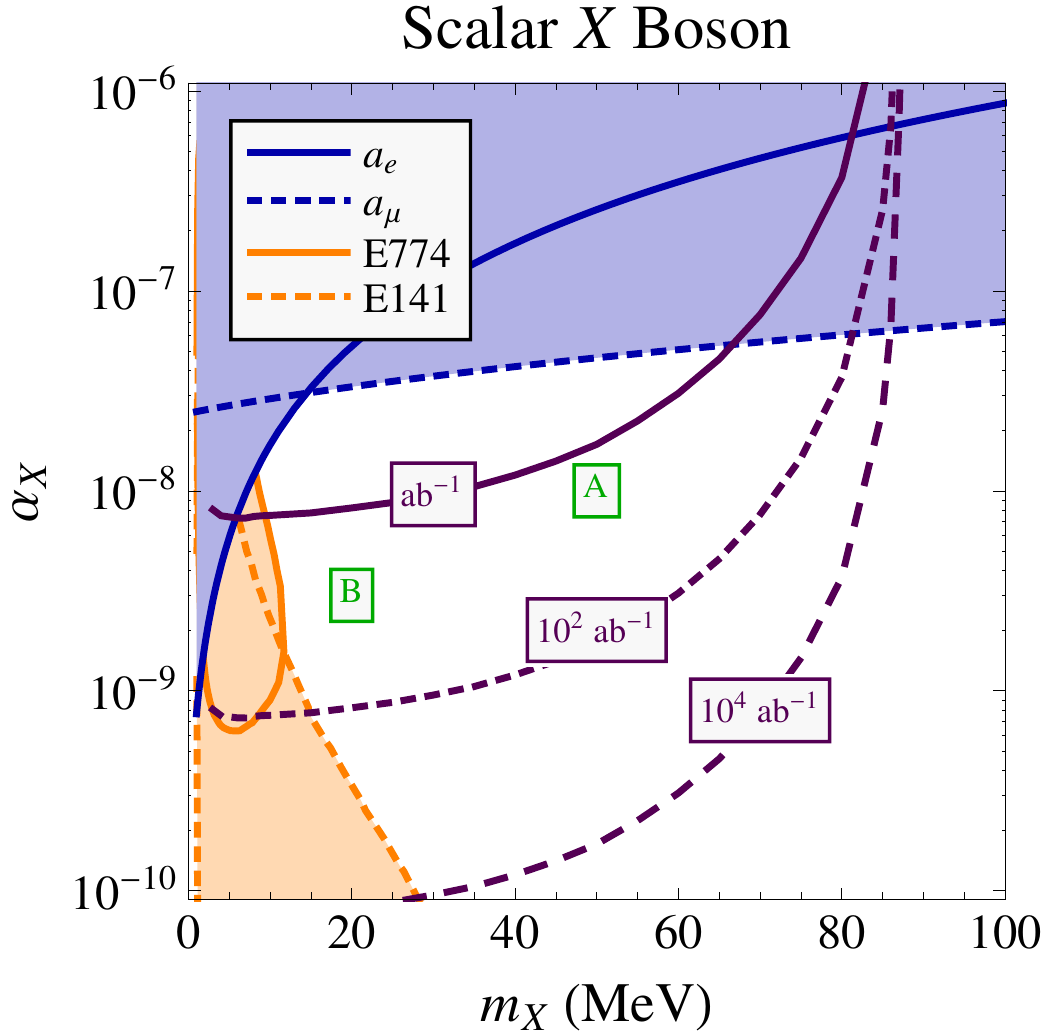} \includegraphics[scale=0.7]{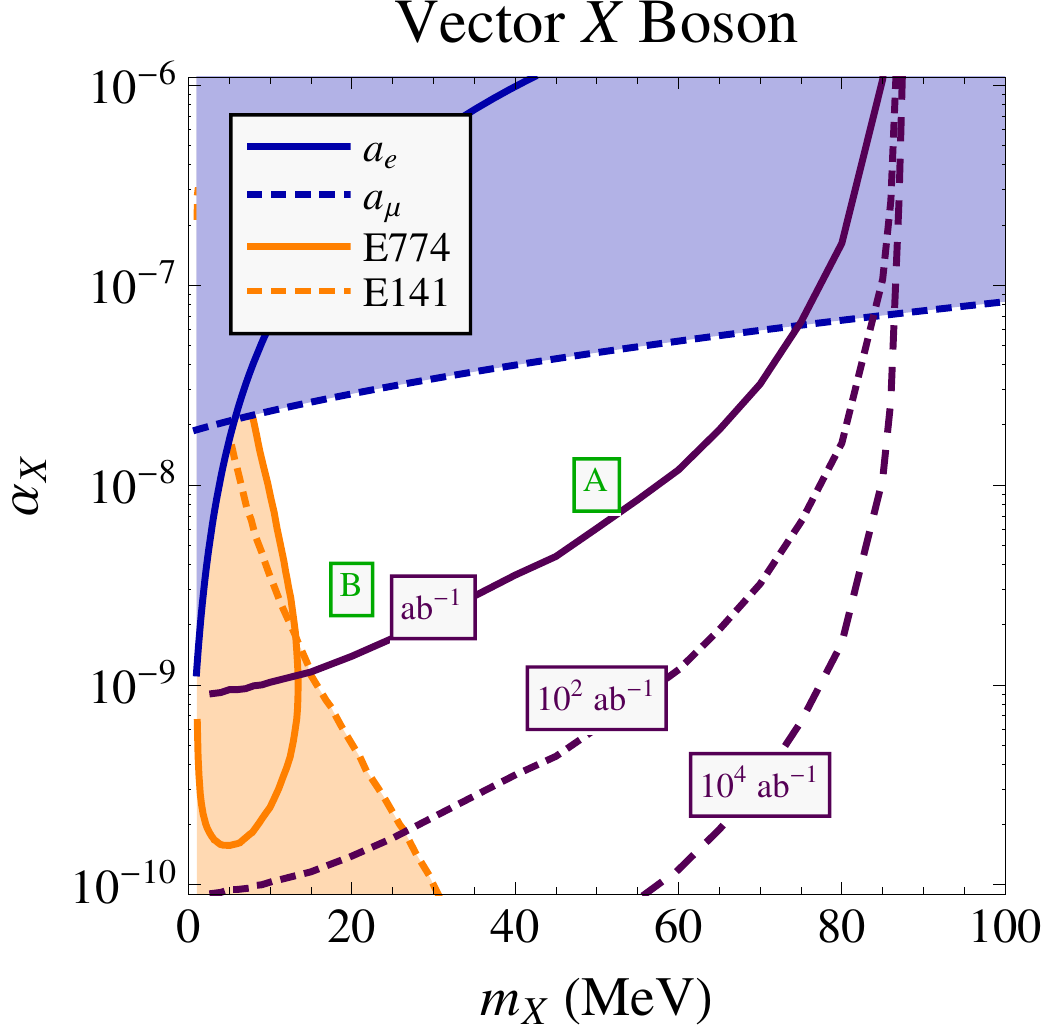}}
\caption{Summary of this study, showing constraints and reach on the $m_X$--$\alpha_X$ plane.  The scalar couplings are on the left, vector couplings on the right.  The top plots are logarithmic in $m_X$, while the bottom plots are linear in $m_X$ to highlight the relevant parameter space.  Blue curves:  bounds from lepton anomalous magnetic moments.  Orange curves: bounds from prior beam dump experiments.  Purple curves:  luminosity necessary to achieve $S/\sqrt{B} = 5$ assuming $m_{e^+ e^-}$ invariant mass resolution of $1 \MeV$, and a detector acceptance of $-2 < \eta < 2$, $\mathrm{KE}_p > 0.5 \MeV$, and $\mathrm{KE}_{e^\pm} > 5 \MeV$.  The reach in the pseudoscalar and axial-vector cases are the same as the scalar and vector cases, respectively, but the anomalous magnetic moment bounds differ.  The points labelled ``A'' and ``B'' correspond to the benchmark scenarios studied in \sec{sec:kinedist}.}
\label{fig:execsummary}
}

\clearpage

\section{Summary}

The key results of our study are summarized in this section.  We take an electron beam with a fiducial energy of $100 \MeV$, incident on a proton target at rest in the lab frame.  At the FEL facility, the proton target is likely to be a hydrogen gas storage cell, which is sufficiently diffuse to allow recirculation of the FEL beam~\cite{Milner:2009--}.  In \fig{fig:execsummary}, we show a plot of the $m_X$--$\alpha_X$ plane for the case that $X$ has scalar or vector couplings to electrons.\footnote{The plots for pseudoscalar and axial-vector couplings appear later in the text.   The reach for the pseudoscalar (axial-vector) is the same as the scalar (vector), but the magnetic moment bounds differ.}  Here, $m_X$ is the mass of the $X$ boson, and $\alpha_X \equiv \lambda_X^2 / 4\pi$ is the coupling of the $X$ boson to electrons, normalized such that it can be roughly compared with $\alpha_{\rm EM} \simeq 1/137$.   

There are three sets of curves shown on the $m_X$--$\alpha_X$ plane:
\begin{itemize}
\item Indirect constraints from anomalous magnetic moments.  An $X$ boson that couples to electrons will radiatively generate $a_e$ and, assuming lepton universality, $a_\mu$.  The shaded region is excluded, as explained in \sec{sec:magmoment}.
\item Direct constraints from beam dump experiments.  When the $X$ boson is sufficiently long-lived, high intensity beam dump experiments are sensitive to the decay $X \rightarrow e^+ e^-$.  The shaded region is excluded, as explained in \sec{sec:beamdump}.
\item Discovery reach in $ep$ scattering.  As detailed in \sec{sec:rawreach}, we use an idealized detector with pseudorapidity coverage $-2 <\eta < 2$ (i.e.\ tracking up to $15.4^\circ$ of the beamline), kinetic energy thresholds of $\mathrm{KE}_{p} > 0.5 \MeV$ and $\mathrm{KE}_{e^\pm} > 5 \MeV$, and invariant mass resolution of 1 MeV.  Assuming no systematic errors, the curves show the integrated luminosity needed to achieve a $5\sigma$ discovery with statistics alone,  i.e.\ $S/\sqrt{B} = 5$ in a 1 MeV resolution bin centered around $m_X$.  If the energy resolution is improved, this integrated luminosity required for discovery improves linearly.  Also, the reach of the experiment can be further improved using a matrix element method, as proposed in \sec{sec:mereach}.  
\end{itemize}
In addition, the points labelled ``A'' and ``B'' indicated the benchmark scenarios considered in \sec{sec:kinedist}.

We see that with an integrated luminosity of 1 ab$^{-1}$, there is a range of $X$ boson masses and couplings that are consistent with known bounds but visible in low energy $ep$ scattering.  With an average current of 10 mA, the FEL beam produces $6 \times 10^6$ electrons per second~\cite{Neil:2005jy}, while a
hydrogen gas target of thickness 10$^{19}$ cm$^{-2}$ is expected to be technically feasible~\cite{Milner:2009--}. Thus, the expected luminosity of the JLab FEL setup is $6 \times 10^{35}~\mathrm{cm}^{-2}~\mathrm{s}^{-1}$, which is approximately 1 ab$^{-1}$/month.  With one month to one year of running, such a facility could probe an interesting range of $X$ boson masses and couplings.

\section{New Light Bosons}
\label{sec:theory}

We consider four kinds of coupling for the new light boson $X$:  scalar, pseudoscalar, vector, and axial-vector.  For simplicity, we assume that $X$ only couples to electrons (and other charged leptons), and show in  \appx{sec:generalize} that proton couplings do not drastically change our conclusions.  In the scalar and pseudoscalar cases, we augment the standard model with a new boson $X$ that couples to the electron field $\psi_e$ as
\be
\label{eq:lagrangescalar}
\mathcal{L}_{s/p} =\overline{\psi}_e \left(\lambda_s + \lambda_p \gamma^5 \right) \psi_e X.
\ee
For the vector and axial-vector cases, we add a massive vector $X_\mu$ with couplings
\be
\label{eq:lagrangevector}
\mathcal{L}_{v/a} = \overline{\psi}_e  \left(\lambda_v \gamma^\mu + \lambda_a \gamma^\mu \gamma^5 \right) \psi_e X_\mu.
\ee
While more exotic operators are possible, \eqs{eq:lagrangescalar}{eq:lagrangevector} cover the generic possibilities for how $X$ can couple to electrons.  

In a complete theory, there is usually some kind of lepton universality, yielding equivalent couplings of $X$ to muons and taus.  Assuming that $X$ does not introduce lepton flavor violation, then the coupling of $X$ to the different leptons will either be approximately equal or proportional to the lepton masses.\footnote{In the special case that $X$ is a pseudo Nambu-Goldstone boson (such as in Ref.~\cite{Nomura:2008ru}), one expects $\lambda_p = m_\ell / f_a$, where $m_\ell$ is the mass of the lepton and $f_a$ is the decay constant.}  As we will see below, though, if the couplings are indeed proportional to the lepton masses, then the constraints from $a_\mu$ exclude any of the interesting region for this study.

For convenience where relevant, we define
\be
\lambda_X \equiv \sqrt{|\lambda_s|^2 + |\lambda_p|^2} \mbox{ or } \sqrt{|\lambda_v|^2 + |\lambda_a|^2}, \qquad \alpha_X \equiv \frac{\lambda_X^2}{4\pi}.
\ee
In the text, we work in the limit $m_e \ll m_X$, and present formulas for finite $m_e$ in \appx{app:finitemass}.

\subsection{Indirect Constraints}
\label{sec:magmoment}

\FIGURE[t]{
  \centerline{\includegraphics[scale=0.7]{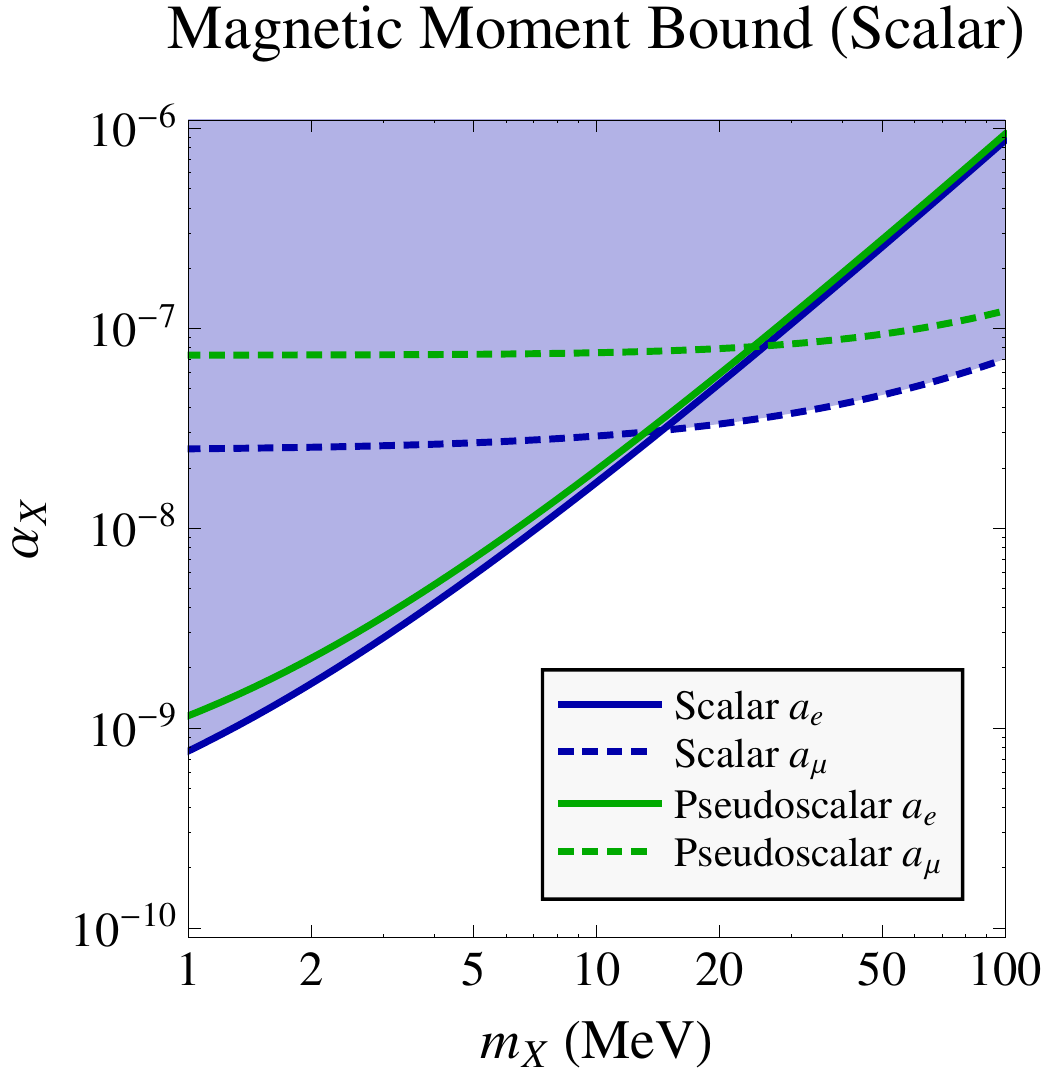} \includegraphics[scale=0.7]{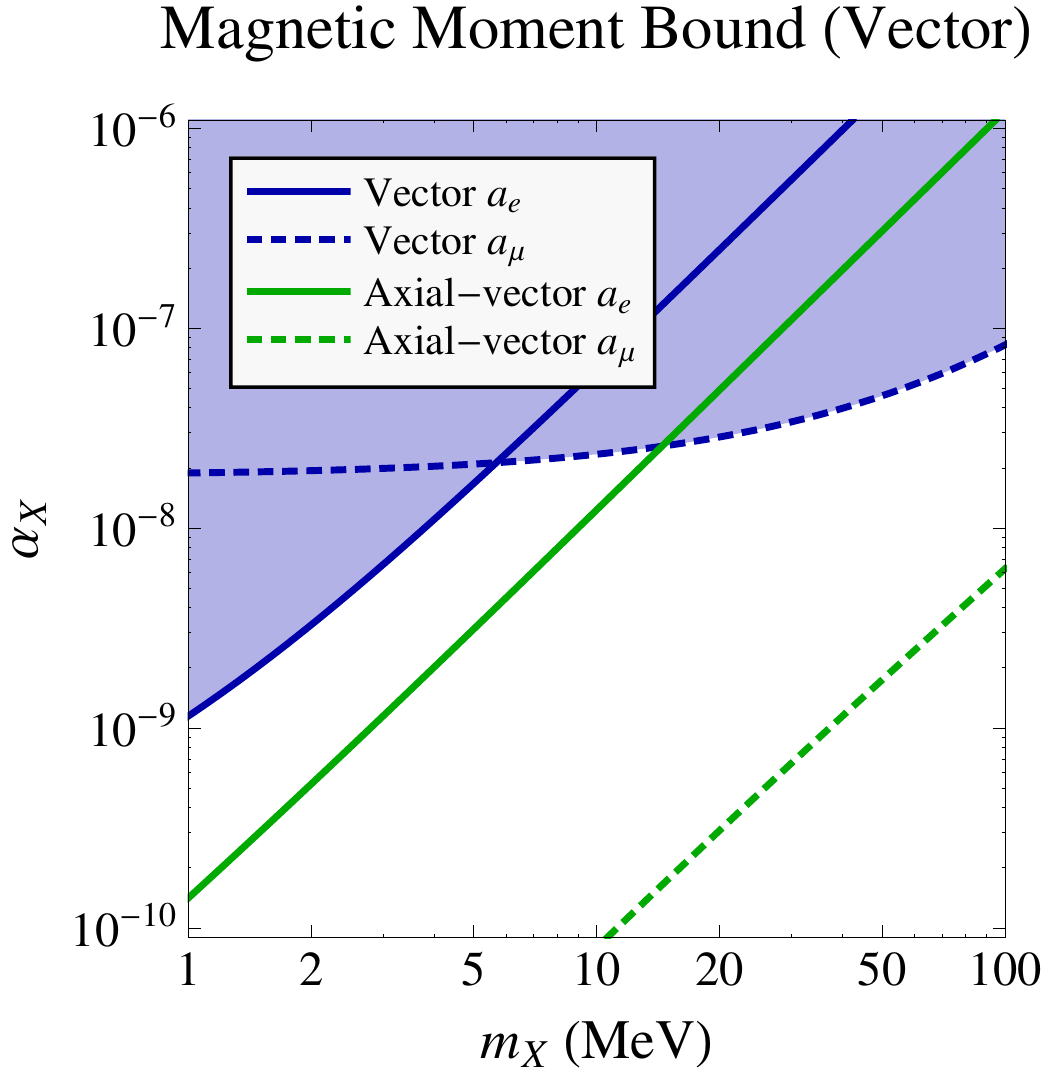}}
  \caption{Constraints on the $X$ boson mass $m_X$ and coupling $\alpha_X$ from anomalous magnetic moments, $a_e = (\frac{g-2}{2})_e$ and $a_\mu = (\frac{g-2}{2})_\mu$.  We assume lepton universality with $\alpha_X^e = \alpha_X^\mu$.}
  \label{fig:magmoment}
}

Previous studies of the indirect constraints on the $X$ boson appear in Ref.~\cite{Fayet:2007ua}. The strongest indirect bounds comes from the effect of the $X$ boson on the anomalous magnetic moments of the electron and muon, $a_e$ and $a_\mu$, arising from $X$ boson loops. We take the limits to be $\delta a_e < 1.7 \cdot 10^{-11}$~\cite{Pospelov:2008zw} and $\delta a_\mu < 2.9 \cdot 10^{-9}$~\cite{Amsler:2008zzb}.  However, it should be noted that in the case of vector or pseudoscalar couplings, the $a_\mu$ bound should not be taken as a hard constraint.  We have chosen a limit such that the addition of the $X$ boson does not significantly change the agreement between experiment and theoretical predictions.  Yet with the currently observed muon magnetic anomaly the agreement between theory and experiment in $a_\mu$ actually improves for vector or pseudoscalar couplings just above the constraint.

If the couplings of the $X$ boson to leptons were proportional to the lepton masses, then the $a_\mu$ constraint would exclude the interesting parameter range for this study.  The reason is that the production rate for $X$ bosons is proportional to $\alpha_X^e$, but the constraint on $a_\mu$ is on $\alpha_X^\mu = \alpha_X^e  (m_\mu / m_e)^2$.  Thus, the effective bound on $\alpha_X^e$ is almost 5 orders of magnitude stronger than if $\alpha_X^e = \alpha_X^\mu$.  For this reason, we focus on the case with lepton universal couplings.

The formula for the anomalous magnetic moment appears in \appx{app:gminus2detail}, mirroring known results from Refs.~\cite{Sinha:1985aw,Leveille:1977rc}. Taking the limit $m_e \ll m_X$,
\begin{align}
  \delta a^e_{s/p} &= \frac{1}{16\pi^2}\frac{m_e^2}{m_X^2}
    \left(\lambda_s^2\left(\log\frac{m_e^2}{m_X^2} - \frac{7}{6}\right) - \lambda_p^2\left(\log\frac{m_e^2}{m_X^2} - \frac{11}{6}\right)\right), \\
  \delta a^e_{v/a} &= \frac{1}{16\pi^2}\frac{m_e^2}{m_X^2}
    \left(\lambda_v^2\frac{4}{3} - \lambda_a^2\frac{20}{3}\right) + \mathcal{O}\left(\frac{m_e^4}{m_X^4} \right).
\end{align}
The calculation in the muon case does not admit a simple approximate form since $m_\mu \approx m_X$ in the range under consideration, so one must use the full formula from the appendix.

The constraints on the coupling of the $X$ boson to electrons and muons are shown in \fig{fig:magmoment}, assuming $\alpha_X^e = \alpha_X^\mu$.   For $m_X \gtrsim 10 \MeV$, the $a_\mu$ bound dominates, giving roughly $\alpha_X \lesssim 10^{-7} \text{ -- } 10^{-8}$, except for the axial-vector case, for which the bound excludes much of the interesting parameter space for this study.  Thus, we see that the coupling of the $X$ boson to leptons must be 5 to 7 orders of magnitude weaker than the electromagnetic coupling $\alpha_{\rm EM} \simeq 1/137$.

\subsection{Direct Constraints}
\label{sec:beamdump}

The strongest direct constraints on the $X$ boson come from beam dump experiments.  As explained below, other direct constraints from rare meson decay modes, production from cosmic rays, or supernova cooling either fall outside the parameter space under consideration or are subsumed by the beam dump and anomalous magnetic moment constraints.

Despite the small coupling of the $X$ boson to electrons, it still has a sizable production rate in beam dump experiments because of their very high luminosity.  As long as the $X$ boson lifetime is sufficiently long, the decay of $X$ to electrons happens at finite displacement from the target.  Using the couplings in \eqs{eq:lagrangescalar}{eq:lagrangevector}, the $X$ boson width has been calculated in \appx{app:lifetimedetail}.  In the limit $m_e \ll m_X$, the width is linear with mass
\be
\Gamma_{s/p} = \frac{\alpha_X}{2}m_X + \mathcal{O}\left(\frac{m_e^2}{m_X^2} \right), \qquad
\Gamma_{v/a} = \frac{\alpha_X}{3}m_X  + \mathcal{O}\left(\frac{m_e^2}{m_X^2} \right).
\ee
Roughly, the beam dump constraints are relevant for lifetimes longer than $\sim 10^{-3}$ cm.

\FIGURE[t]{
  \centerline{\includegraphics[scale=0.7]{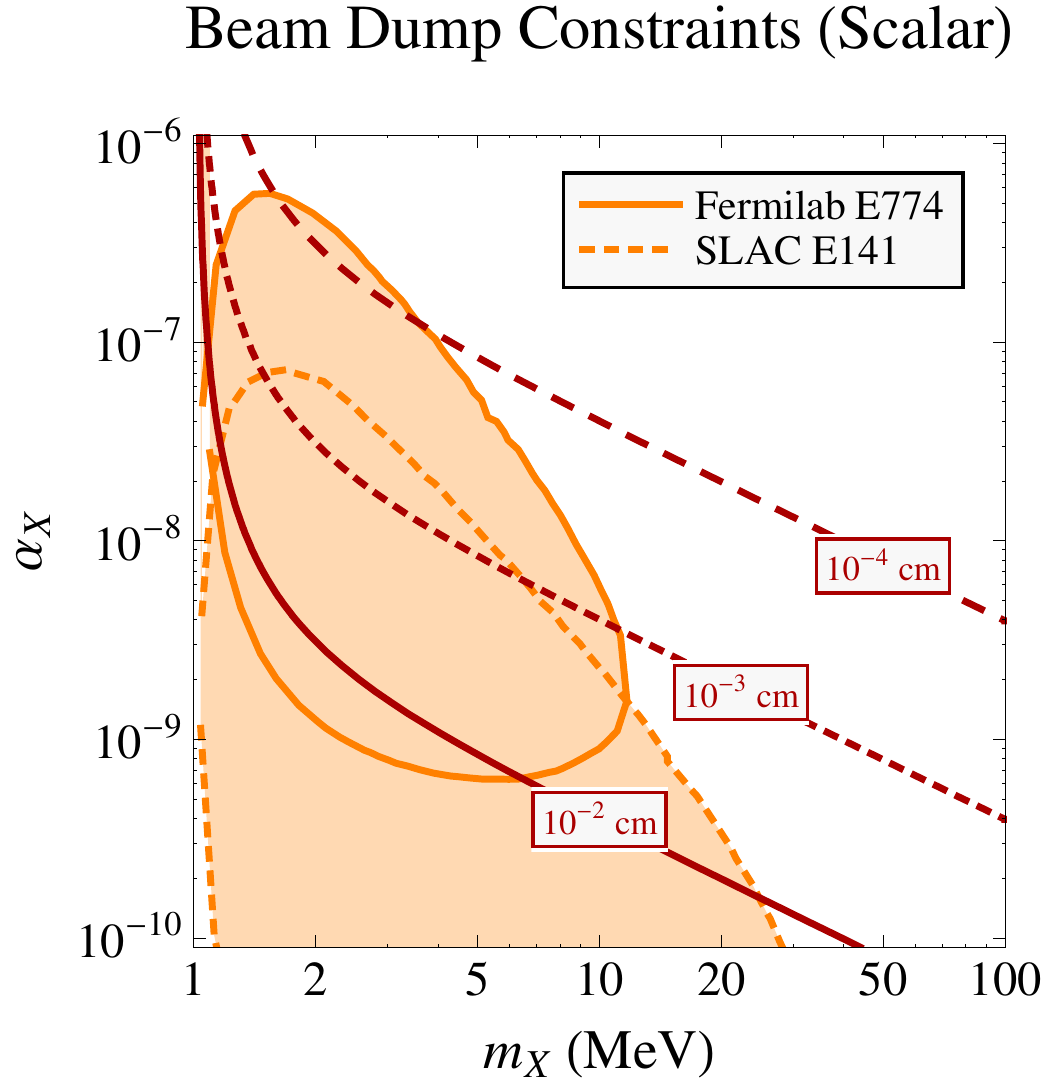} \includegraphics[scale=0.7]{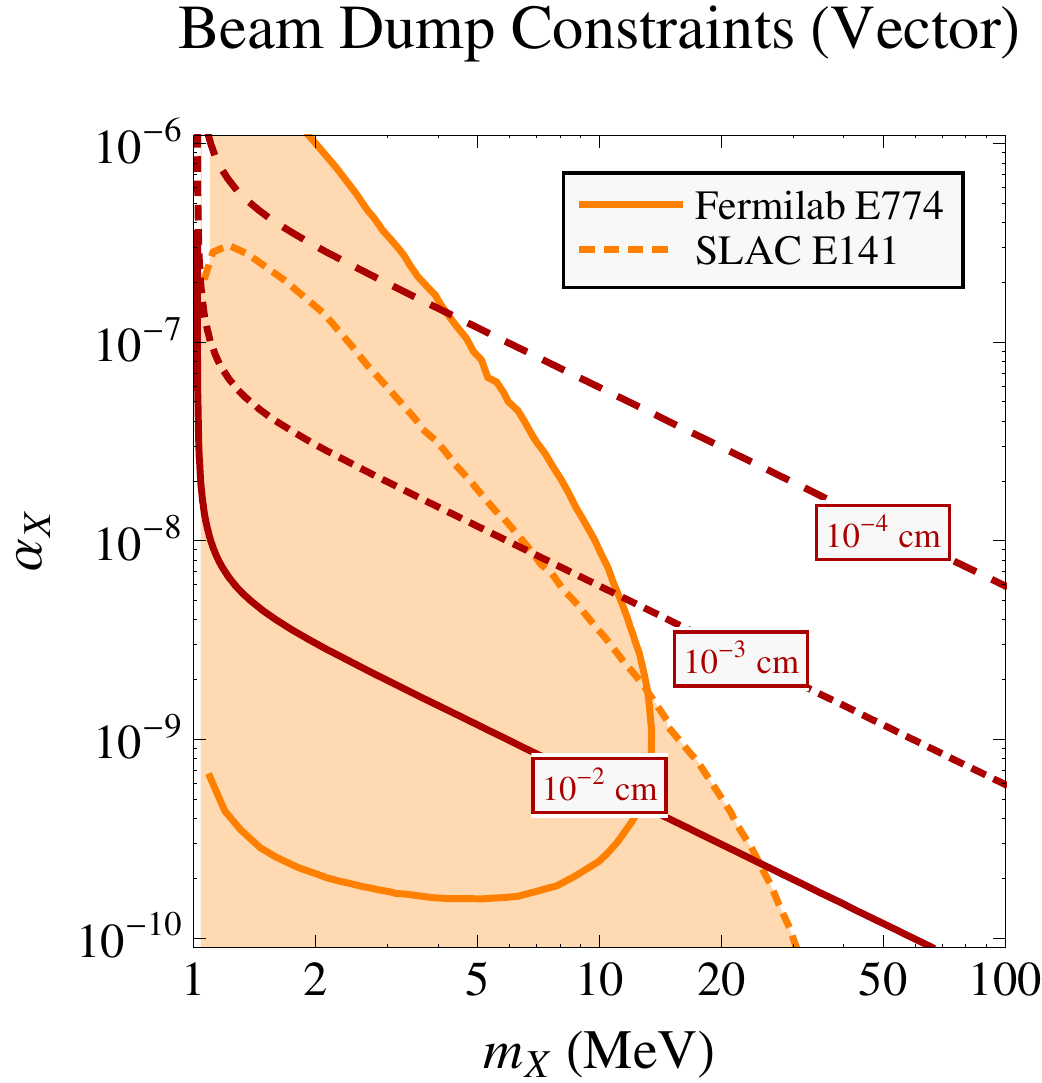}}
  \caption{Constraints on the $X$ boson mass $m_X$ and coupling $\alpha_X$ due to prior beam dump experiments, E774 at Fermilab and E141 at SLAC.   The red lines correspond to $X$ boson lifetimes.   The lifetime (and corresponding beam dump constraints) of the pseudoscalar and axial-vector are nearly identical to the scalar and vector, respectively, except in the very low mass region.}
  \label{fig:beamdump}
}

As recently discussed in Ref.~\cite{Bjorken:2009mm}, the two experiments relevant in the parameter space of interest are Fermilab's E774 and SLAC's E141 electron beam dumps, both originally searches for MeV-scale axions.  E774 consisted of a dump of $5.2 \cdot 10^9$ electrons at 275 GeV on a 19.6-cm tungsten target, with a 20-cm wide detector 7.25 meters away.  With the trigger requiring a decay product with energy of 2.75 GeV, 17 events would qualify for discovery \cite{Bross:1989mp}. The E141 beam dump consisted of a 9 GeV source with $2 \cdot 10^{15}$ electrons incident on a 12-cm tungsten target with a detector 35 m away down a 7.5-cm pipe. The trigger consisted of a decay dumping 4.5 GeV on the detector. Given the background, 1000 events would constitute discovery, with the greater number of events necessary due to the lack of veto counters on this experiment \cite{Riordan:1987aw}.

To determine the excluded parameter space, we mirror the analysis of the vector case from Ref.~\cite{Bjorken:2009mm} and extend it to the other $X$ boson couplings.  In particular, we use an approximate formula for the cross section and kinematics for $X$ boson production and decay in order to model the coherent nuclear effects involved in the beam dump experiments. The excluded regions are shown in \fig{fig:beamdump}, with $X$ boson lifetimes overlayed.  The upper diagonal boundary of the excluded region corresponds to when the $X$ boson lifetime is sufficiently short that $X$ decays to electrons within the shielding of the beam.  Naively, one would then expect the upper boundary to be along lines of constant lifetime. The discrepancy is due to the fact that lower-mass bosons are created with higher average Lorentz factors, allowing more potential $X$ boson events to be seen in the detector downstream past the shielding.  The lower boundary of the excluded region is determined by the rate of $X$ boson production.  If the typical $X$ boson decays past the detector, then this boundary would be approximately horizontal.  The reason is that for an $X$ decay length $\ell_X$ and distance to detector $L$, the fraction of decayed $X$ bosons, approximately $L/\ell_X$, cancels the $1/m_X$ mass dependence in the production rate.  When the decay length becomes less than the distance to the detector, the lower boundary rises at a diagonal.

Note that the beam dump constraints exclude any of the parameter space accessible in this search where $X$ is a very long-lived particle.  This is important, because if $X$ were too long-lived, then it would decay outside of a typical detector volume, and be inaccessible in the $ep$ scattering experiment under consideration.  Regions of moderate $X$ lifetime are still allowed, so the $X$ might decay with a displaced vertex in $ep$ scattering.  We will return to this possibility in \sec{sec:displacedpossibility}.

We can also consider direct constraints that rely on model-dependent assumptions on hadronic couplings. Most of these end up being either irrelevant or superfluous in the region of interest.  For example, $B$-factories provide constraints on the $X$ boson coupling constant though rare $\Upsilon$ decays, but such constraints currently only apply for $m_X \ge 2m_\mu$, and are thus beyond the range of interest~\cite{:2009cp}.   A recent BaBar analysis~\cite{Aubert:2009pw} does constrain dark force models, but relies on a model-dependent signature present only in non-Abelian dark sectors.  Some proton beam dumps, such as CHARM at CERN~\cite{Bergsma:1985qz} do cover some of the parameter space in question, but are already subsumed by the electron beam dumps.   There also exist bounds from the rare pion decay mode $\pi^0 \to e^+ e^-$~\cite{Kahn:2007ru}, which can place some additional constraints in a small region close to where the electron and muon anomalous magnetic moment bounds intersect. 

There are also potential bounds that end up being irrelevant since the $X$ boson is short-lived in the parameter space of interest.  For example, $X$ bosons could potentially be observed due to cosmic rays interacting with the earth, such that $X$ bosons could be seen at detectors such as AMANDA and ICE-CUBE.  However, these experiments rule out couplings that are smaller (equivalently, lifetimes that are longer) than those we are trying to observe here \cite{Bjorken:2009mm}.  Additionally, neutrino searches such as LSND and MiniBooNE can also be used to place constraints on MeV scale bosons, but again at much lower couplings \cite{Batell:2009di}.  Another constraint comes from supernovae, where $X$ boson production could lead to additional cooling of the core. However, the $X$ boson would have to travel at least $\mathcal{O}(10\text{ km})$ in order to escape the core. Thus, such constraints require lifetimes for the boson to be several orders of magnitude longer than those considered in this search.  Detailed calculations done for the axion case in Ref.~\cite{Turner:1987by} agree with this rough estimate. 

Assuming hadronic couplings, strong bounds on the $X$ boson might be obtained from data mining the extant pion decay data set from the KTeV collaboration.\footnote{We thank Maxim Pospelov for bringing this possibility to our attention.} Over the course of two runs, approximately $1.8\cdot 10^6$ decays of $\pi^0 \to e^+ e^- \gamma$ with invariant mass $m_{e^+ e^-} > 65\MeV$ have been reconstructed~\cite{Abouzaid:2006kk}. Assuming this can be extended to the whole invariant mass range, this should give a data set of close to $6\cdot 10^7$ decays.  One can then look for an $X$ boson by looking for a small peak in the $m_{e^+ e^-}$ spectrum, in a search much like the $e^+ e^- \to e^+ e^- \gamma$ search described later in \sec{sec:compexp}.  Using an approximate formula for the reach from Ref.~\cite{Reece:2009un}, a search looking for $X$ bosons should be able to detect couplings as low as $\alpha_X \sim \text{few} \times 10^{-8}$, an improvement on current bounds, but not matching the reach for the search proposed in this paper.

\section{Electron-Proton Scattering Below the Pion Threshold}
\label{sec:reach}

\FIGURE[t]{
\centerline{\includegraphics[scale=0.6]{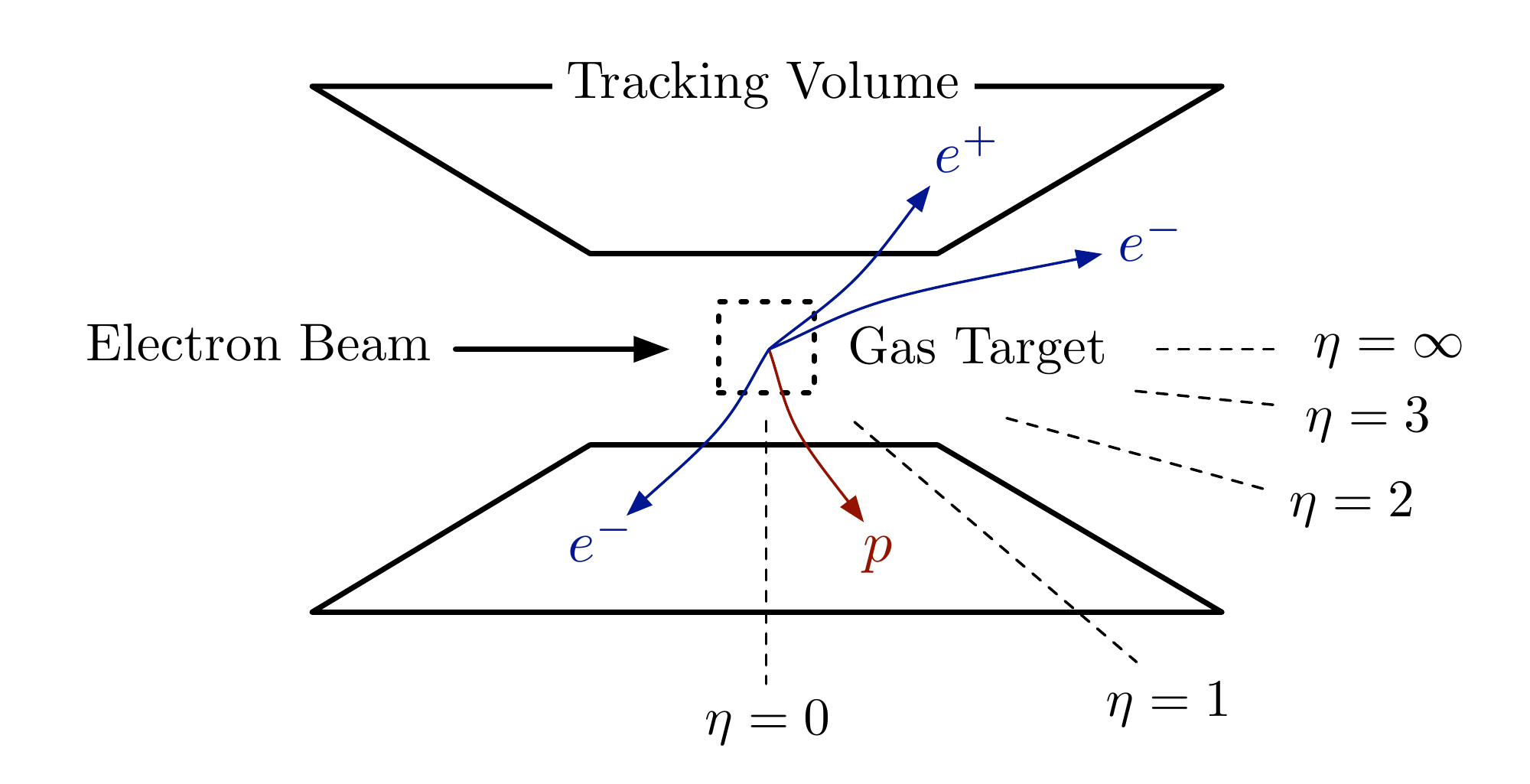}}
\caption{The envisioned experimental setup to probe $e^-p \to e^-p \, e^+e^-$ scattering, where a high intensity electron beam is incident on a diffuse hydrogen gas target.  In order to reconstruct all four outgoing fermions, a large tracking volume is needed, and we will consider $-2 < \eta < 2$ coverage as a benchmark.  While this is a fixed target experiment where most of the tracks go in the forward direction, we take a detector symmetric around $\eta = 0$ for simplicity.  For reference, $\eta = 1$ is an angle to the beamline of $\theta = 40.4^\circ$, $\eta = 2$ is $\theta = 15.4^\circ$, $\eta = 3$ is $\theta = 5.7^\circ$, and $\eta = 4$ is $\theta = 2.1^\circ$.}
\label{fig:detector}
}

We now describe a promising venue for a light $X$ boson search: low-energy $ep$ scattering.   Despite the above constraints on $\alpha_X$, the cross section for $X$ boson production in $ep \to ep+X$ is quite large, on the order of picobarns.  The $X$ would then decay promptly to $e^+ e^-$, yielding an $e^-p \,e^+e^-$ final state where the electron/positron pair reconstructs a narrow $X$ resonance.  The envisioned experimental setup is sketched in \fig{fig:detector}, where a high intensity electron beam is incident on a hydrogen gas target.  Assuming a sufficiently large tracking volume, all four outgoing fermions can in principle be reconstructed.

As we will see, the primary challenge for this search is the large irreducible QED background, roughly four orders of magnitude larger than the signal, making this a background limited search.  In this section, we summarize the signal and background calculations and present the reach for the $X$ boson, first using simple cuts on the detector geometry and then including full information about the event kinematics through a matrix element method.

\subsection{Cross Section Calculations}

In the range of couplings allowed by the anomalous magnetic moment bounds, the width of the $X$ boson is an eV or smaller.  Therefore, to calculate the signal rate for $e^-p \to e^-p \, e^+e^-$, we can safely use the narrow width approximation.  Since the angular distribution of the decay $X \rightarrow e^+ e^-$ is relevant for understanding the vector couplings, we maintain full polarization information in the signal process $e^-p \to e^-p \, e^+e^-$ as explained in \appx{sec:signal}.   We used a custom matrix element/phase space generator to calculate the signal cross sections, and checked the results with \texttt{CompHEP 4.5.1}~\cite{Boos:2009un}.  In particular, we used \texttt{CompHEP} to verify that any interference between the signal and background processes is a subdominant effect given the narrowness of the $X$ boson.

For the couplings in \eqs{eq:lagrangescalar}{eq:lagrangevector}, the diagrams that contribute to the signal cross section appear in \fig{fig:signal2t3}.   In the limit that $m_e \rightarrow 0$, the cross section for the pseudoscalar (axial-vector) case is identical to the scalar (vector) case.  While we keep finite $m_e$ effects in our calculations, we will only show reaches for the scalar and vector cases, since the finite $m_e$ effects are small.  While there could be contributions to the signal from $X$ boson couplings to the proton, we argue in \appx{sec:generalize} that such effects can be ignored.  In reconstructing the $X$ resonance, there is combinatoric confusion about which electron to pair with the positron, and this confusion is included in our plots.  We neglect the electromagnetic form factor of the proton, which is a fair approximation since we are considering incoming electron energies $E_e \ll m_{\rm proton}$.  

\FIGURE[t]{
\centerline{\includegraphics[scale=0.4]{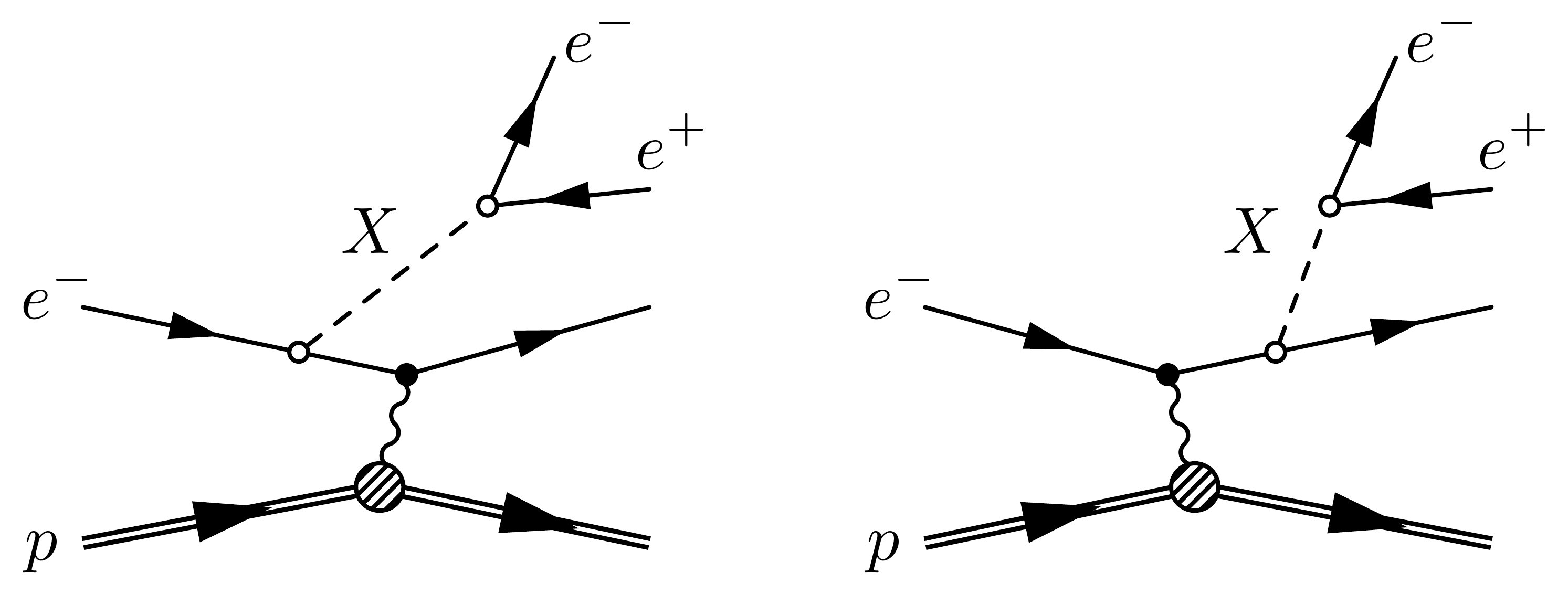}}
\caption{Diagrams contributing to the $X$ boson signal.  Here, the $X$ boson propagator is evaluated in the narrow-width approximation.}
\label{fig:signal2t3}
}
 
The background to $e^-p \to e^-p\, e^+e^-$ is due to QED radiative processes $e^-p \to e^-p + \gamma^*$ with $\gamma^* \rightarrow e^+ e^-$ and to the Bethe-Heitler trident process, shown in \fig{fig:background}.  Details of these backgrounds appear in \appx{sec:background}, where we again ignore the proton form factor.    We calculated the background cross sections using a custom phase space generator interfaced with the stand-alone version of \texttt{MadGraph 4.4.17}~\cite{Alwall:2007st}, and checked the results using \texttt{CompHEP}.

\FIGURE[t]{
\centerline{\includegraphics[scale=0.35]{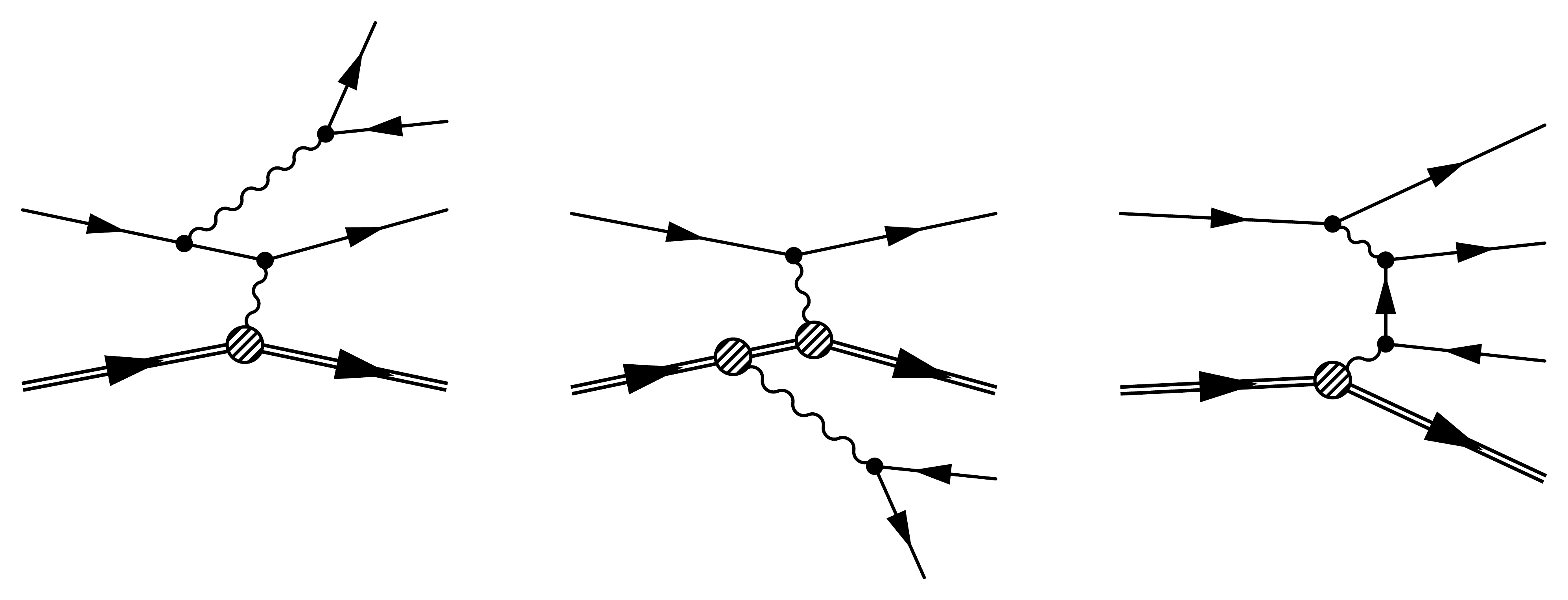} }
{\caption{Representative diagrams contributing to the radiative QED background.  With indistinguishable outgoing electrons, there are 12 diagrams in total, consisting of $\gamma^*$ emission off the incoming/outgoing electron/proton lines and the Bethe-Heitler process.  In this study, we ignore the electromagnetic form factor of the proton, which contributes at most a 5\% correction.}
\label{fig:background} }
}

Because we are considering $m_X \sim E_e$, thinking of the $\gamma^*$ as coming from initial or final state radiation is not a good approximation to the background process.   In the region of phase space where an $e^+ e^-$ pair fakes an $X$ resonance, the photon is far off-shell relative to the energy scales involved, so we are far away from soft-collinear limit.  In particular, it should be noted that considerable (constructive) interference increases the background above naive expectations from the Weizs\"{a}cker-Williams approximation.  To give a sense of how important this interference is, one sees changes on the order of $10\%$ in cross sections with $e^- p$ collisions versus $e^- \bar{p}$ collisions, whereas the sign of the proton charge would be irrelevant in the Weizs\"{a}cker-Williams picture.

In principle, there is another background we should consider.  Since the proposed JLab FEL experiment is really an electron-hydrogen gas collider, one might be concerned about backgrounds from $e^-e^-$ collisions.  In fact, this is only an issue for very low values of $m_X$.  For an electron beam with energy $100 \MeV$, the center-of-mass energy of $e^-e^-$ collisions is around $10 \MeV$, so radiative M{\o}ller scattering is only relevant for $m_X \lesssim 10 \MeV$.  We saw in \sec{sec:beamdump} that such light bosons are already ruled by direct constraints.  Moreover, given the fact that we imagine using the recoiling proton and electron momentum as a handle on the collision process, particle identification should be sufficiently robust to distinguish $e^-e^-$ from $ep$ collisions.  Finally, there are important experimental backgrounds, including event pileup and photon conversion, which we do not address in this study.

\subsection{Resonance Reach}
\label{sec:rawreach}

\FIGURE[t]{
\centerline{\includegraphics[scale=0.7]{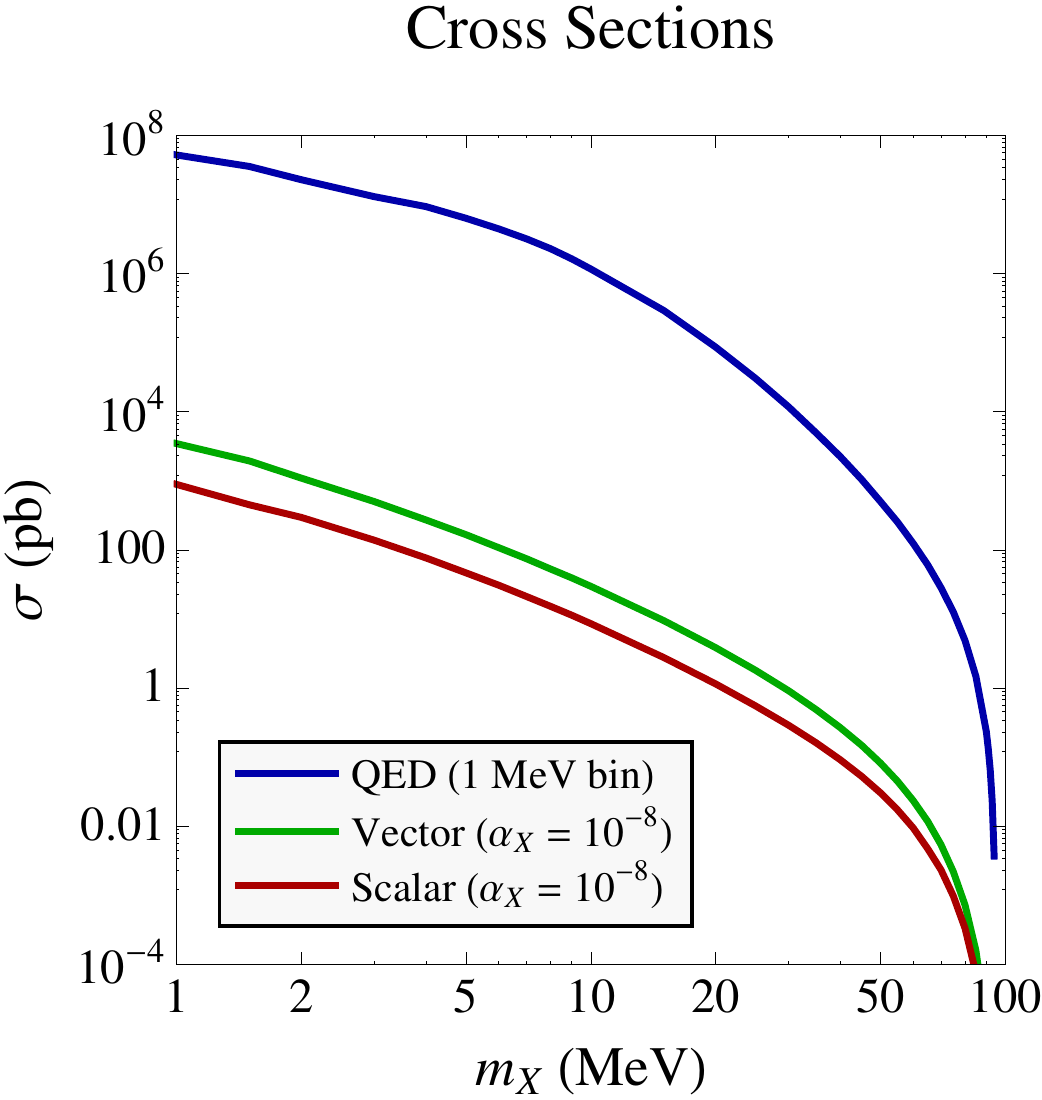}}
\caption{Cross section for the $X$ boson signal with  $\alpha_X = 10^{-8}$ and the QED background imposing that at least one $e^+e^-$ pair is in the $m_X \pm 0.5 \MeV$ invariant mass bin.  The signal cross section scales linearly with $\alpha_X$.  Generically, the expected signal is four orders of magnitude smaller than the background.}
\label{fig:rawxsec}
}

Since the $X$ boson is very narrow, with infinitely good mass resolution and the relatively large $X$ boson production rate, one could easily produce the few events required in the same $e^+e^-$ invariant mass bin to conclude the existence of $X$.  In practice, though, one must take into account finite experimental resolution.  In \fig{fig:rawxsec}, we plot the signal cross sections for fixed $\alpha_X = 10^{-8}$ as a function of $m_X$, and compare it to the QED background, imposing a cut that at least one $e^+e^-$ pair within a 1 MeV mass bin around $m_X$.  One can see that the signal cross section is in the range $10^{-2}$ to $10^2$ pb, but the background size is generically four orders of magnitude large than signal.   Thus, one will need a very large integrated luminosity to establish the signal over statistical fluctuations in the background.  

To assess the reach of experiment precisely, one would need to know the true resolution, efficiency, and dimensions of the detector.  As a rough approximation to the detector geometry, we assume full azimuthal coverage, and consider a detector with pseudorapidity coverage of  
\be
\label{eq:etacuts}
-2 < \eta < 2, \qquad \eta \equiv - \ln \left( \tan \frac{\theta}{2}  \right)
\ee
 (i.e.\ a tracking system that covers angles as close as $15.4^\circ$ to the beam line).  As we will see later in \fig{fig:reachsweepeta}, a more aggressive $-3 < \eta < 3$ coverage (i.e.\ tracking up to $5.7^\circ$) actually has a comparable reach. We also impose a constraint on the kinetic energy of the outgoing particles:
 \be
 \label{eq;kineticenergycuts}
 \mathrm{KE}_p > 0.5 \MeV, \qquad \mathrm{KE}_{e^\pm} > 5 \MeV.
 \ee
 While we assume the detector is symmetric about $\eta = 0$, recall that we are considering a fixed target geometry, so the tracks dominantly appear in the forward part of the detector.  Since the QED background has a large forward peak, the pseudorapidity restriction does improve the signal to background ratio compared to what is shown in \fig{fig:rawxsec}.  We only keep events where all four outgoing fermions are contained in the tracking volume.

As a baseline, we assume that the $e^+e^-$ invariant mass resolution is 1 MeV of the target $m_X$ value.  We then calculate the value of $\alpha_X$ such that for a given luminosity, one can achieve a $5\sigma$ discovery with statistical uncertainties alone, meaning we find where $S/\sqrt{B} = 5$ in a 1 MeV mass bin centered on a candidate $X$ mass.  Since the background is relatively smooth over the kinematic range of interest, the required luminosity for discovery scales inversely with the mass resolution:
\be
\mathcal{L}(x \text{ MeV resolution}) = \frac{1}{x} \mathcal{L}(1 \text{ MeV resolution}).
\ee  
The reach for 1 MeV resolution was shown above in \fig{fig:execsummary}, taking $3 \MeV < m_X < 100 \MeV$.   

\FIGURE[t]{
\centerline{\includegraphics[scale=0.7]{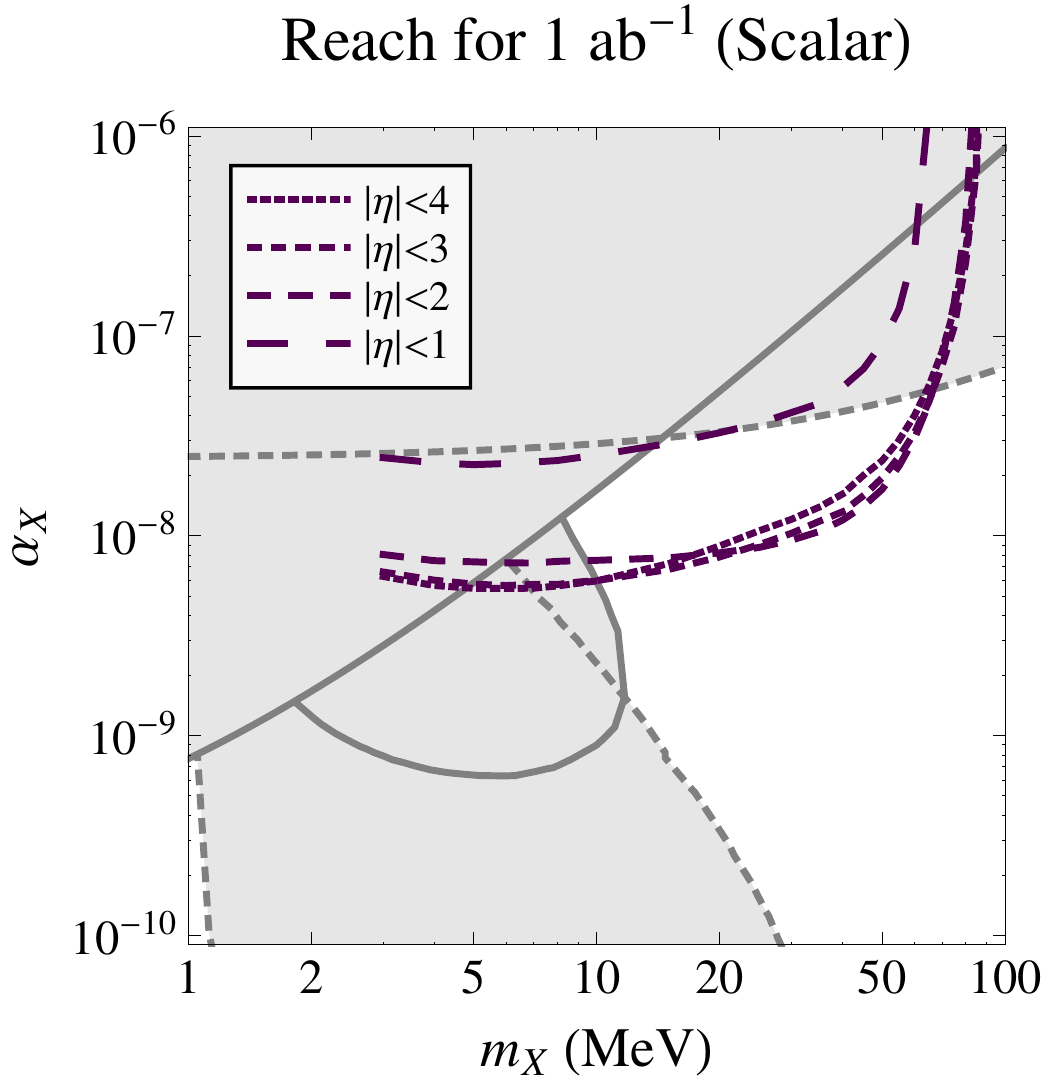} \includegraphics[scale=0.7]{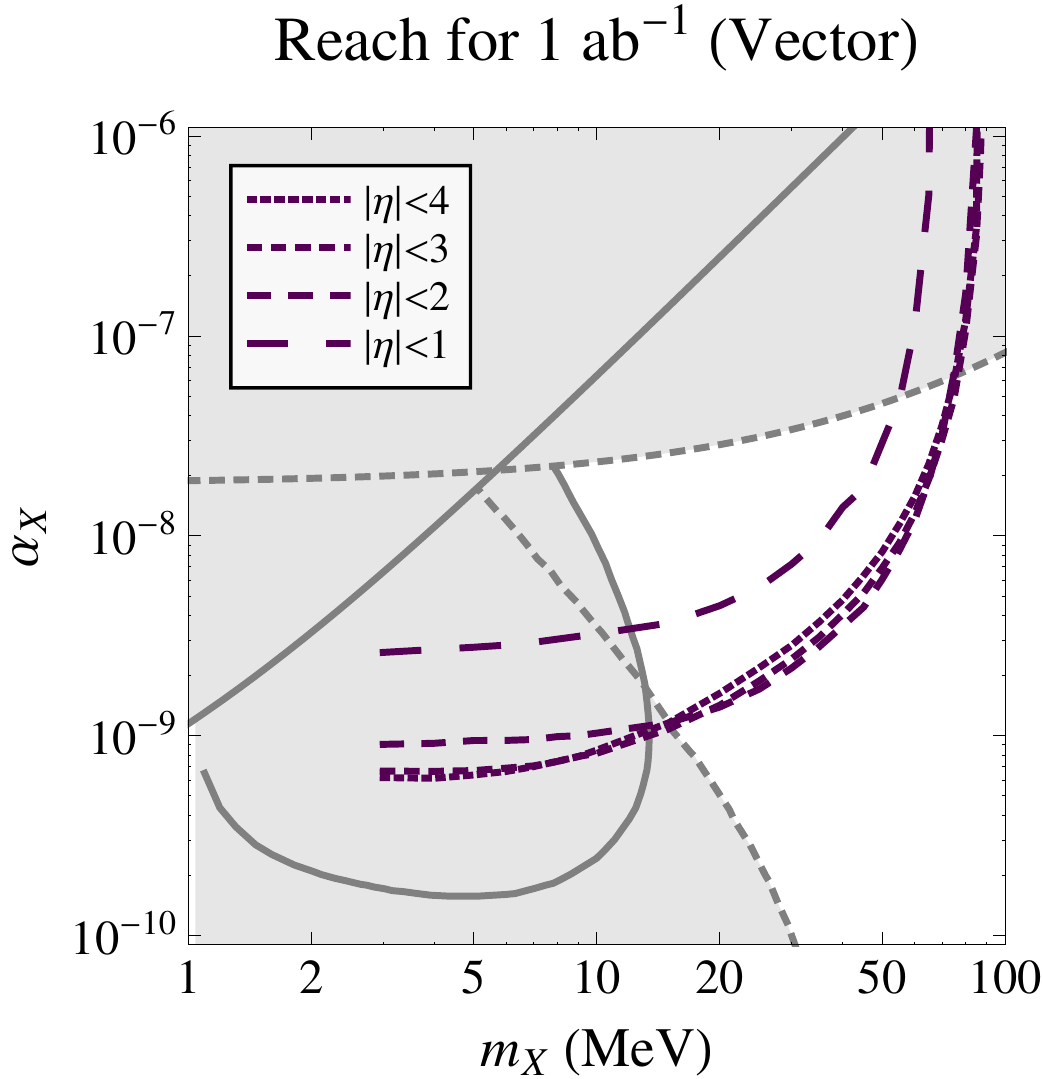}}
\caption{Reach plots for variable detector angular acceptance $\eta$.  The grey shaded regions correspond to the indirect and direct constraints from \sec{sec:theory}.  In all cases, we impose the criteria $\mathrm{KE}_p > 0.5 \MeV$ and $\mathrm{KE}_{e^\pm} > 5 \MeV$.  We again take an integrated luminosity of 1 ab$^{-1}$ and assume $1 \MeV$ $m_{e^+e^-}$ resolution.   For small values of $\eta$, the reach deteriorates because the signal efficiency decreases.  For large values of $\eta$, the reach deteriorates because the background has a large forward peak.
\label{fig:reachsweepeta}}}

In \fig{fig:reachsweepeta}, we show how the reach changes as the angular acceptance is varied.  Going from $|\eta| < 2$ to $|\eta| < 3$ slightly improves the reach for smaller values of $m_X$, though the effect is mild.  As we will discuss more below, the reason the $|\eta| < 2$ geometry is so effective is that by cutting out the phase space close to the beam line, we decrease the background rate without sacrificing much on signal acceptance. 

This $S/\sqrt{B}$ procedure to establish reach is only a crude estimate of the true sensitivity.  In practice, the actual background distribution would have to be fit from the data using some kind of sidebanding procedure (see \sec{sec:kinedist}), and one also must pay a trials factor in looking for an invariant mass peak since the $X$ boson could be anywhere.  With those caveats, we see that with 1 ab$^{-1}$ of data, one begins to probe the interesting parameter regime for the $X$ boson.

\subsection{Reach with Matrix Element Method}
\label{sec:mereach}

The reach plots in \fig{fig:execsummary} do not include any kinematic cuts apart from the $-2 < \eta < 2$ cut on detector acceptance and the kinetic energy restriction from \eq{eq;kineticenergycuts}.    As we will see in \sec{sec:kinedist}, the kinematic distributions for signal events do differ from the background, so one might hope that a set of optimized kinematic cuts might improve the reach for the $X$ boson.  Here we show that a factor of 3 improvement in the reach is in principle possible by using complete kinematic information via a matrix element method~\cite{Abulencia:2005uq,Abazov:2004cs}.  

The matrix element method is often described in terms of a discriminant function~\cite{Abazov:2006bd}, but the essential statistics can be understood by considering a weighted measurement.  For a very narrow resonance $X$, the signal and background matrix elements for $e^-p \to e^-p\,e^+e^-$ are essentially functions of $\tilde{\Phi}_4$, which is the four-body final state phase space $\Phi_4$ with an additional restriction that one of the electron/positron pairs reconstructs a given value of $m_X$.  For simplicity, we will use the notation $\Phi$ to refer to $\tilde{\Phi}_4$.  

For a differential signal cross section times luminosity $S(\Phi)$ and differential background times luminosity $B(\Phi)$, the naive reach calculation is equivalent to integrating over all of $\Phi$ with unit weight:
\be
S = \int \df \Phi \,  S(\Phi), \qquad B = \int \df \Phi \,  B(\Phi).
\ee
The reach is determined by calculating $S/\delta B = S/\sqrt{B}$, where $\delta$ refers to the statistical uncertainty in the measurement.    Now consider a weighted measurement
\be
S_{\rm eff} = \int \df \Phi \,  S(\Phi) w(\Phi), \qquad B_{\rm eff} = \int \df \Phi \,  B(\Phi) w(\Phi),
\ee 
where $w(\Phi)$ is some weight function.  For example, a weight function corresponding to hard kinematic cuts is one where $w(\Phi)$ equals either 0 or 1.  However, more general weight functions still give well-defined measurements.  

\FIGURE[t]{
\centerline{\includegraphics[scale=0.7]{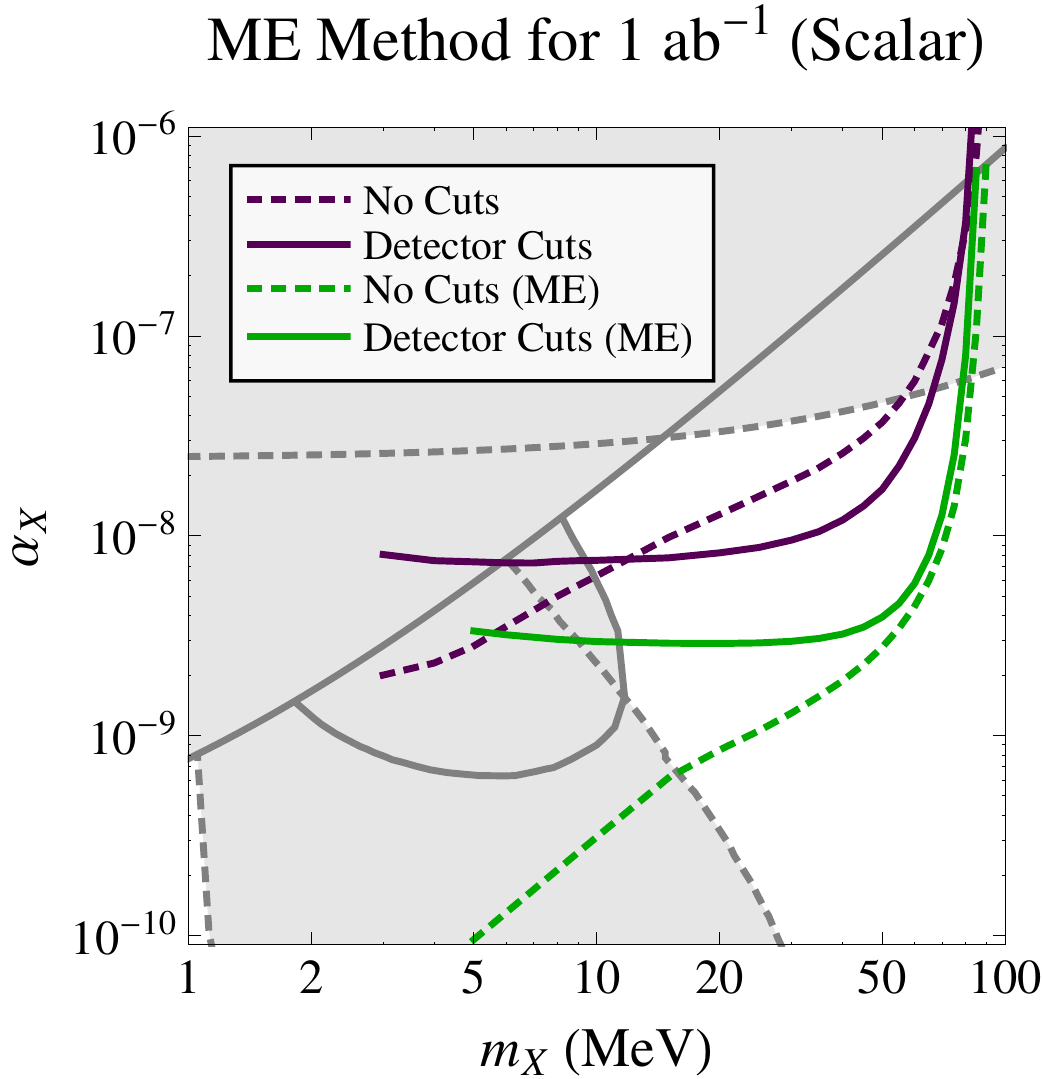} \includegraphics[scale=0.7]{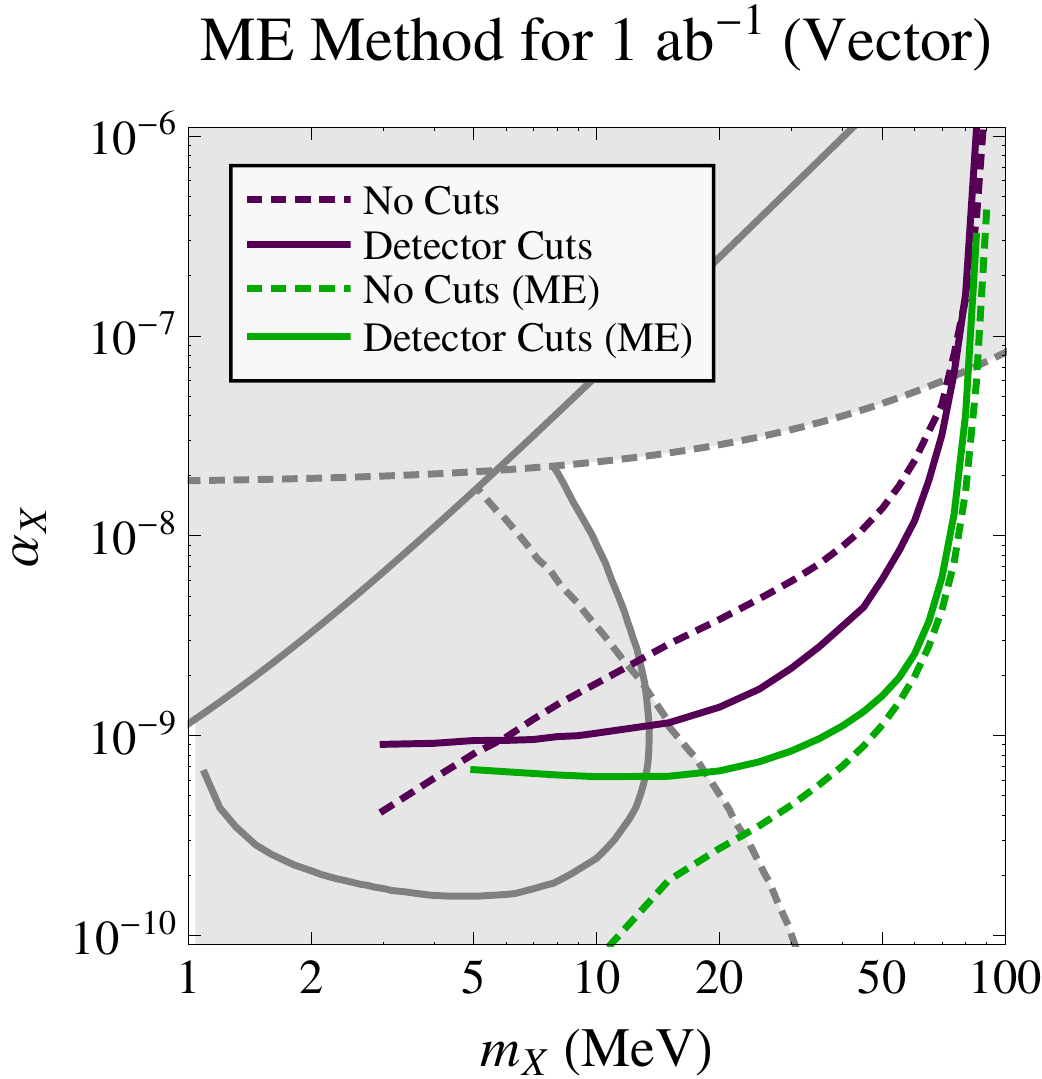}}
\caption{Reach plots using the matrix element method.  The solid curves include the detector acceptance cuts $-2 < \eta < 2$, $\mathrm{KE}_p > 0.5 \MeV$, and $\mathrm{KE}_{e^\pm} > 5 \MeV$, while the dashed lines have no acceptance cuts.  The green curves indicate the reach with the matrix element method, and the purple curves without.  In all cases, we take an integrated luminosity of 1 ab$^{-1}$, and the reach corresponds to $S/\sqrt{B} = 5$ assuming $m_{e^+ e^-}$ resolution of $1 \MeV$.  The matrix element method can yield around a factor of 3 improvement in the reach.  Note also that the detector geometry alone does act like a primitive matrix element method by cutting out the forward region.
\label{fig:reachwithME}}}

The matrix element method calculates the optimal kinematic observable to discriminate signal from background, which corresponds to choosing the optimal function $w(\Phi)$ that maximizes $S_{\rm eff}/\delta B_{\rm eff}$.  As derived in \appx{sec:memethod}, the ideal weighting function is
\be
\label{eq:bestweight}
w_{\rm best}(\Phi) = \frac{S(\Phi)}{B(\Phi)},
\ee 
which yields
\be
\left[\frac{S_{\rm eff}}{\delta B_{\rm eff}} \right]_{\rm best} = \sqrt{\int \df \Phi \, \frac{S(\Phi)^2}{B(\Phi)}}.
\ee
We can therefore recalculate the reach for the $X$ boson using this ideal value for $S_{\rm eff}/\delta B_{\rm eff}$, and the results are shown in \fig{fig:reachwithME}.  As advertised, there is potential factor of 3 improvement in the reach by using the full kinematic information in the signal and background distributions.

Of course, the matrix element method assumes that the $w_{\rm best}(\Phi)$ function is calculated using the true signal and background distributions, and this is not possible in practice, due to both theoretical uncertainties in the matrix elements and detector effects.  Still, one might still hope to improve the reach by doing hard kinematic cuts that approximate $w_{\rm best}(\Phi)$.  As an example of this, consider \fig{fig:reachwithME}. There one sees noticeable improvement in $S/\sqrt{B}$ just from applying the fiducial detector geometry. In this way, the detector geometry does act like a primitive $w(\Phi)$.  In \sec{sec:kinedist}, we will look at $w_{\rm best}(\Phi)$ in more detail to see what other kinds of hard cuts could be most helpful in teasing out the signal. In principle, by using a polarized electron beam, one could obtain additional information from the full $ep \to ep + X$ matrix element, but we will not consider polarized beams in this paper.

\section{Comparison to Other Searches}
\label{sec:compexp}

We argued that low energy $ep$ scattering with at least 1 ab$^{-1}$ of data was a promising venue for looking for a light, weakly coupled $X$ boson.  Unlike beam dump experiments where the proton recoil spectrum is not measurable, a high intensity electron beam on a diffuse gas target allows for full event reconstruction.  

However, $ep$ scattering is certainly not the unique choice of experiment with full reconstruction potential, and electron-electron scattering or electron-position scattering also have large $X$ boson production rates.  Here, we will argue that for the same integrated luminosity, $ep$, $e^-e^-$, and $e^+e^-$ colliders all offer comparable search power.  But given the very high instantaneous luminosity achievable at the JLab FEL, we believe that $ep$ scattering is favored for $X$ bosons in the range $10 \MeV - 100 \MeV$.

Consider the following four scattering processes:
\begin{align}
e^- p \to e^- p + X&  \quad (\mbox{fixed target}),\nn\\
e^- e^- \to e^- e^- + X& \quad (\mbox{colliding beams}),\nn \\
e^+ e^- \to e^+ e^- + X& \quad (\mbox{colliding beams}),\nn \\
e^+ e^- \to \gamma + X& \quad (\mbox{colliding beams}).
\end {align}
The first one is the $ep$ scattering experiment in this paper, the next two are the equivalent processes for $e^- e^-$ and $e^+ e^-$ colliders, and the final search channel is only available for $e^+ e^-$.  The distinction between ``fixed target'' and ``colliding beams'' is only needed to determine the relation between the lab frame and the center-of-mass frame of the experiment.  We assume the same detector technology for all four experiments, with pseudorapidity coverage in the lab frame of $-2 < \eta < 2$, the kinetic energy restriction from \eq{eq;kineticenergycuts}, and 1 MeV invariant mass resolution. For the $e^+ e^- \to \gamma + X$ search, one would also need to impose a cutoff on photon energy, but as the photons are monochromatic for a given beam energy, this would merely correspond to no reach at all once the cutoff is reached and is thus not shown.

In \fig{fig:reachcomp}, we show the reach in $\alpha_X$  as a function of the available beam energy
\be
E_{\rm eff} = \sqrt{s} - m_1 - m_2,
\ee
where $\sqrt{s}$ is the center-of-mass energy of the collider and $m_i$ are the beam masses.  (For $ep$ scattering with a $100 \MeV$ electron beam, $E_{\rm eff} \simeq 95 \MeV$.)  We take a fixed $m_X = 50 \MeV$ and fixed integrated luminosity $\mathcal{L} = 1 \ab^{-1}$.  The signal and background cross sections were calculated using the same method as in \sec{sec:reach}.

\FIGURE[t]{
  \centerline{\includegraphics[scale=0.7]{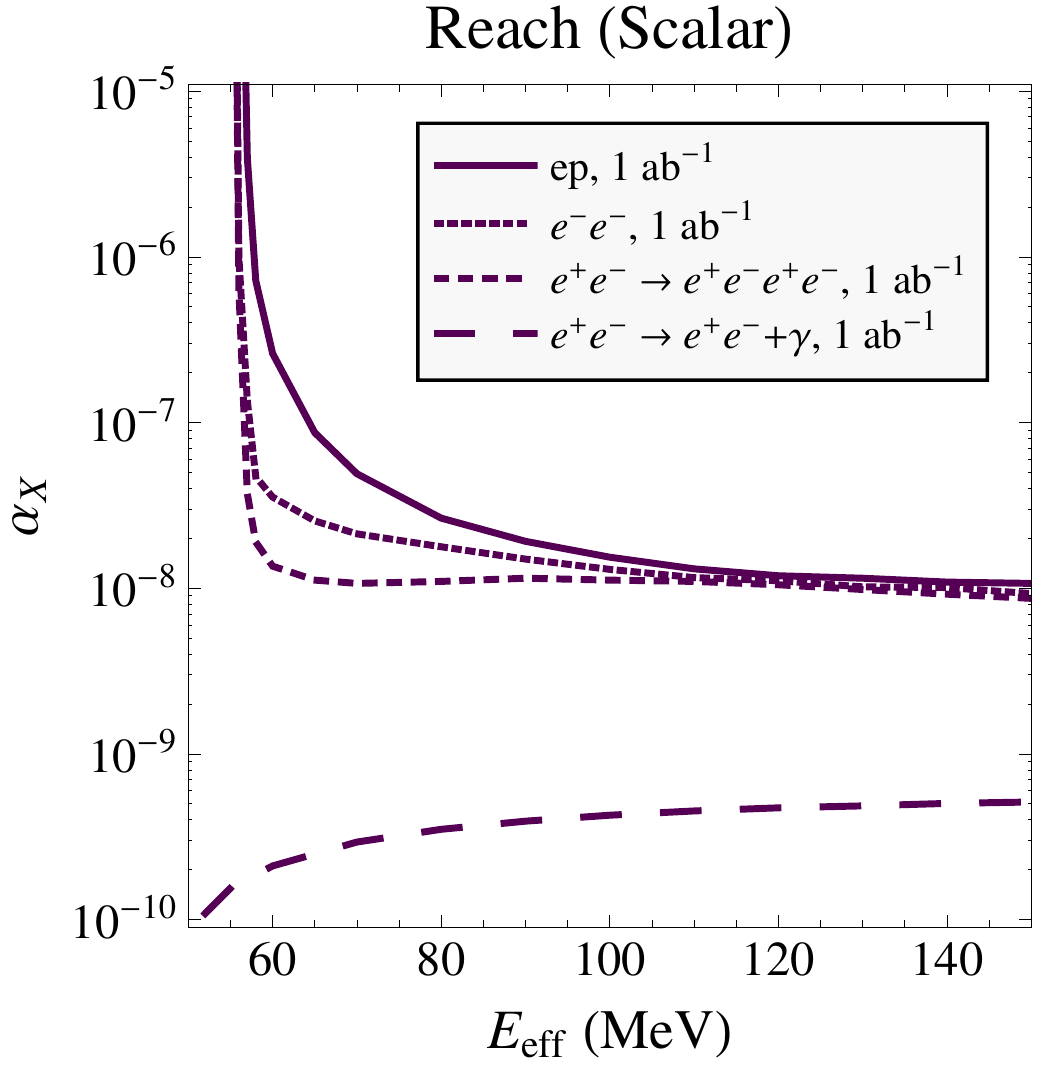} \includegraphics[scale=0.7]{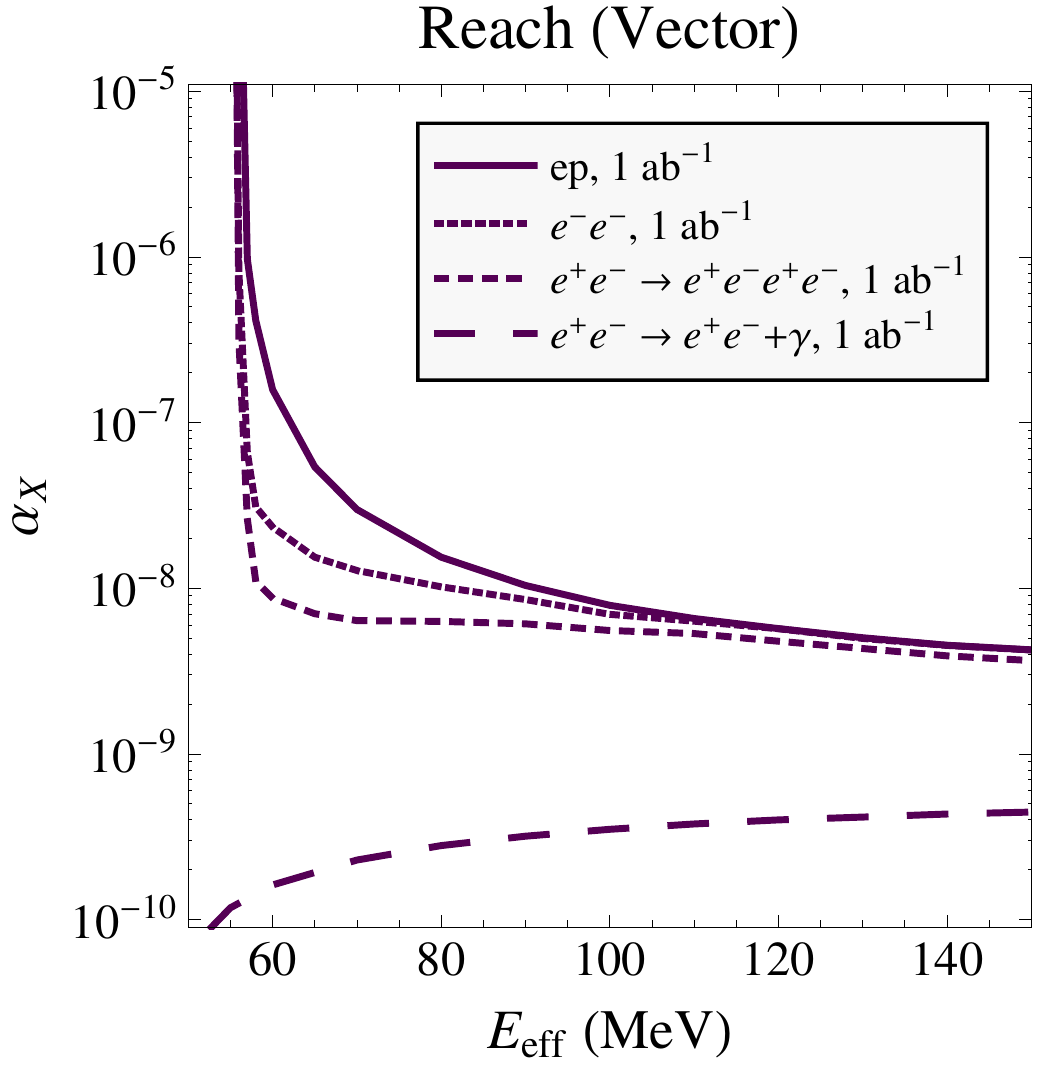}}
  \caption{Reach for an 50 MeV $X$ boson, varying $E_\text{eff} = \sqrt{s} - m_1 - m_2$.  We compare the present $ep$ scattering proposal with three alternative searches: $e^- e^- \rightarrow e^-e^- + X$, $e^+ e^- \rightarrow e^+e^- + X$, and $e^+ e^- \rightarrow \gamma  + X$.  In all cases, we assume a detector acceptance of $-2 < \eta < 2$, $\mathrm{KE}_p > 0.5 \MeV$, and $\mathrm{KE}_{e^\pm} > 5 \MeV$.  While the in principle reach in $e^+ e^- \rightarrow \gamma + X$ appears to be better than the other three, we discuss the challenges of that search in the text.  For the same integrated luminosity and large enough $E_\text{eff}$, the three searches with four outgoing fermions have comparable reach, which favors the $ep$ scattering proposal where high luminosity is more readily achievable.
}
  \label{fig:reachcomp}}

Given the same integrated luminosity and high enough values of $E_\text{eff}$, the reach for the searches involving four final state fermions are within a factor of 2 of each other.  However, getting 1 ab$^{-1}$ of data in the proposed FEL experiment requires only 1 month of data taking, while the maximal luminosity currently achieved in colliding beam experiments is $1.7\cdot 10^{34}~\mathrm{cm}^{-2}~\mathrm{s}^{-1} \simeq 0.5 \ab^{-1} / \mathrm{yr}$.  Since high luminosity is critical to probe the parameter space of interest, this favors a fixed target experiment for the four fermion final states.  For the same luminosity, the $e^+ e^- \rightarrow \gamma + X$ search is 1 to 2 orders of magnitude more sensitive than the $ep$ search, and we will comment more on this search below.

It should be noted that existing collider experiments, such as BaBar and Belle, already have data sets with integrated luminosities of $\sim 1 \ab^{-1}$.  However, the collisions there occur at much higher energies. At those energies and for the same search strategy, many additional backgrounds, both reducible and irreducible, are present, increasing the difficulty of the analysis.  In fact, studies of these detectors typically focus on the case with $m_X > 2m_\mu$, as the decay of the $X$ boson to muons is more easily reconstructed than the decay to electrons~\cite{Essig:2009nc,:2009cp}.

On the other hand, even at the higher energies of existing data sets, the $e^+e^- \to \gamma + X$ channel does look quite promising for an $X$ boson search due to the lack of hadronic backgrounds.  That said, two factors lead this channel to be more complimentary than competitive with the search we propose.  Our naive estimate of the $\gamma + X$ reach assumed perfect reconstruction of every event meeting the detector geometry cuts.  This is significantly more difficult with a search using on-shell photons as there is no longer tracking information for every particle.  In order to identify the single energetic photon in the event with high accuracy, it is necessary to put tighter cuts on the photon angle to get farther away from the beam pipe.  Estimates in Ref.~\cite{Borodatchenkova:2005ct} indicate that when such cuts are put in place, the reach in coupling actually becomes comparable to that of the $ep$ search.  Additionally, the search for $\gamma + X$ is complicated by photon conversion in the tracking volume from the much larger $e^+ e^- \to \gamma \gamma$ process. While this can be offset by cutting on displaced vertices, such an approach becomes difficult for $e^+ e^-$ mass bins below 50 MeV, and impossible in mass bins below 20 MeV~\cite{Kolomensky:2009--}.  For $X$ bosons above 50 MeV, one should be able to extend the $\gamma + X$ search already done for $X \to \mu^+\mu^-$~\cite{Aubert:2009pw} to $X \to e^+ e^-$, making the search complementary to $ep$ scattering by filling in the mass range between $\sim 50\MeV$ and $2m_\mu$.

\FIGURE[t]{
\centerline{\includegraphics[scale=0.7]{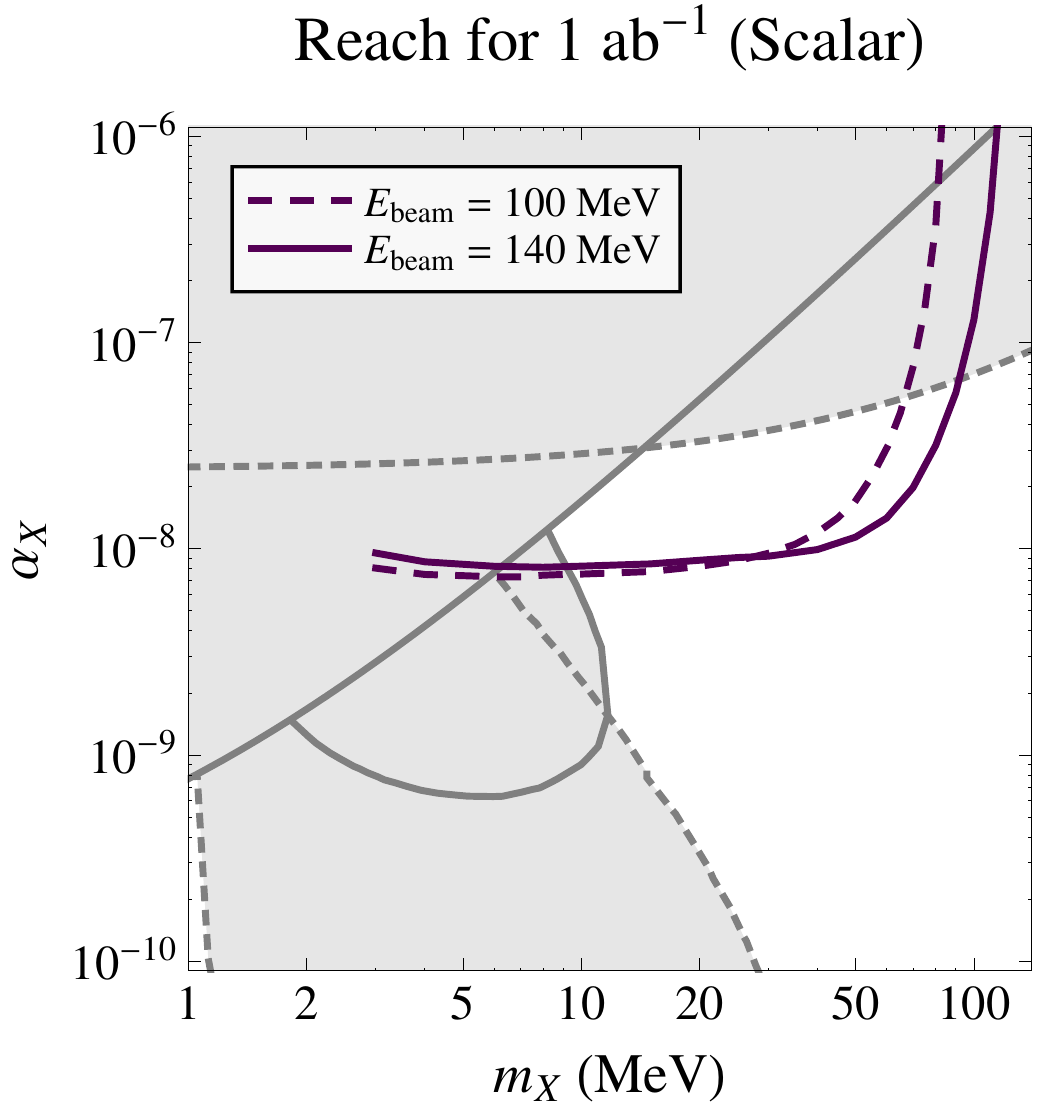} \includegraphics[scale=0.7]{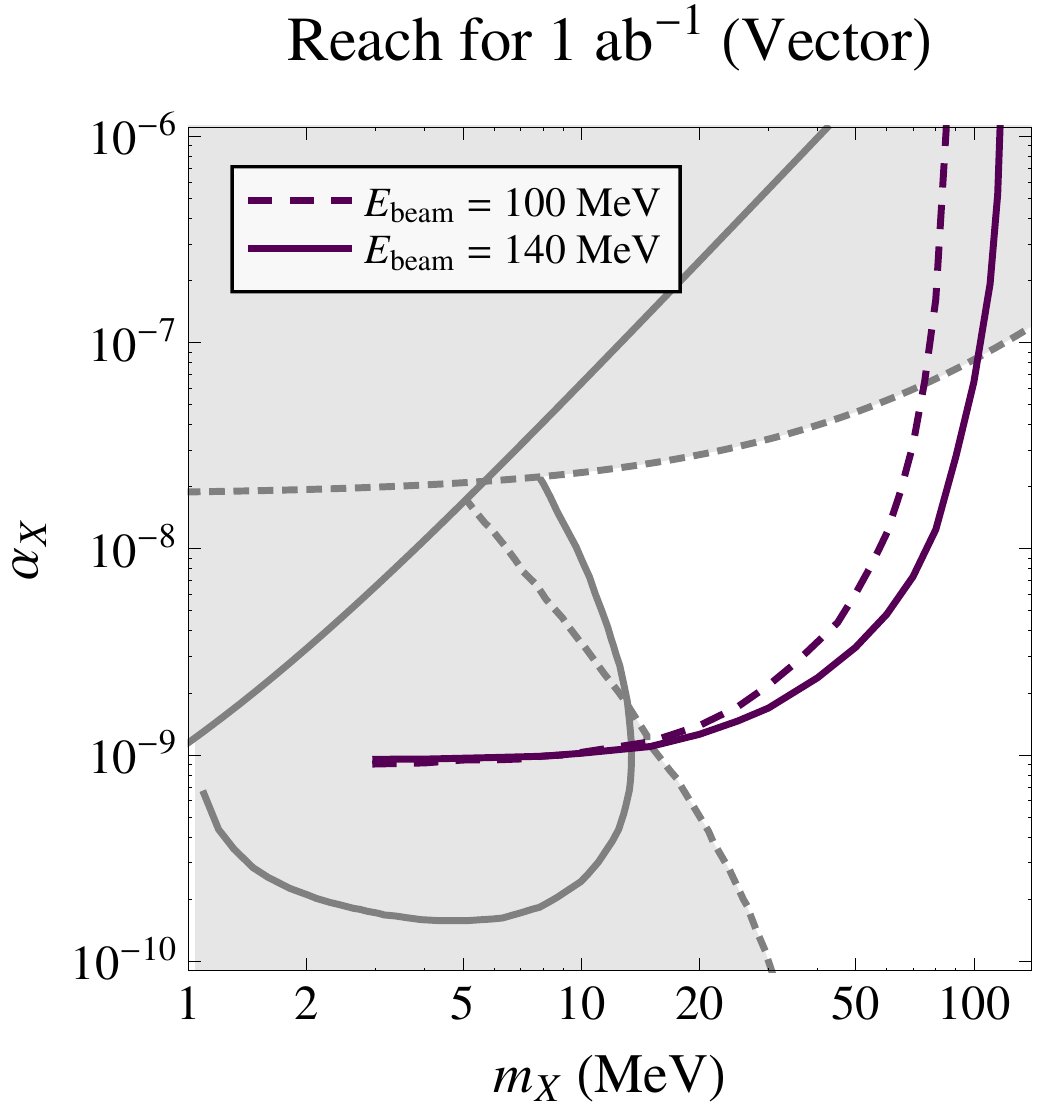}}
\caption{Reach plots for $ep$ collisions, increasing the electron beam energy from $E_e = 100 \MeV$ to $E_e = 140 \MeV$, the maximal sustainable beam energy available at the JLab FEL.  Assumptions about detector geometry, integrated luminosity, and energy resolution are the same as the previous figures.
\label{fig:reach140}}
}

The preceding discussion of the benefits of $ep$ scattering applies for $X$ bosons in the mass range $10 \MeV < m_X < 100 \MeV$.  For higher values of $m_X$, the JLab FEL simply does not have the kinematic reach achievable in other colliders.  The maximum sustainable FEL beam energy is around $140 \MeV$ \cite{Milner:2009--}, limiting the in-principle reach to $m_X \lesssim 131 \MeV$.  Moreover, even if one were able to get higher energy electron beams, it is no longer clear whether $ep$ scattering would pose any advantage, since for electron beam energies above the pion mass, inelastic scattering channels open up, increasing the number of tracks per beam crossing.

Below the pion mass threshold, though, the reach in $ep$ collisions can be improved in going to somewhat higher $E_\text{eff}$ as seen in \fig{fig:reachcomp}.  A more detailed look is shown in \fig{fig:reach140}, which compares the reach for $E_e = 100 \MeV$ to the maximum sustainable FEL energy of $E_e = 140 \MeV$.  For $X$ boson masses close to the kinematic limit, the higher energy electron beam gives improved sensitivity, though at low masses, most of which have already been ruled out by beam dump constraints, the reach gets slightly worse.

\section{Benchmark Studies}
\label{sec:kinedist}

In this section, we consider two benchmark $X$ boson scenarios, indicated on \fig{fig:execsummary},
\begin{align}
\mathrm{A}: &\qquad m_X = 50 \MeV \qquad \alpha_X = 10^{-8},\\
\mathrm{B}: &\qquad m_X = 20 \MeV \qquad \alpha_X = 3 \cdot 10^{-9},
\end{align}
with both scalar and vector couplings in each case.  These points were chosen to be roughly close to the 1 ab$^{-1}$ reach lines in \fig{fig:execsummary}.  We will show an example analysis strategy for $X$ boson signal extraction and then show various kinematic distributions to highlight which parts of phase space are most sensitive to $X$ boson production.

\FIGURE[t]{
\centerline{\includegraphics[scale=0.55]{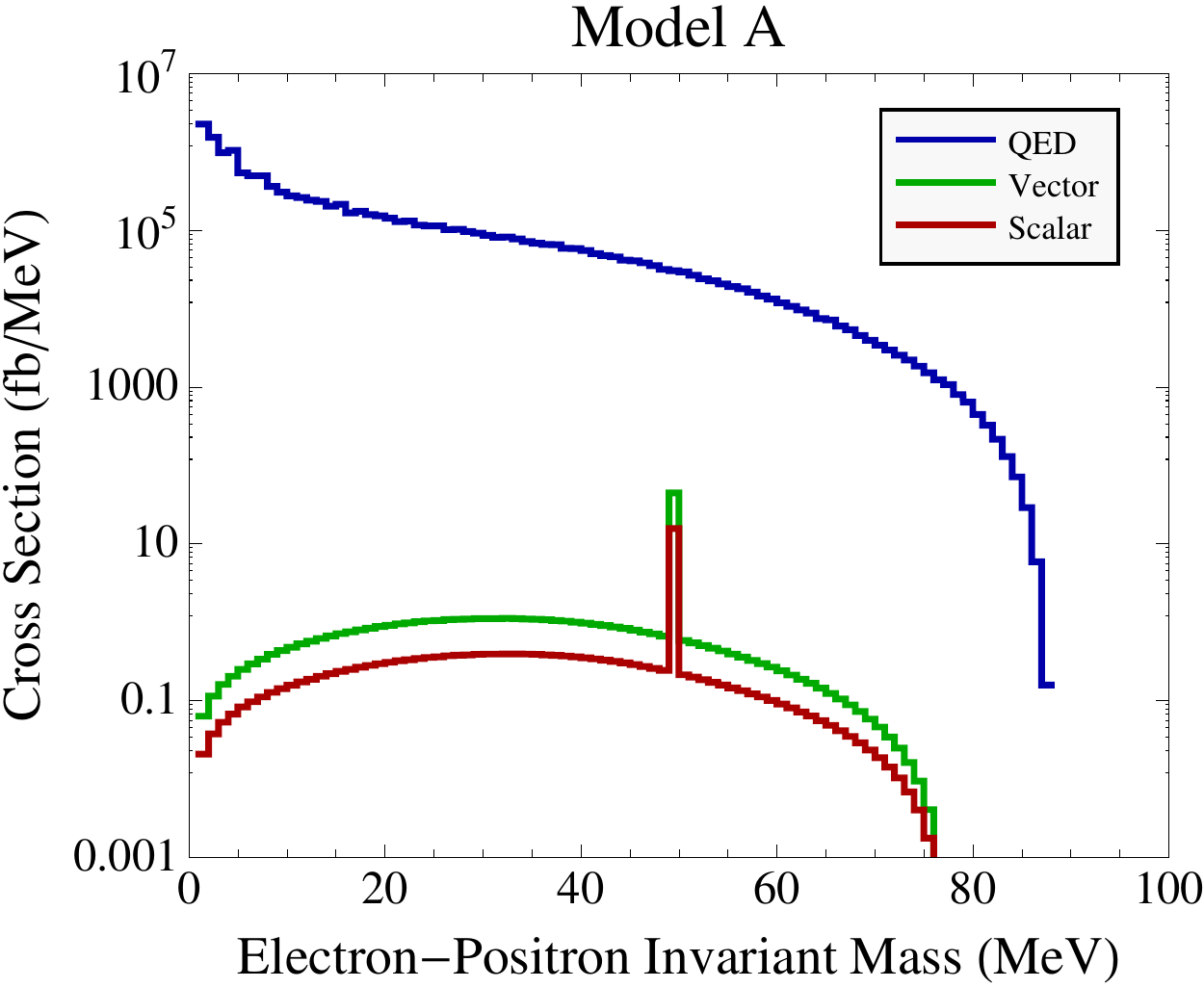} $\quad$ \includegraphics[scale=0.55]{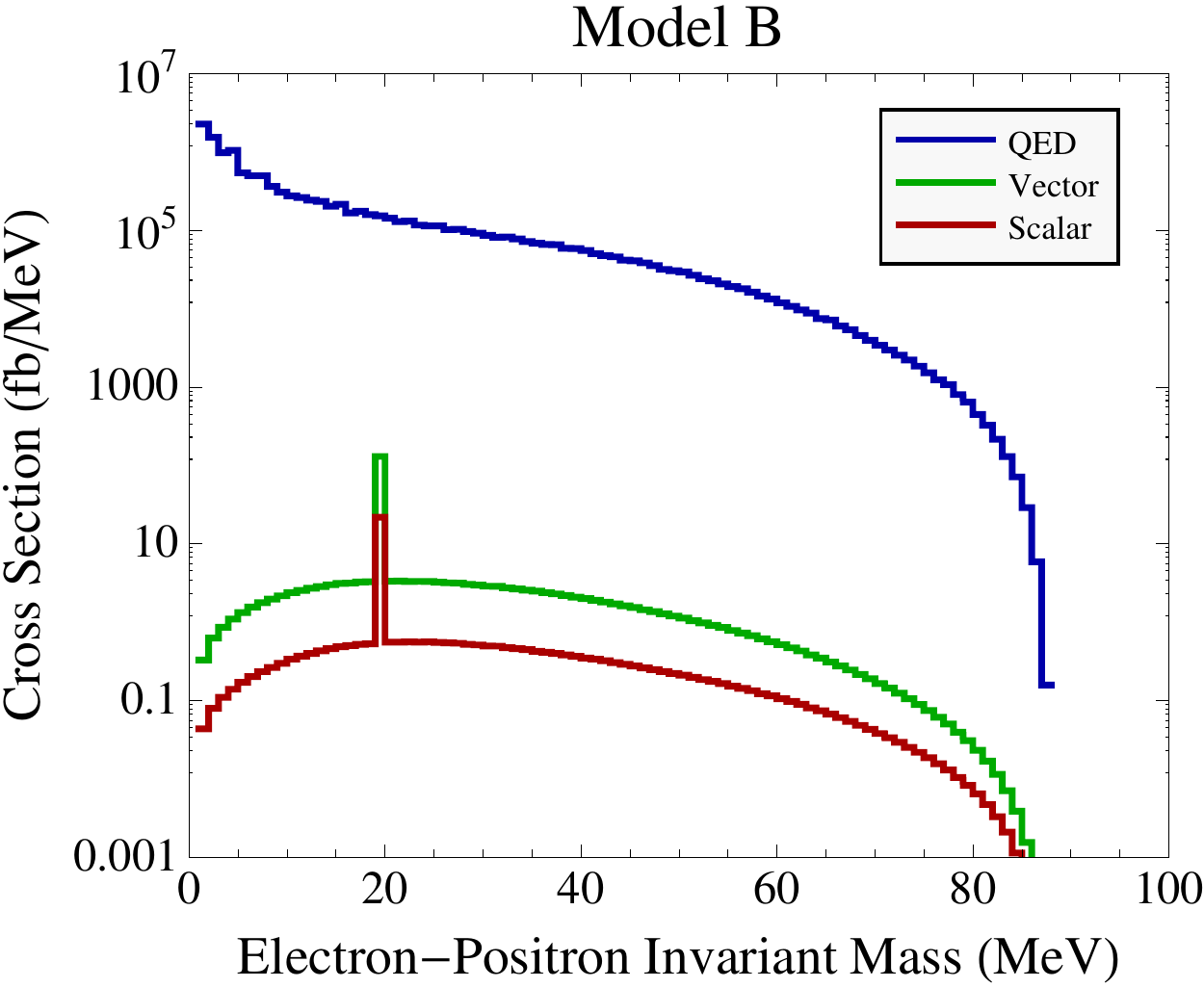}}
\caption{Invariant mass distribution for $m_{e^+e^-}$, comparing the QED background to benchmark models A (left) and B (right).  The fluctuations in the background distribution at low $m_{e^+e^-}$ are from Monte Carlo statistics, and are not indicative of expected statistical fluctuations.  The signal includes the combinatoric background from pairing the ``wrong'' electron with the positron.  These plots include the detector acceptance criteria $-2 < \eta < 2$, $\mathrm{KE}_p > 0.5 \MeV$, and $\mathrm{KE}_{e^\pm} > 5 \MeV$.  Note the four orders of magnitude difference between the expected signal and background.
\label{fig:eeinvmass}}
}

To begin, consider reconstruction of the $X$ boson resonance.  In \fig{fig:eeinvmass}, we show the invariant mass distribution for $e^+e^-$ pairs, taking the fiducial detector acceptance from \eqs{eq:etacuts}{eq;kineticenergycuts}.   Since the final state electrons are indistinguishable, the plot includes two histogram entries per event.  In the case of the signal distribution, there is the expected spike at $m_X$ accompanied by a combinatoric background from pairing the positron with the ``wrong'' electron.  We see that there is a four orders of magnitude difference between signal and background, consistent with \fig{fig:rawxsec} which did not include any detector acceptance effects.

\FIGURE[t]{
\centerline{\includegraphics[scale=0.55]{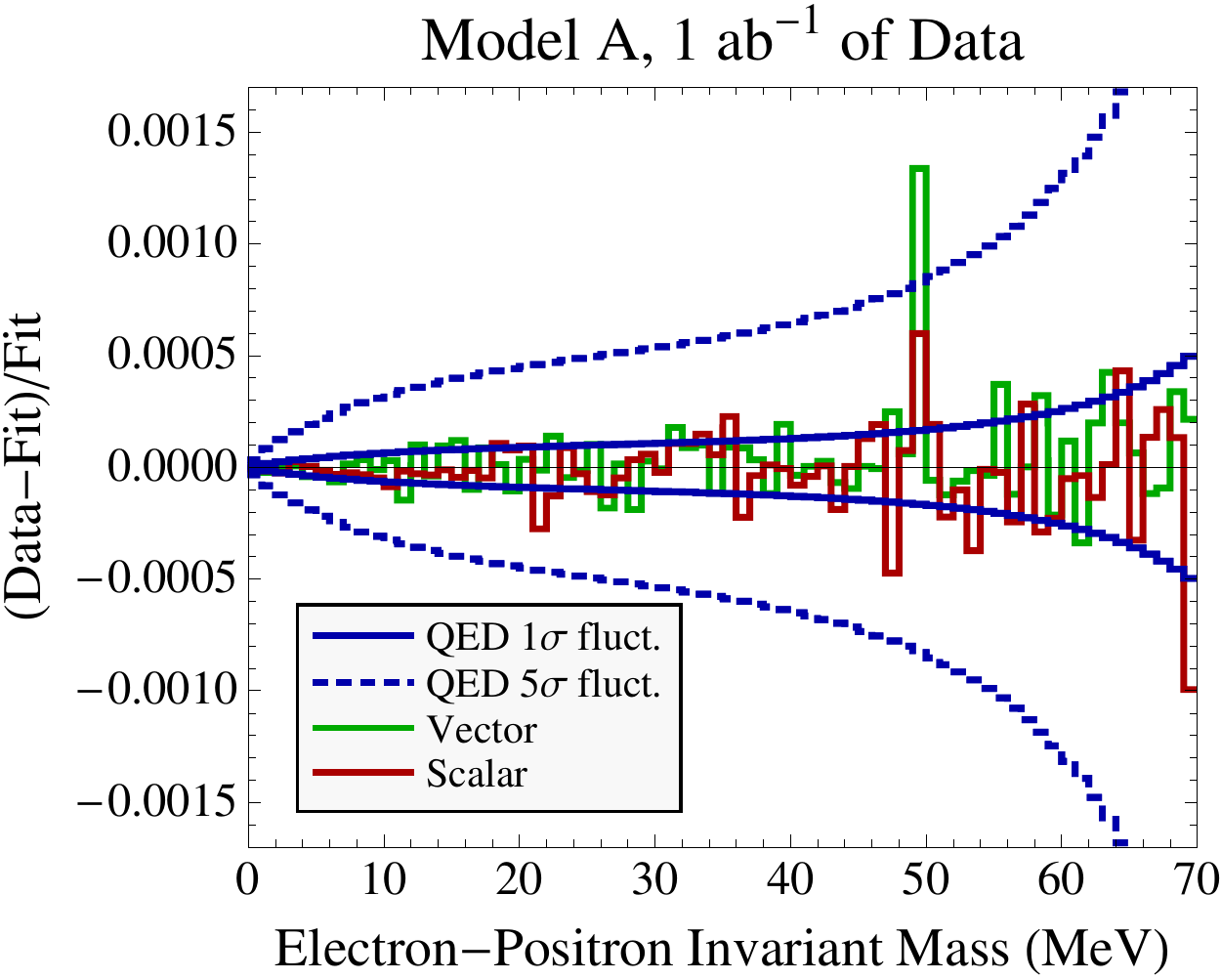} $\quad$ \includegraphics[scale=0.55]{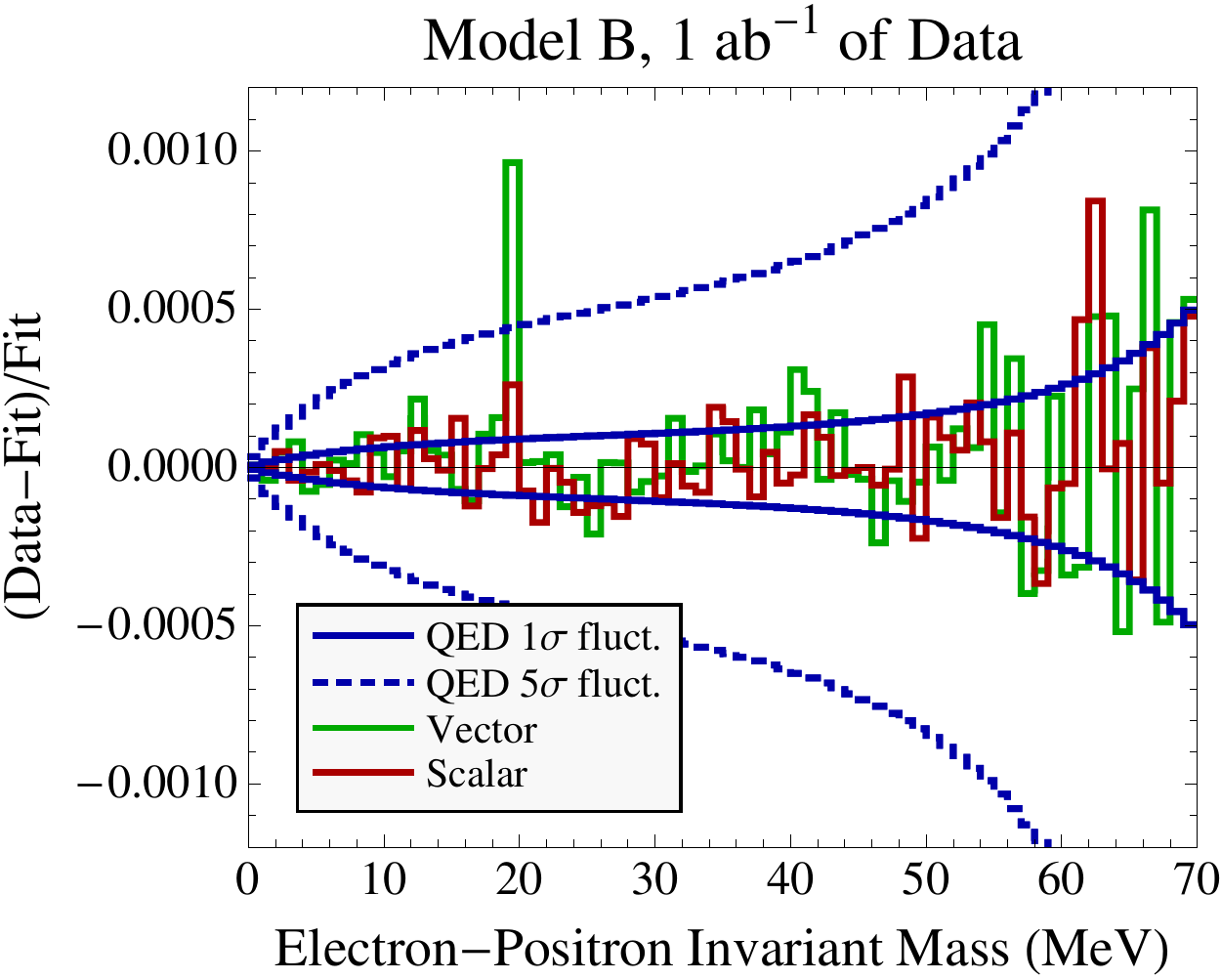}}
\caption{Simulated $m_{e^+e^-}$ distributions compared to a background fit.  This plot was made with 1 ab$^{-1}$ of signal and background pseudo-data, assuming 1 MeV invariant mass resolution.   The blue curves show the expected statistical uncertainties in the background, $1\sigma$ (solid) and $5\sigma$ (dashed).  For the vector couplings (green), both model A (left) and B (right) show a prominent bump in the dilepton invariant mass distribution, which is expected since these benchmark points lie above the 1 ab$^{-1}$ reach line.  For the scalar couplings (red), a bump cannot be seen, since these benchmark points lie below the 1 ab$^{-1}$ reach line.
\label{fig:simdistro}}}

To show how the $X$ boson resonance could be seen despite the large background, \fig{fig:simdistro} shows a simulated distribution of $m_{e^+ e^-}$ created as follows.  First, we generate 1 ab$^{-1}$ of background pseudo-data for the $m_{e^+ e^-}$ distribution and add it to 1 ab$^{-1}$ of signal pseudo-data.  We take the combined signal plus background distribution, and fit it to an ad hoc functional form:
\be
\label{eq:fitfuncform}
\frac{\df \sigma_{\rm fit}}{\df m} = N (m)^a (m_{\max} - m)^b (e^m)^c,
\ee
where $N$, $a$, $b$, and $c$ are fit coefficients, and $m_{\max}$ is the maximum kinematically allowed value for $m_{e^+ e^-}$.  We then plot the fractional difference between the pseudo-data and the final fit function in \fig{fig:simdistro}.  For the case of the vector couplings where the benchmark points lie above the 1 ab$^{-1}$ reach line, a peak at the $m_{e^+ e^-}$ distribution at $m_X$ is indeed visible above the expected statistical fluctuations in the background, showing that a sideband procedure for extracting the background is feasible.  For the scalar benchmarks, no such peak is visible, as expected since these benchmarks lie below the 1 ab$^{-1}$ reach line.

\FIGURE[!p]{
\centerline{\includegraphics[scale=0.45]{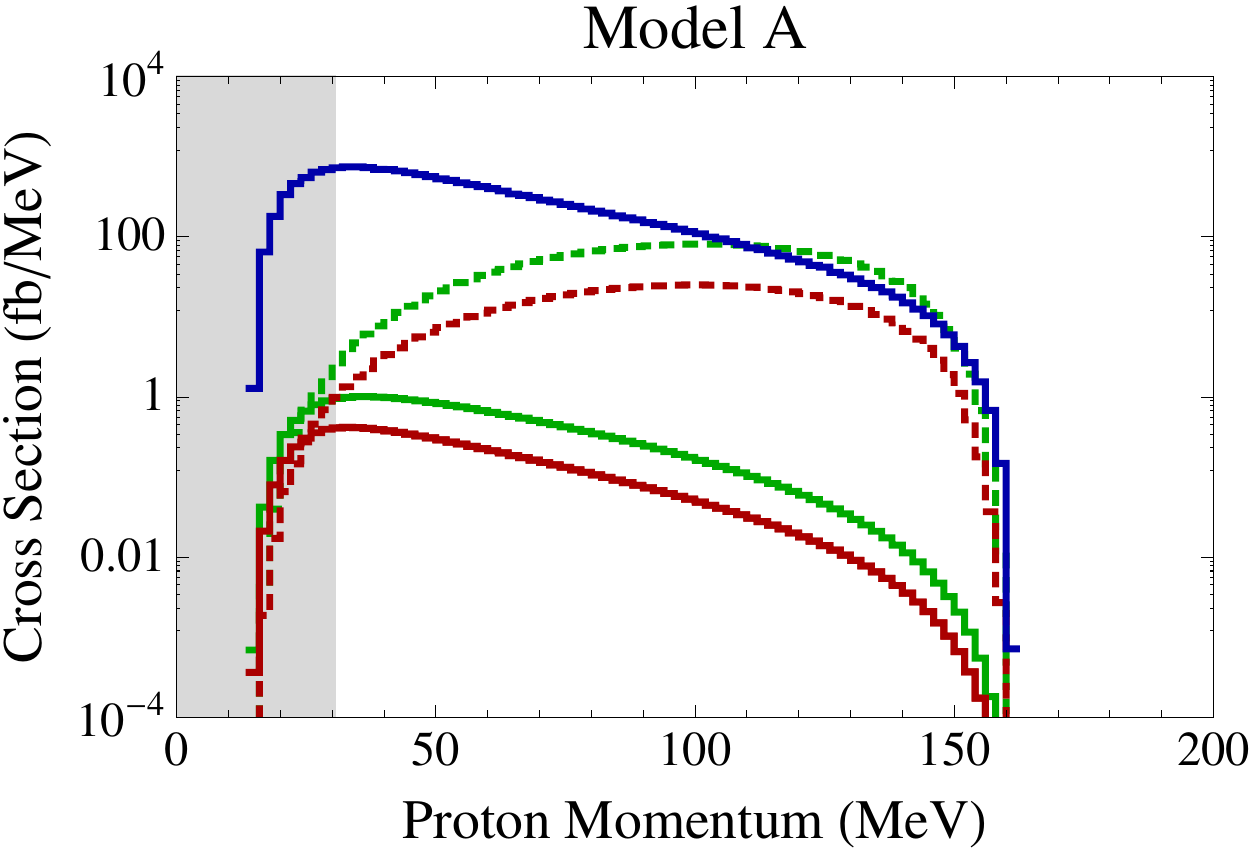} $\qquad$ \includegraphics[scale=0.45]{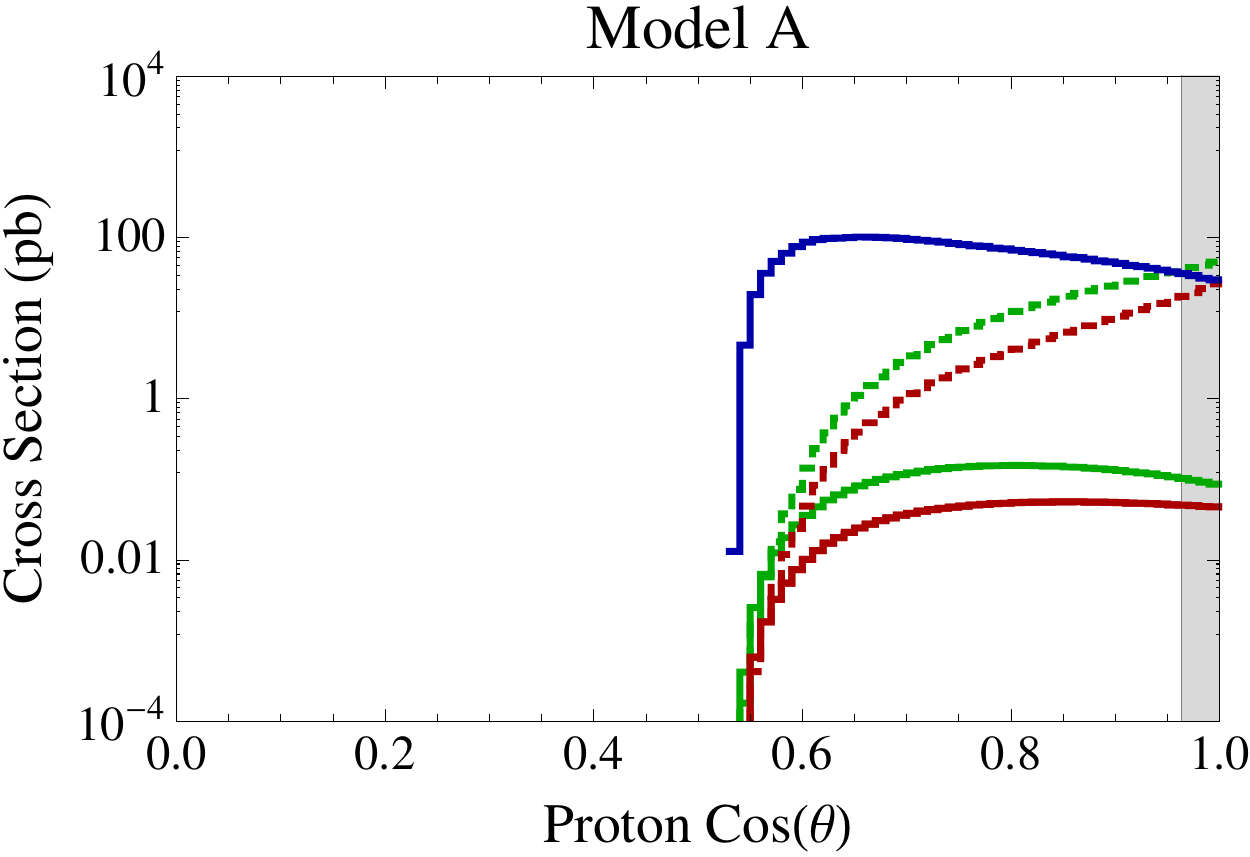}}
~\\
\centerline{\includegraphics[scale=0.45]{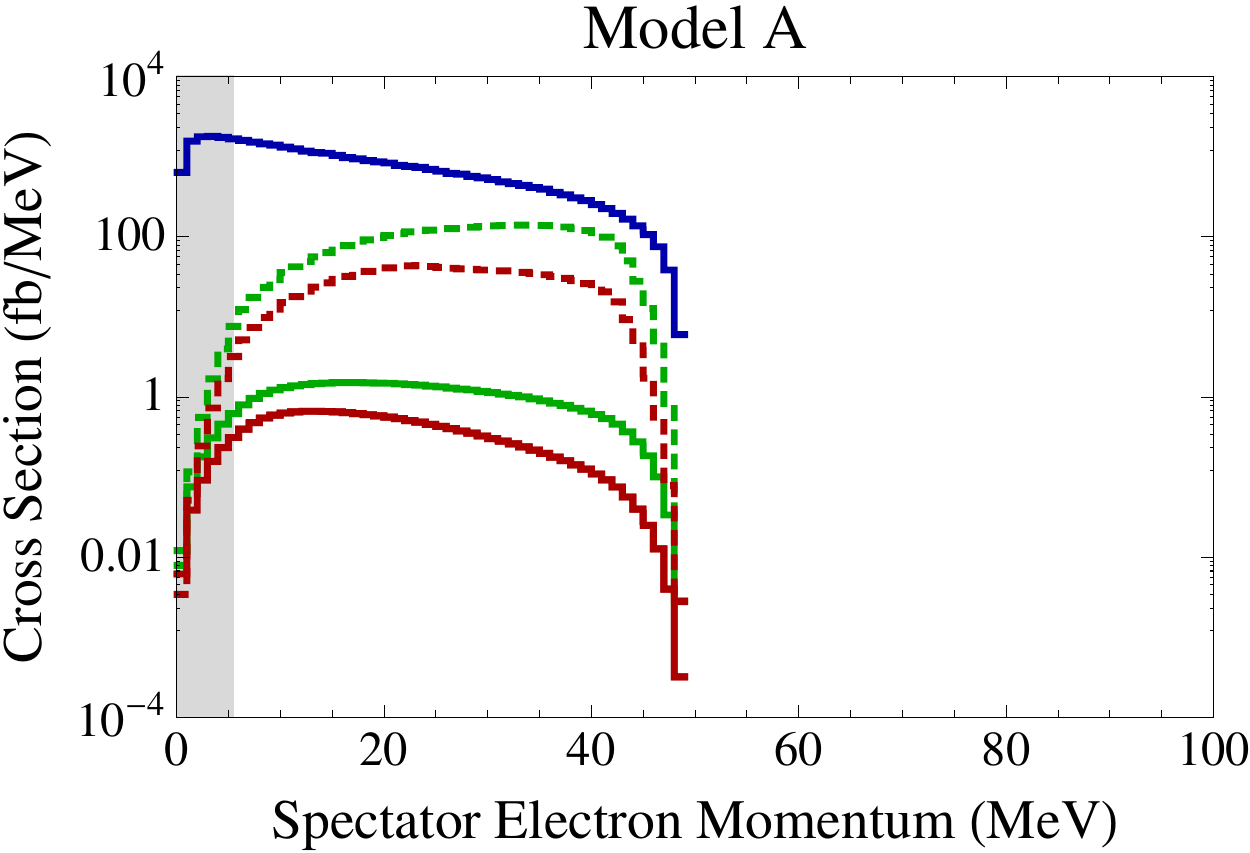} $\qquad$ \includegraphics[scale=0.45]{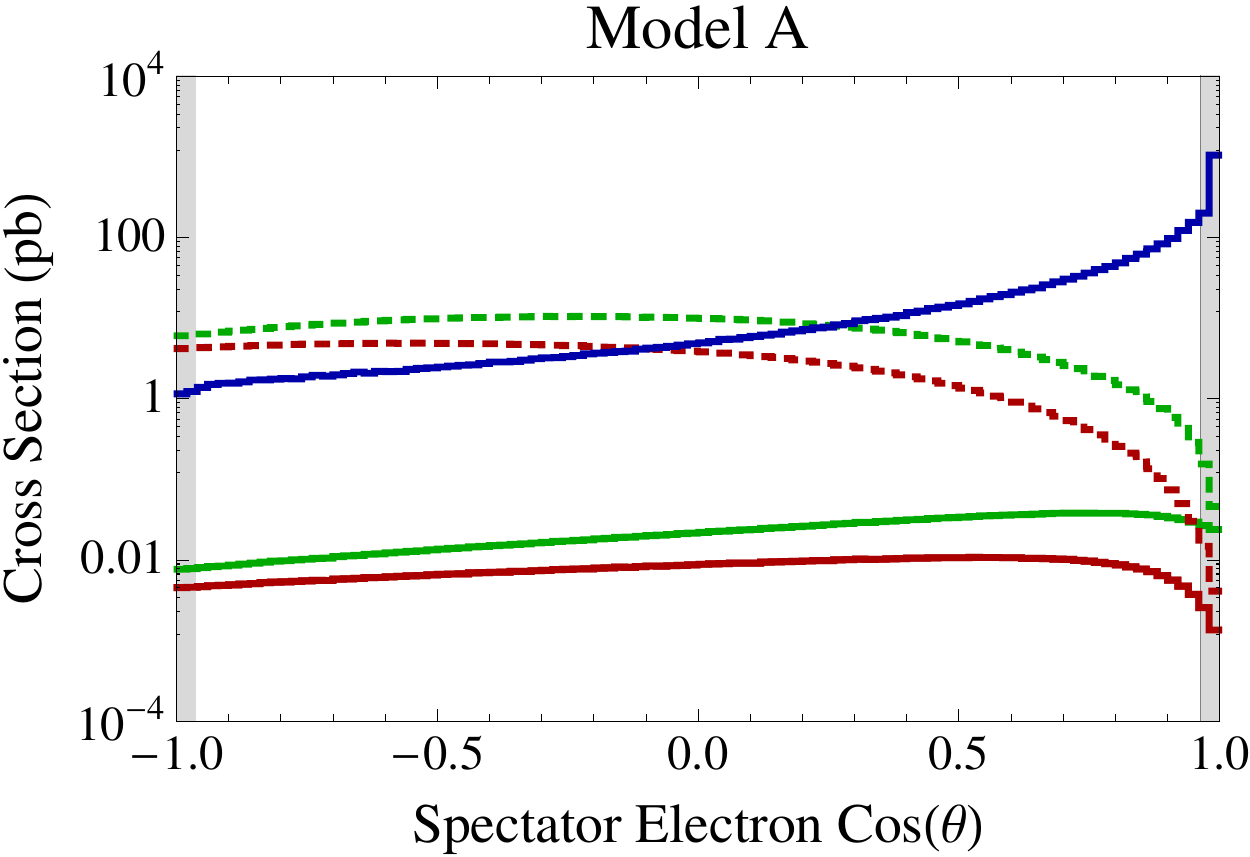}}
~\\
\centerline{\includegraphics[scale=0.45]{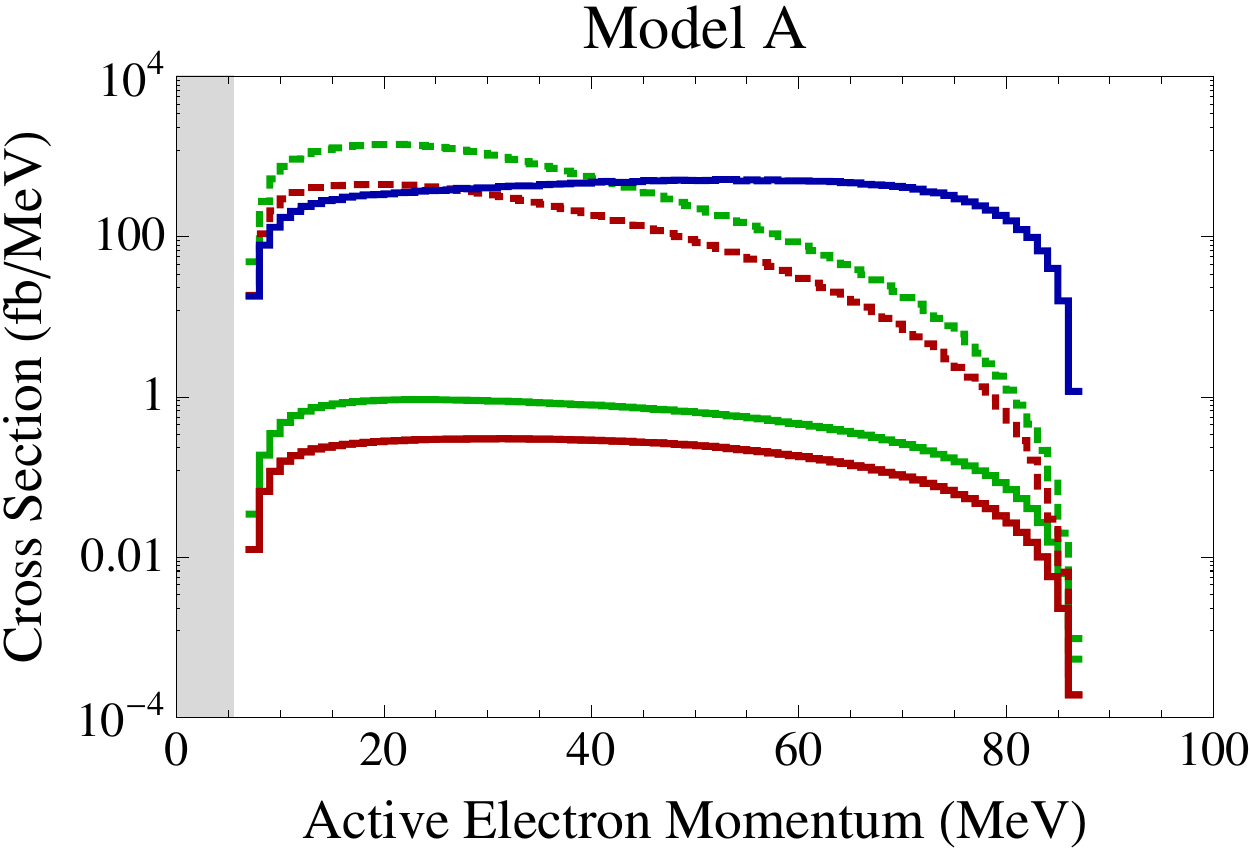} $\qquad$ \includegraphics[scale=0.45]{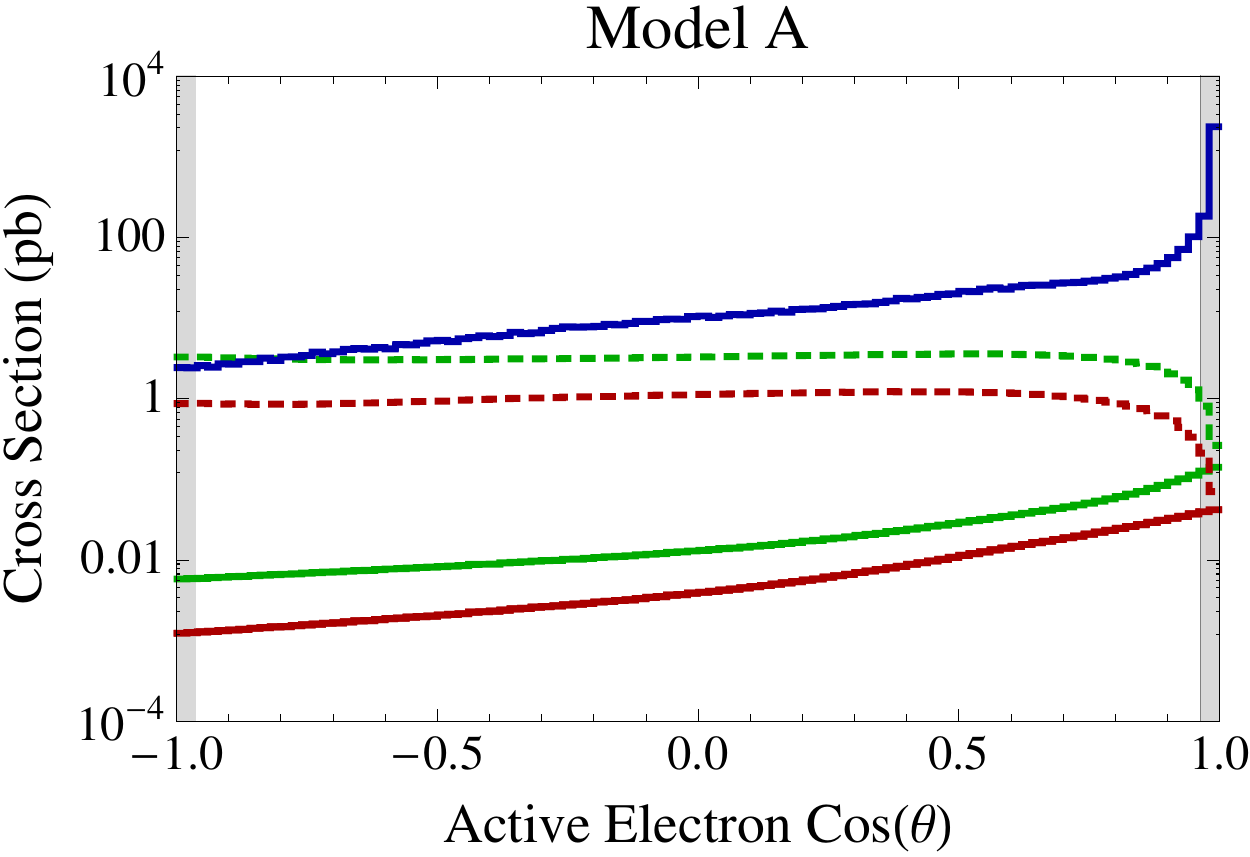}}
~\\
\centerline{\includegraphics[scale=0.45]{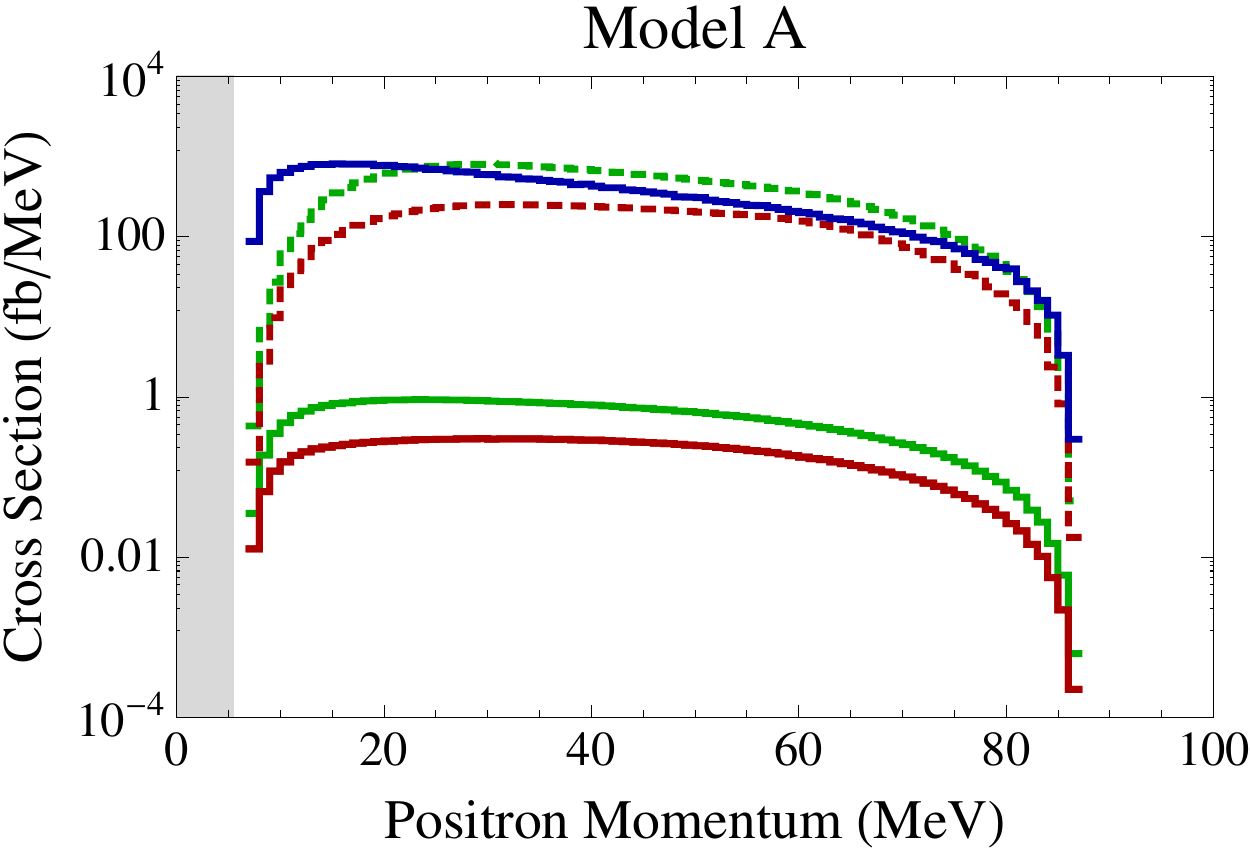} $\qquad$ \includegraphics[scale=0.45]{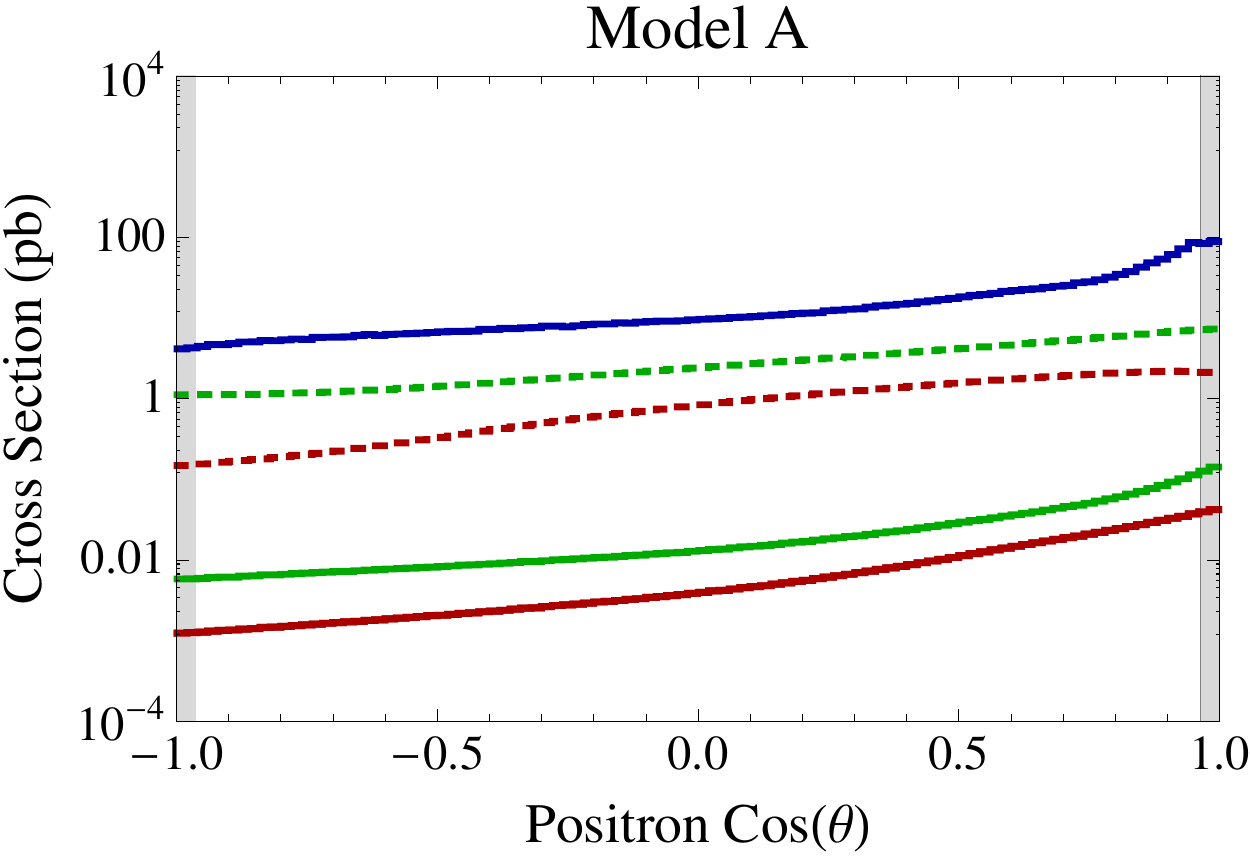}}
~\\
\centerline{\includegraphics[scale=0.55]{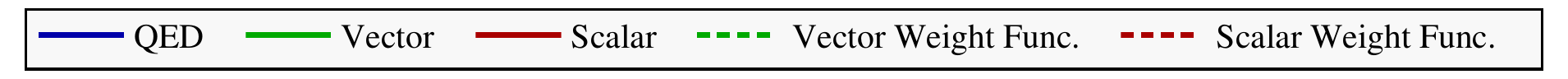}}
\caption{Momentum and angular distributions for model A, with $m_X = 50 \MeV$ and $\alpha_X = 10^{-8}$.   The QED background is restricted to have one $e^+e^-$ pair reconstruct $m_X$, and the corresponding electron is called the active electron while the other is the spectator electron.  These plots include detector acceptance cuts, but the cut corresponding to the plotted distribution is indicated by shading.  The solid blue curves are the QED background, and the solid red (green) curves are the scalar (vector) signal.  The dashed red (green) curves are the ideal weighting functions for the scalar (vector) case with arbitrary normalization, which are large in the region of phase space most sensitive to $X$ boson production.
}
\label{fig:modelAscalar1}}

\FIGURE[!p]{
\centerline{\includegraphics[scale=0.45]{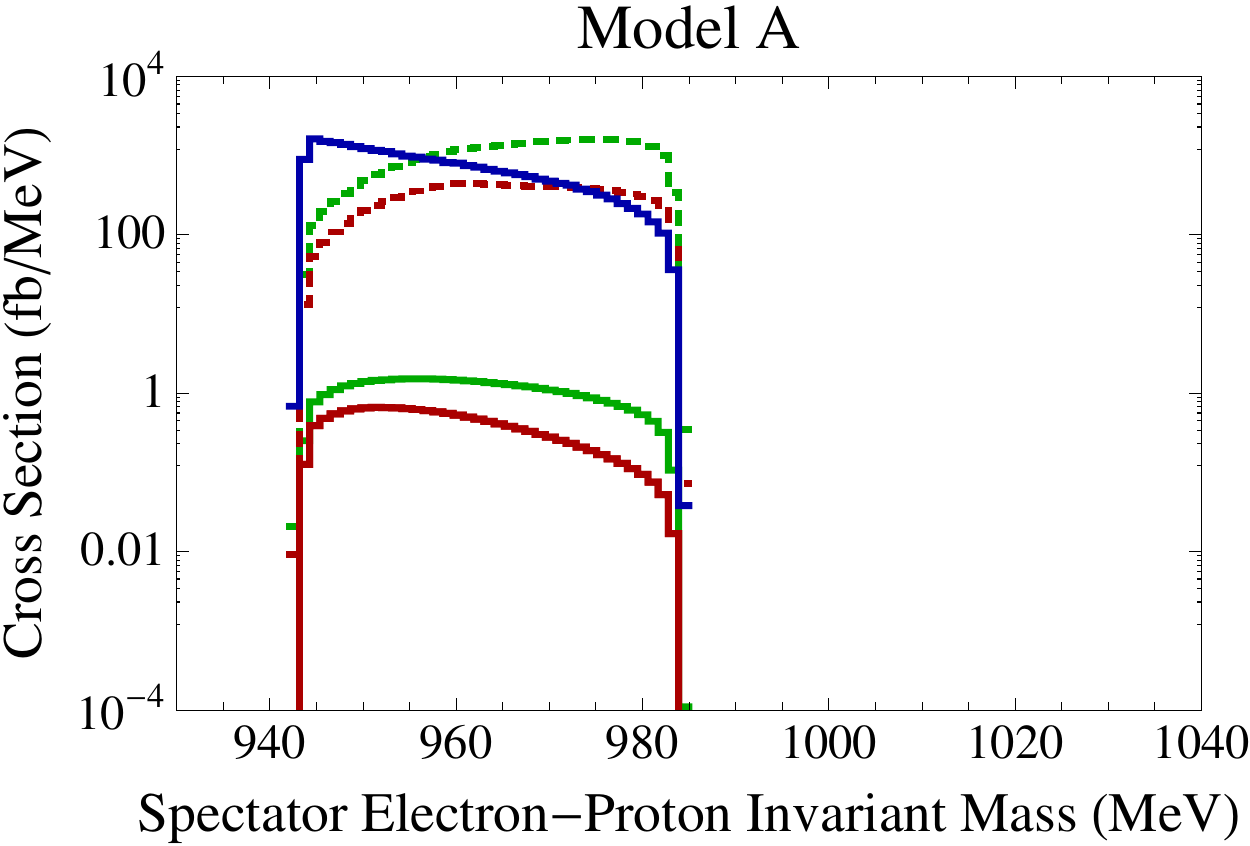} $\qquad$  \includegraphics[scale=0.45]{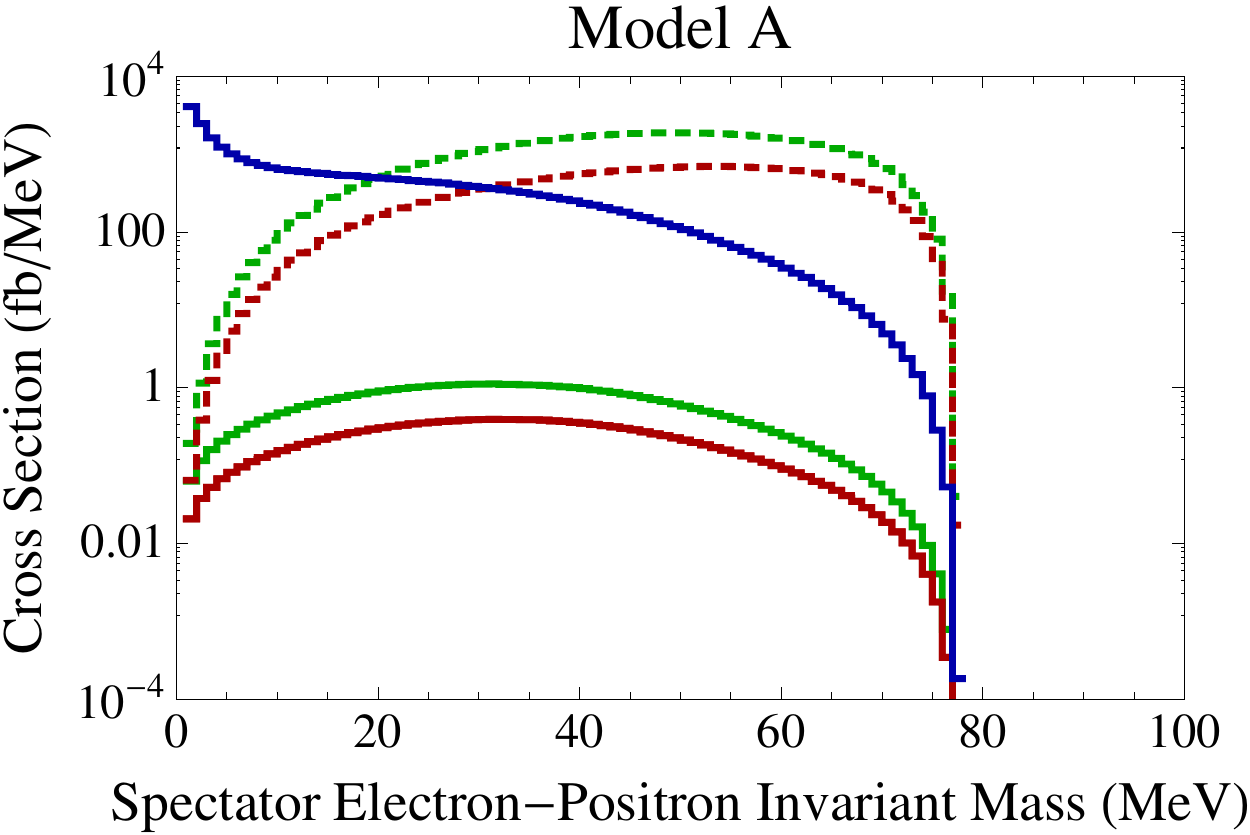}}
~\\
\centerline{\includegraphics[scale=0.45]{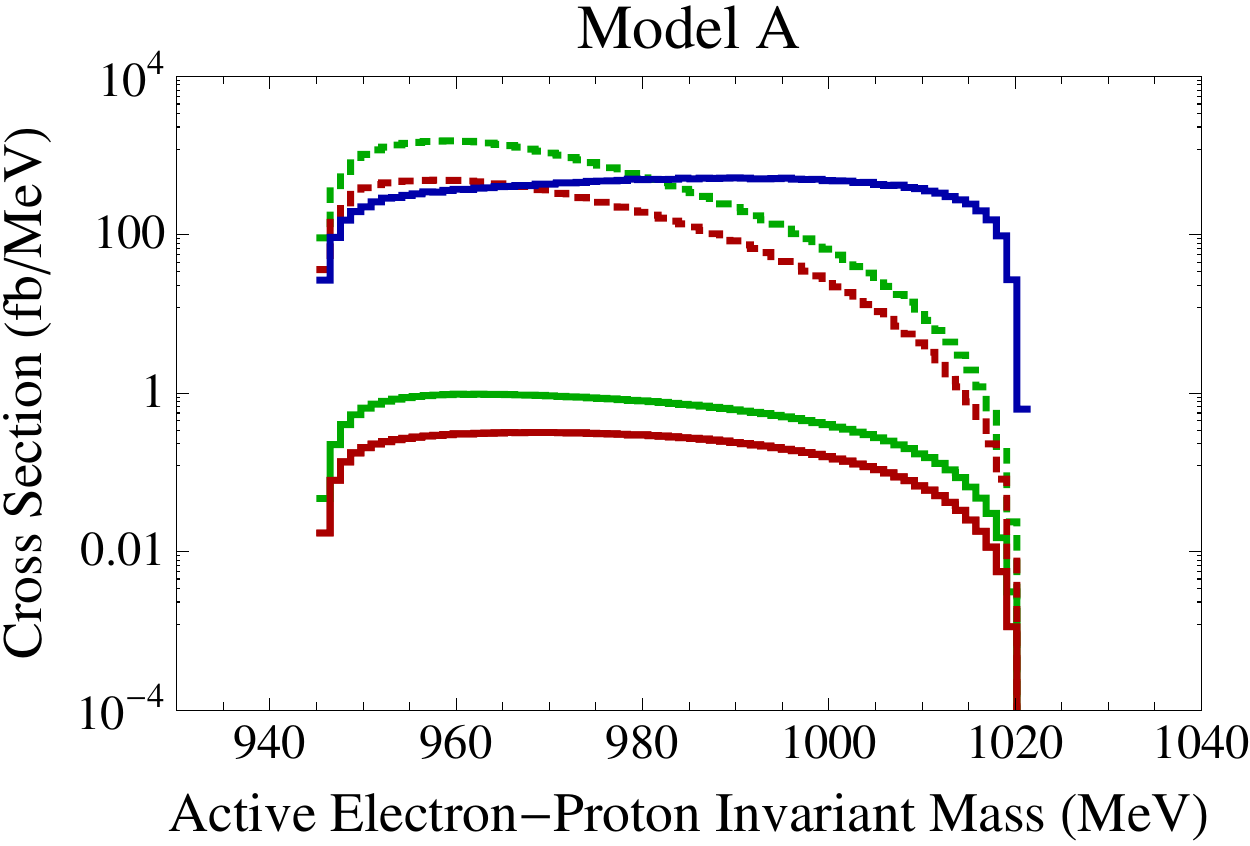}  $\qquad$ \includegraphics[scale=0.45]{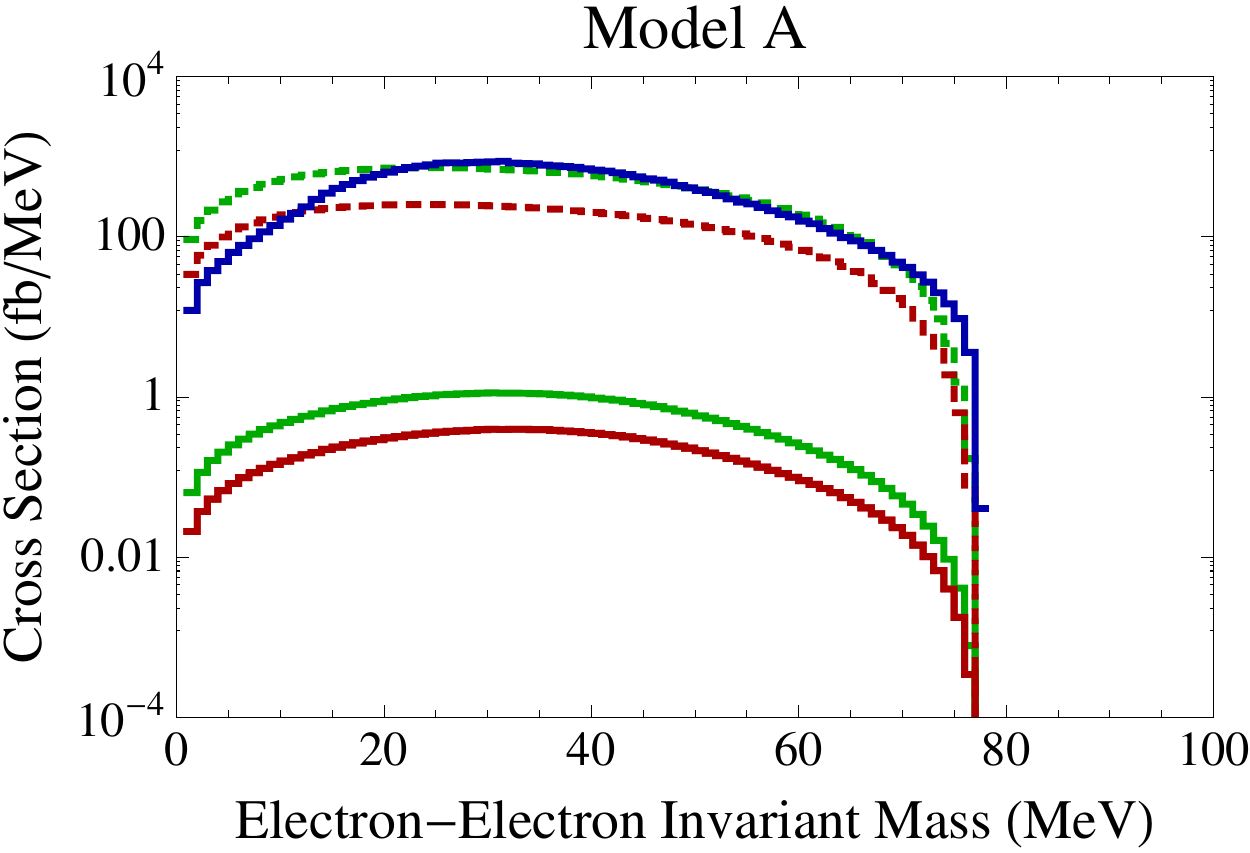}}
~\\
\centerline{\includegraphics[scale=0.45]{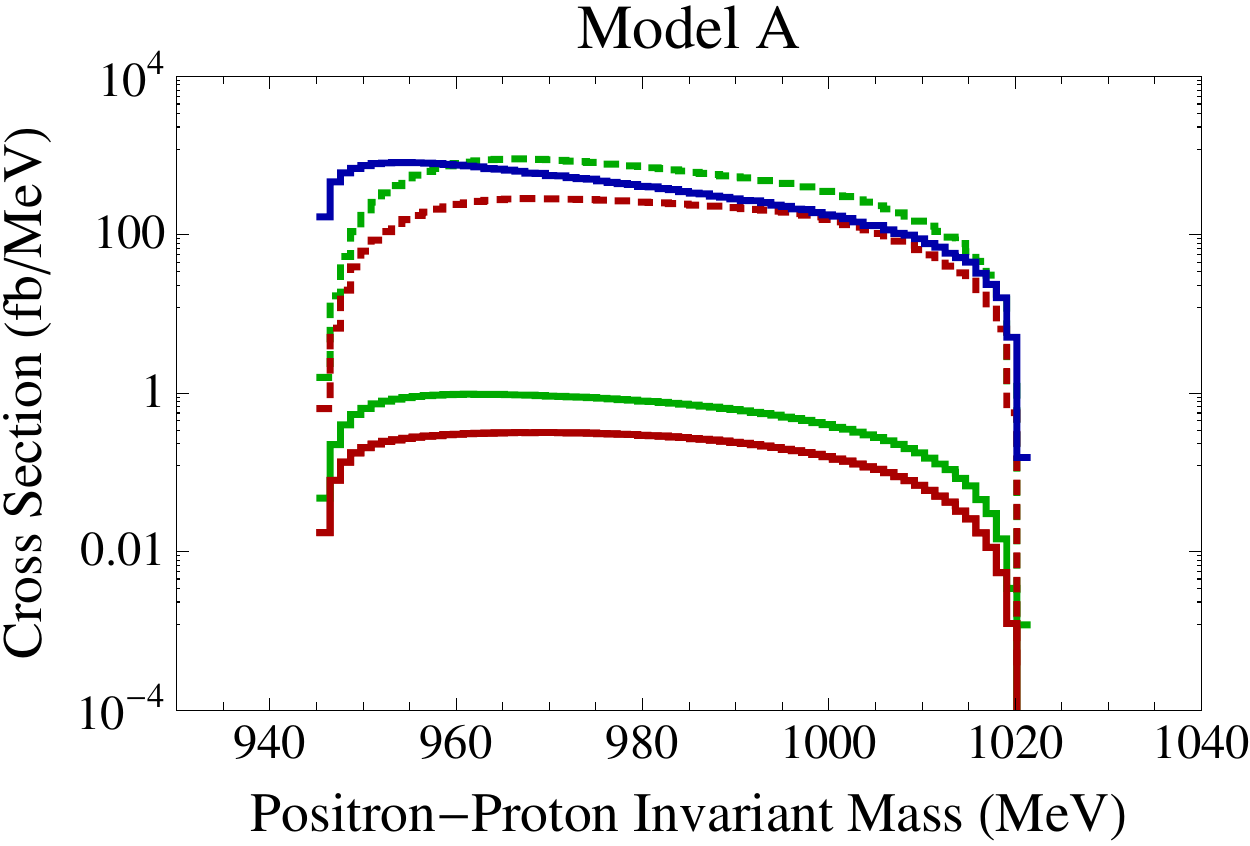} $\qquad$  \includegraphics[scale=0.45]{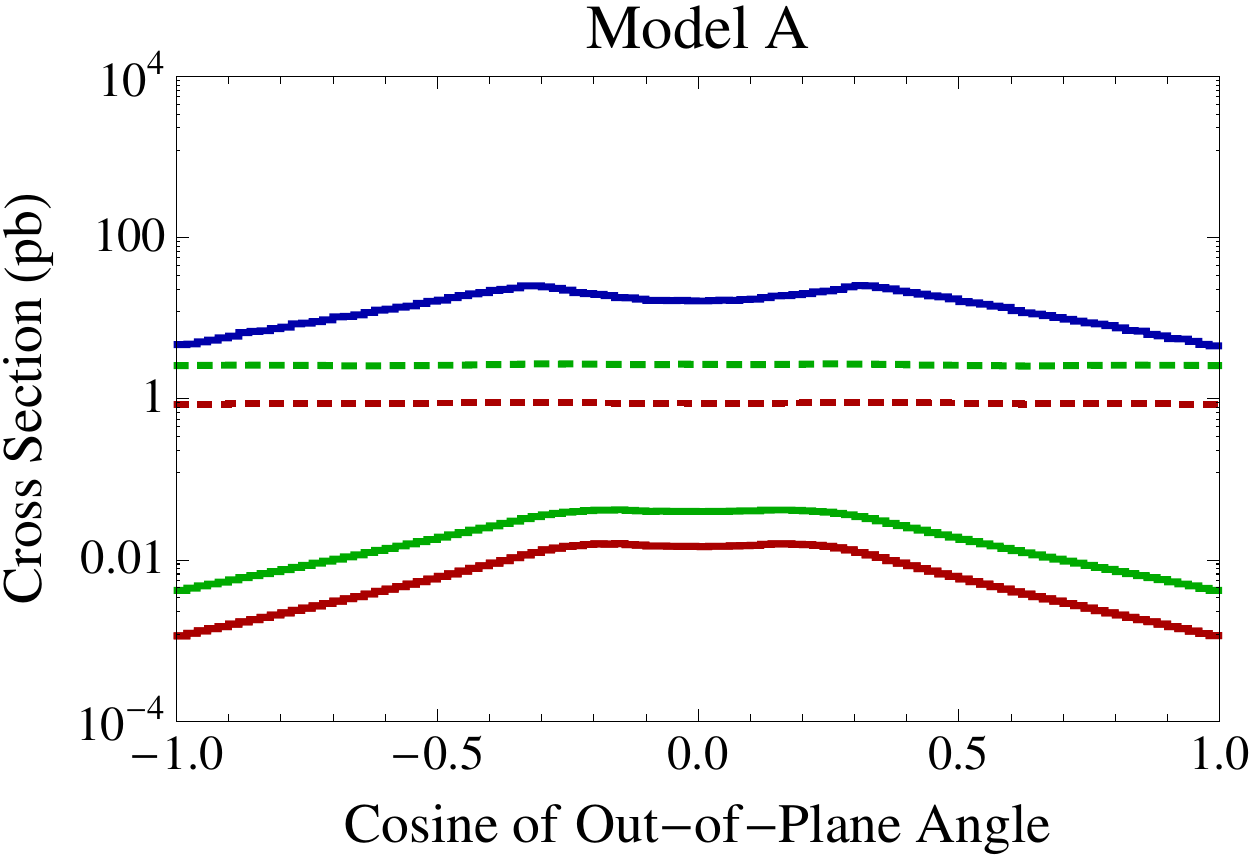}}
~\\
\centerline{\includegraphics[scale=0.45]{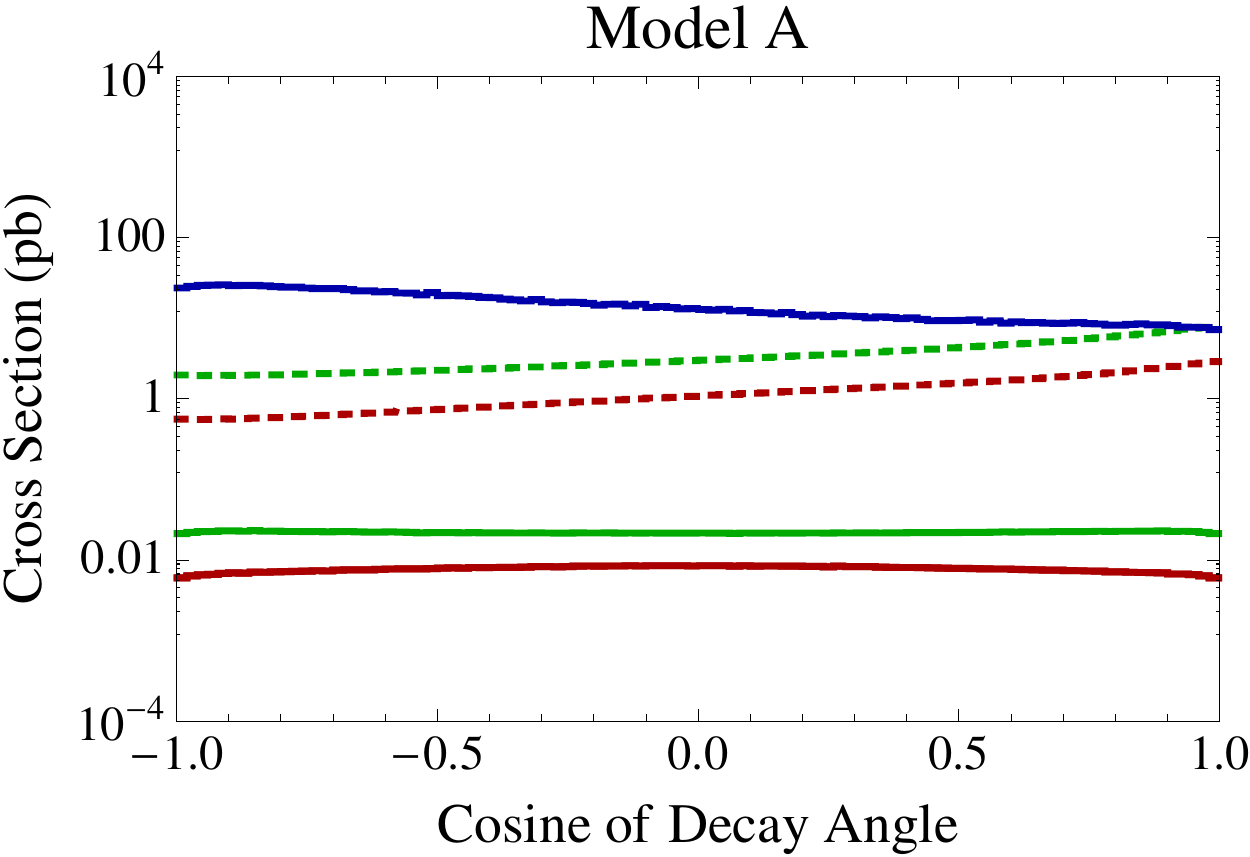} $\qquad$  \includegraphics[scale=0.45]{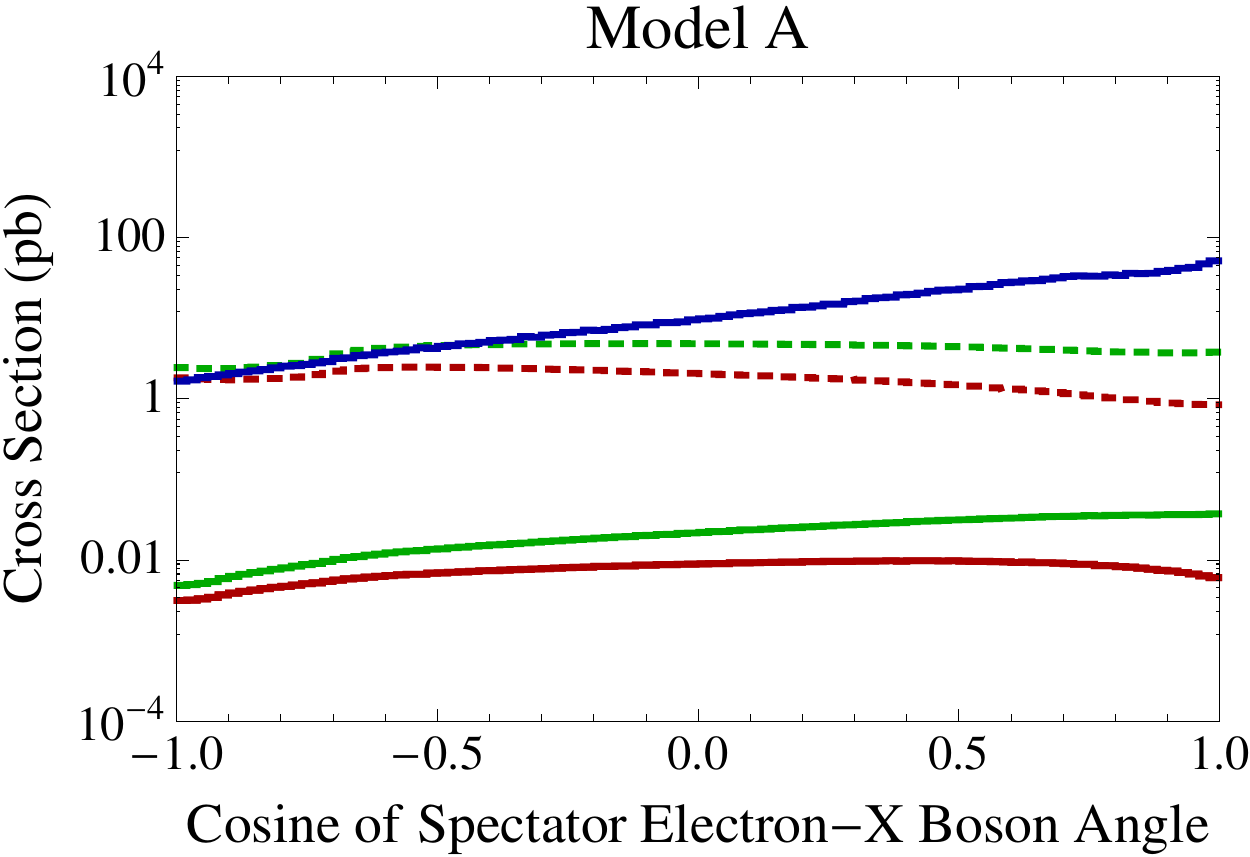}}
~\\
\centerline{\includegraphics[scale=0.55]{figures/KinematicLegend.pdf}}
\caption{Additional kinematic distributions for model A, with the same criteria and labeling as \fig{fig:modelAscalar1}.  Shown are five pairwise invariant mass distributions (the active electron-positron invariant mass would of course just give a peak at $m_X$).  The out-of-plane angle is the between the reconstructed $X$ boson and the incoming electron/spectator electron plane.  The decay angle is the angle of the $X$ boson decay products relative to the $X$ boson momentum, measured in the $X$ rest frame.  Also shown is the angle between the spectator electron and the reconstructed $X$ boson.
}
\label{fig:modelAscalar2}}

\FIGURE[!p]{
\centerline{\includegraphics[scale=0.45]{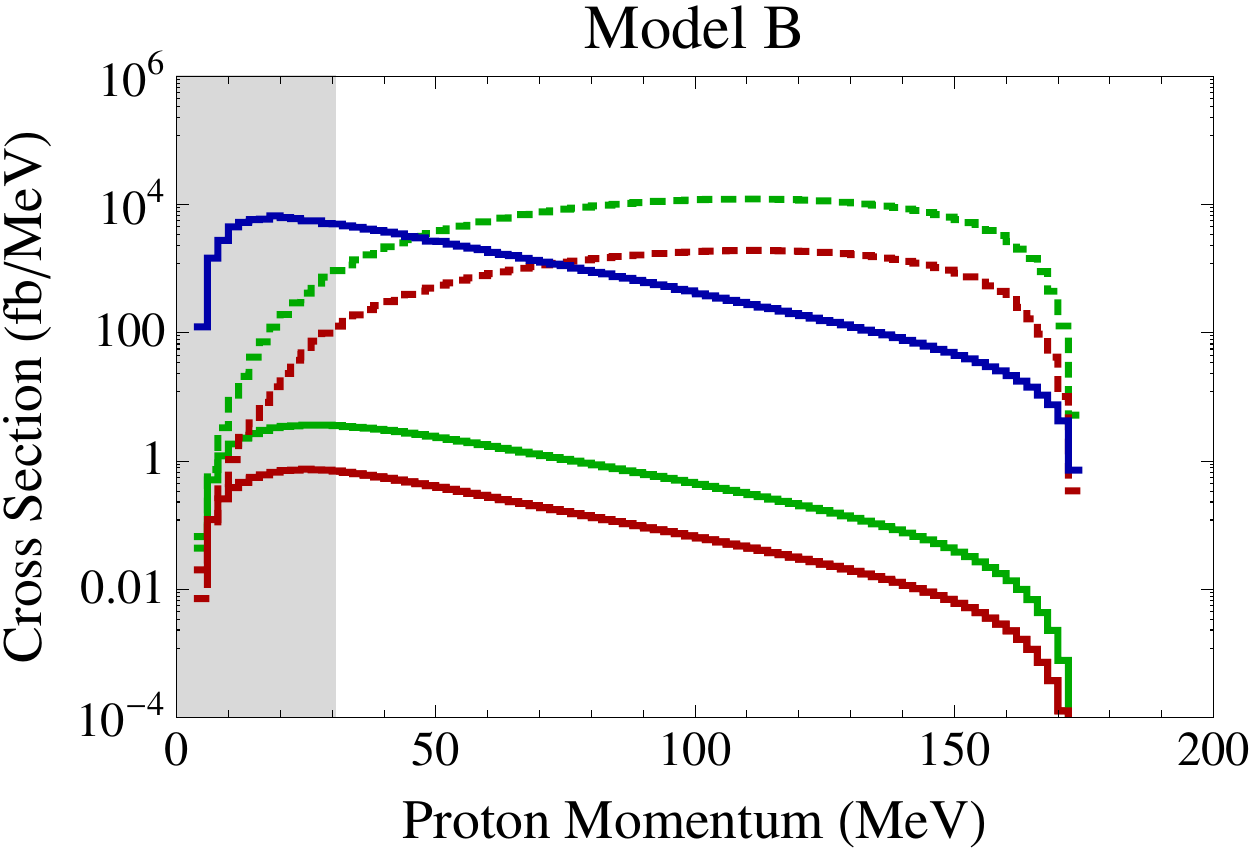} $\qquad$ \includegraphics[scale=0.45]{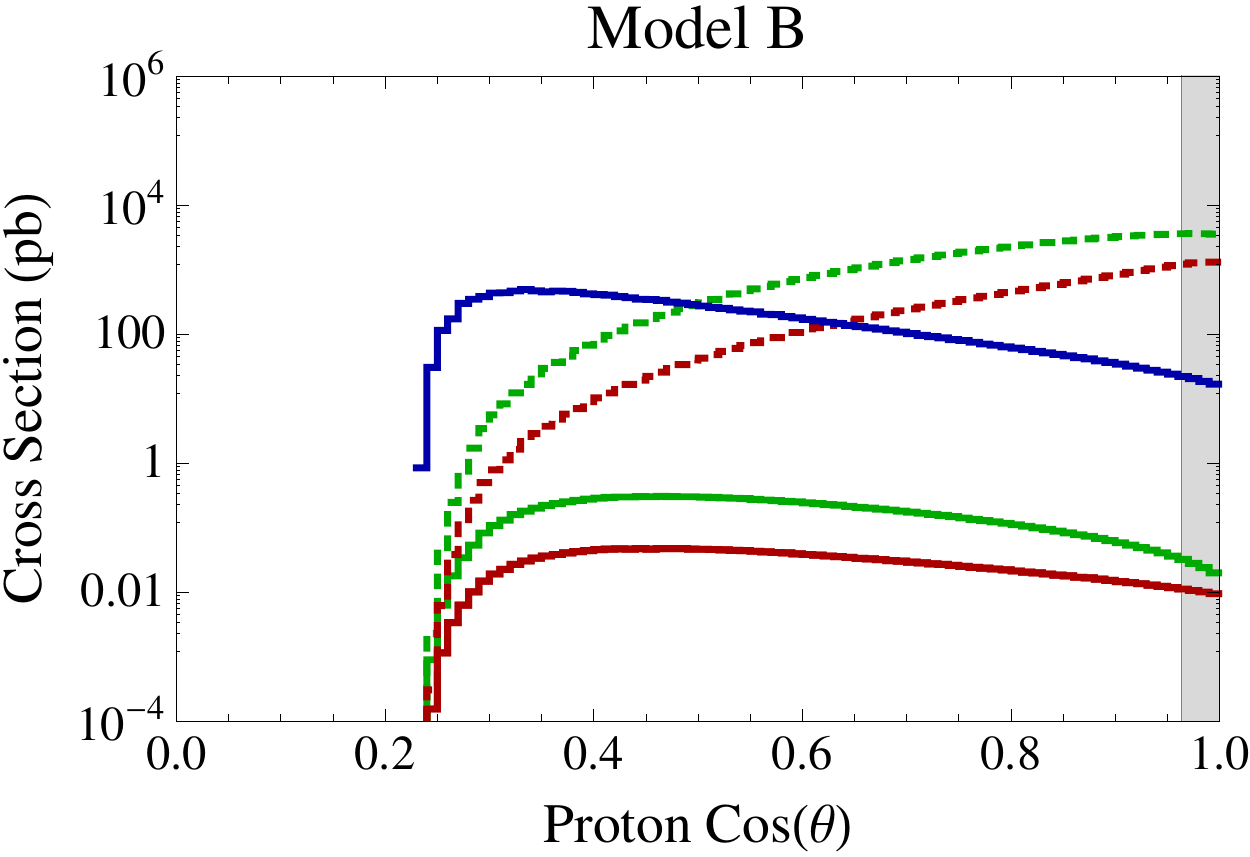}}
~\\
\centerline{\includegraphics[scale=0.45]{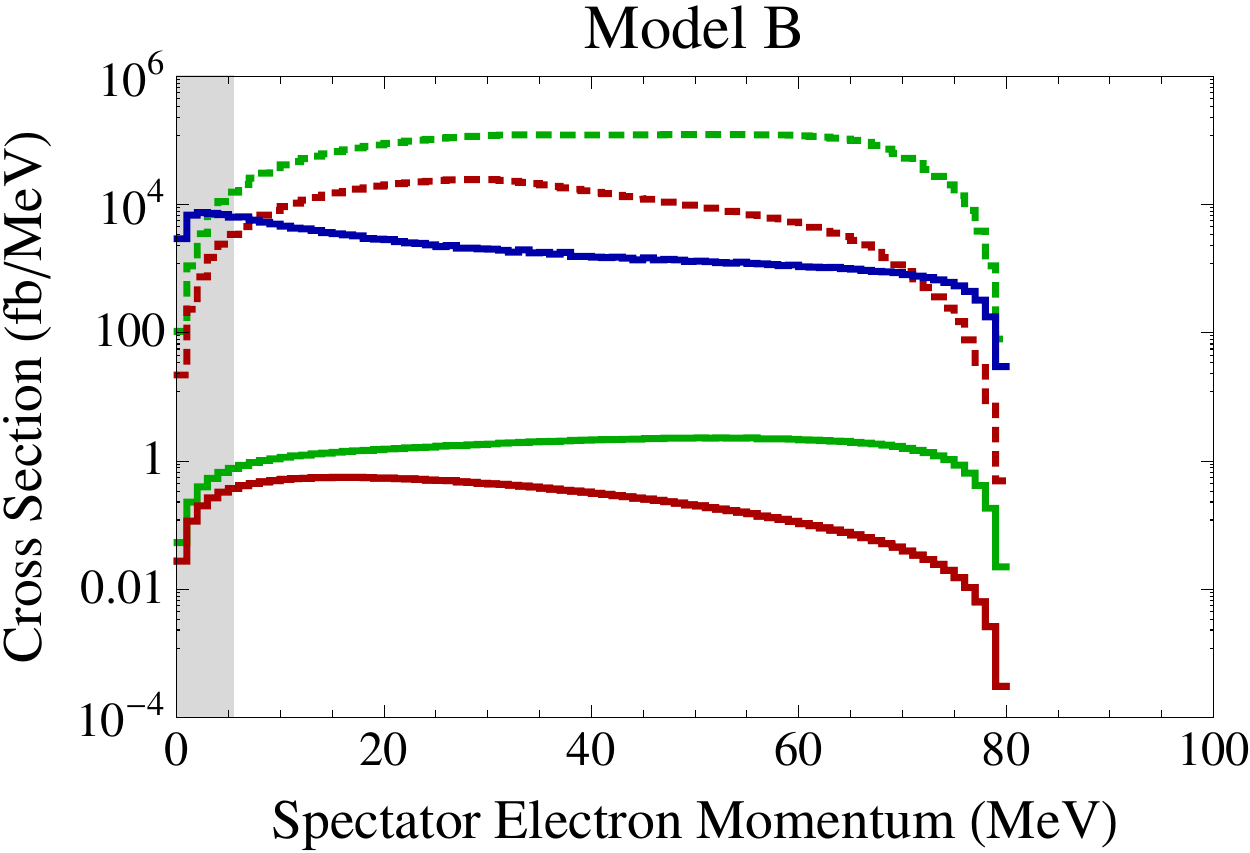} $\qquad$ \includegraphics[scale=0.45]{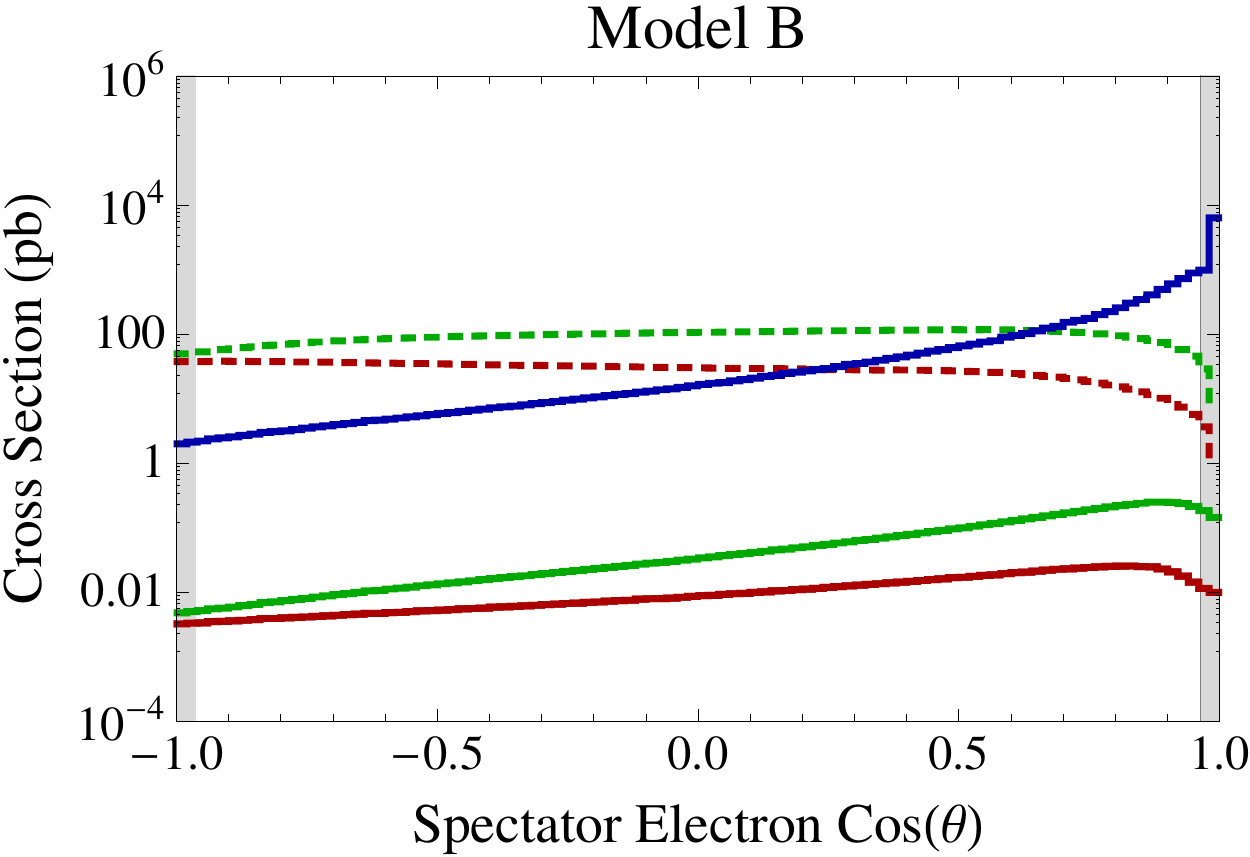}}
~\\
\centerline{\includegraphics[scale=0.45]{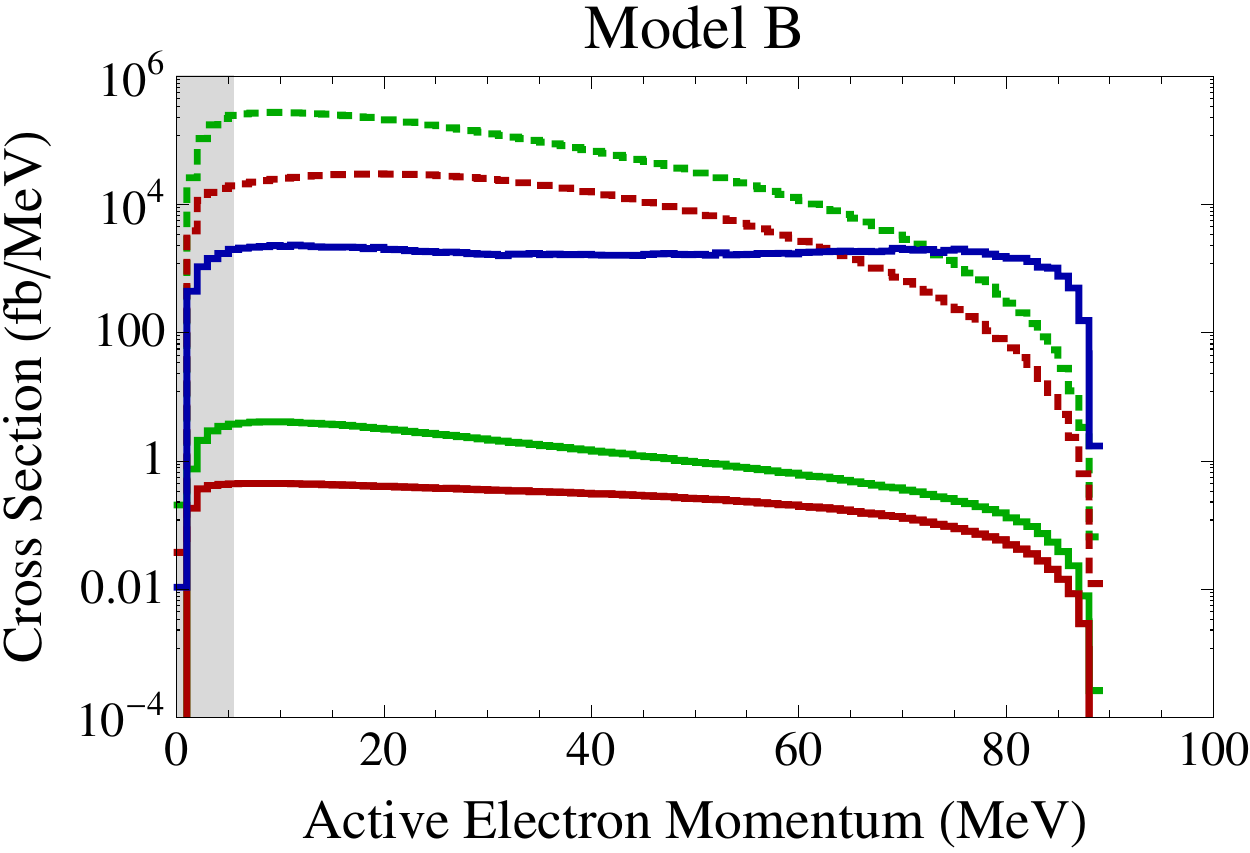} $\qquad$ \includegraphics[scale=0.45]{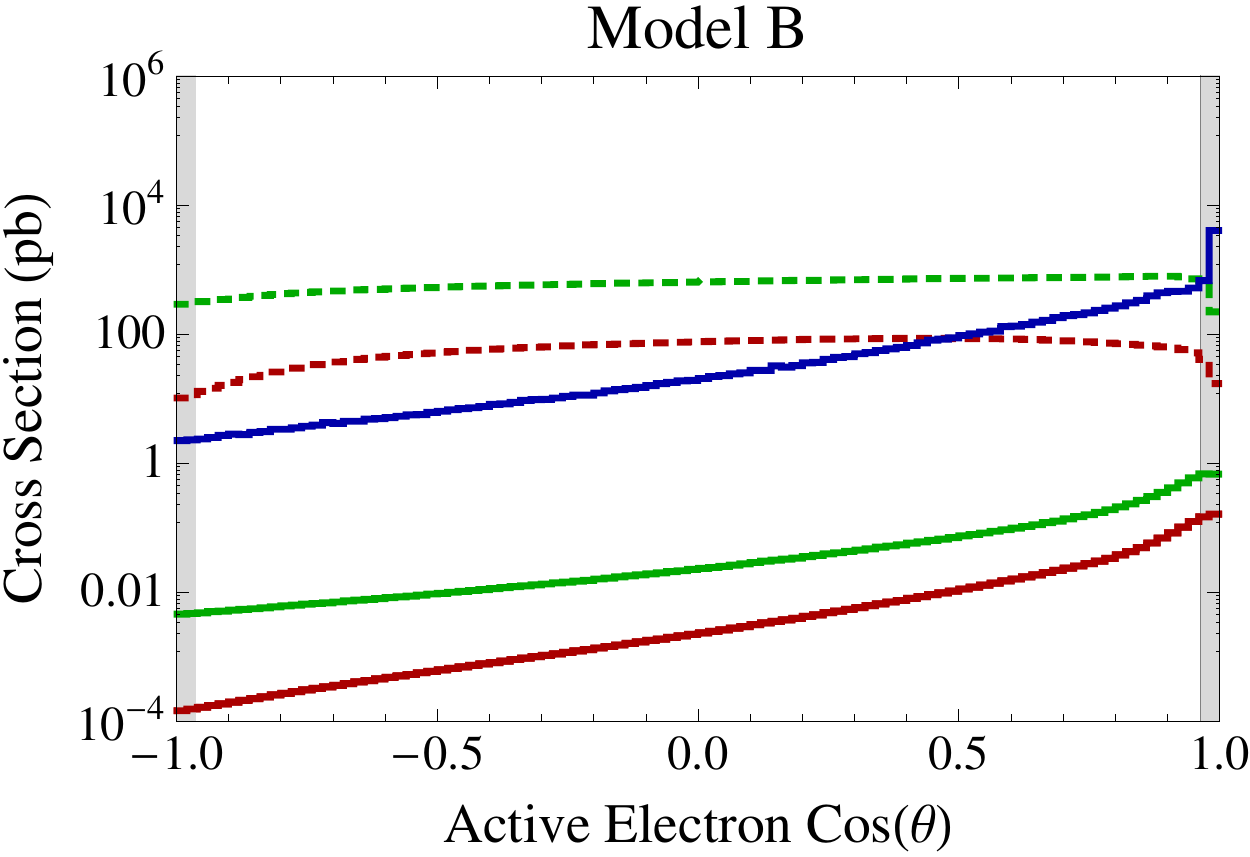}}
~\\
\centerline{\includegraphics[scale=0.45]{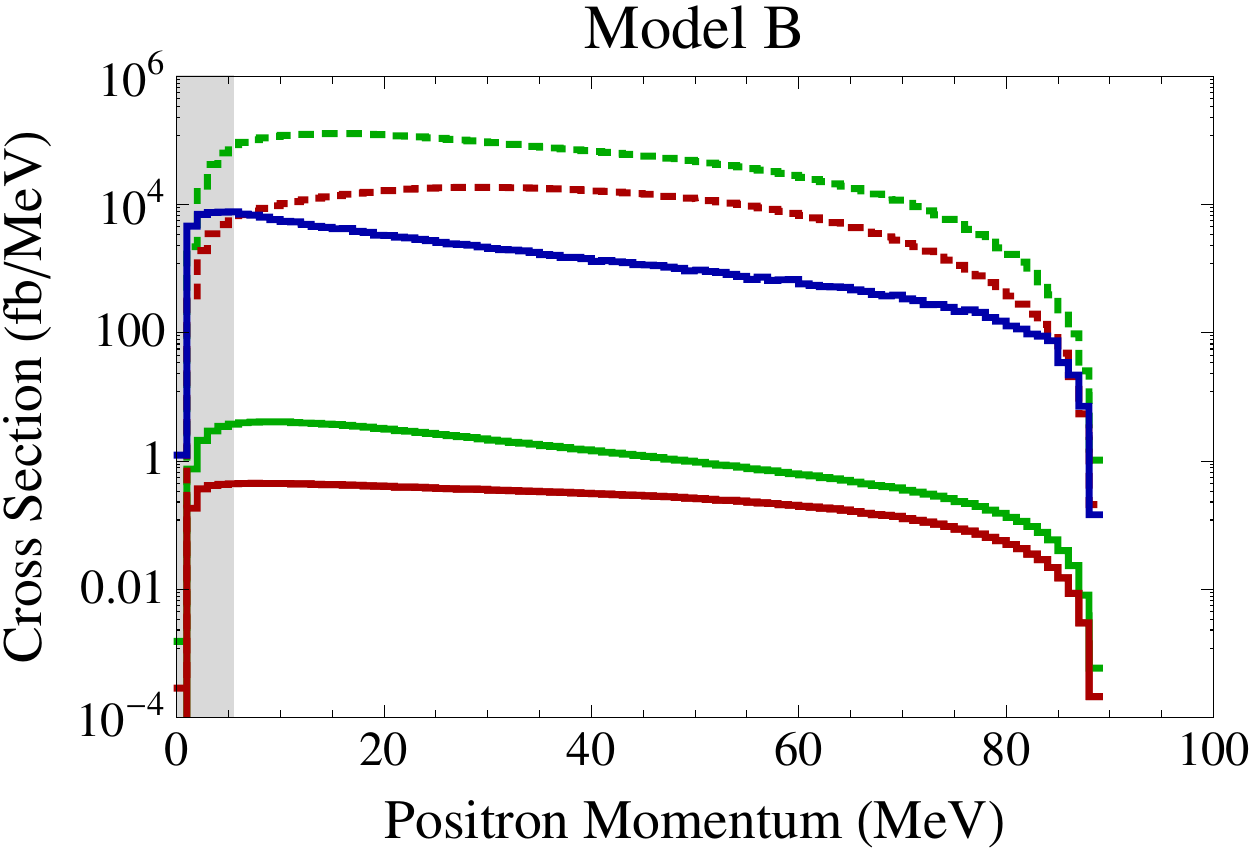} $\qquad$ \includegraphics[scale=0.45]{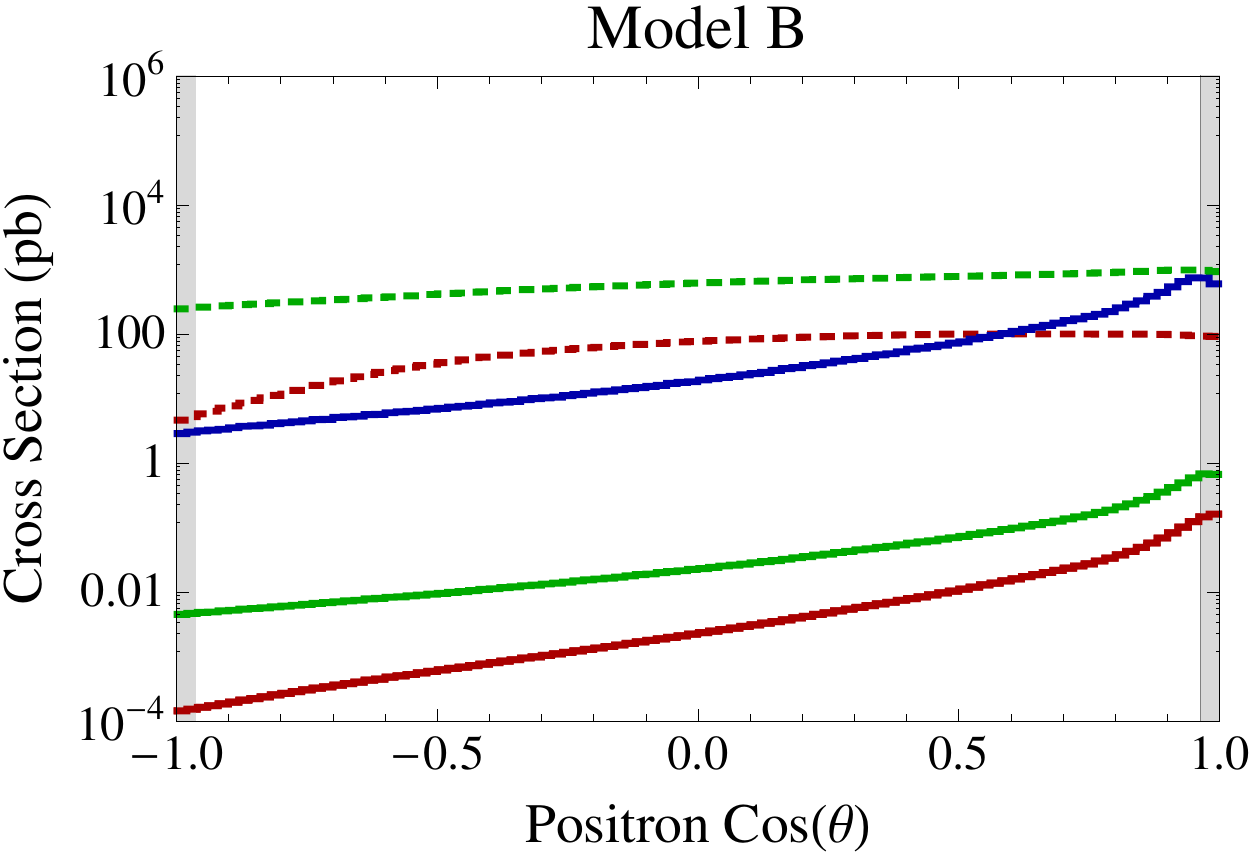}}
~\\
\centerline{\includegraphics[scale=0.55]{figures/KinematicLegend.pdf}}
\caption{Momentum and angular distributions for model B, with $m_X = 20 \MeV$ and $\alpha_X = 3 \cdot 10^{-9}$, analogous to \fig{fig:modelAscalar1}.
}
\label{fig:modelBscalar1}
}
\FIGURE[!p]{
\centerline{\includegraphics[scale=0.45]{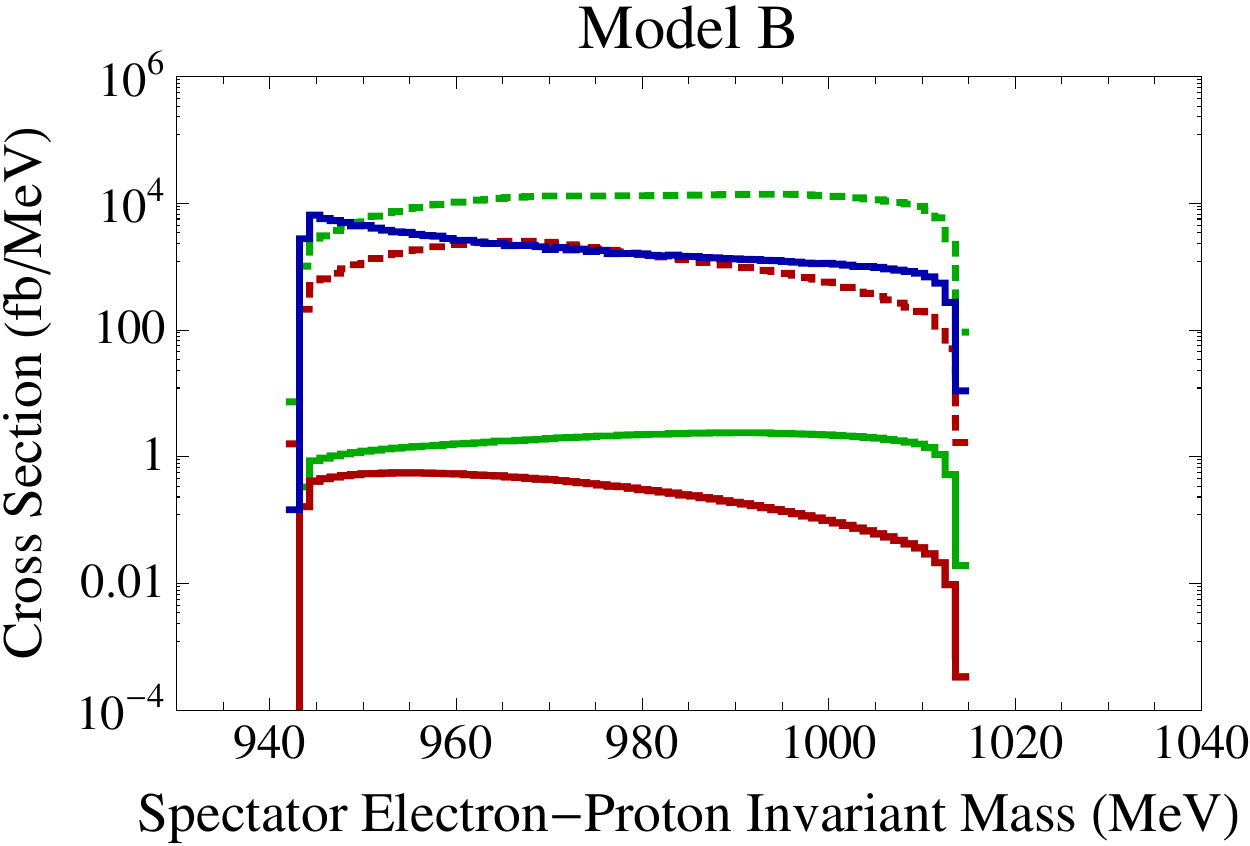} $\qquad$ \includegraphics[scale=0.45]{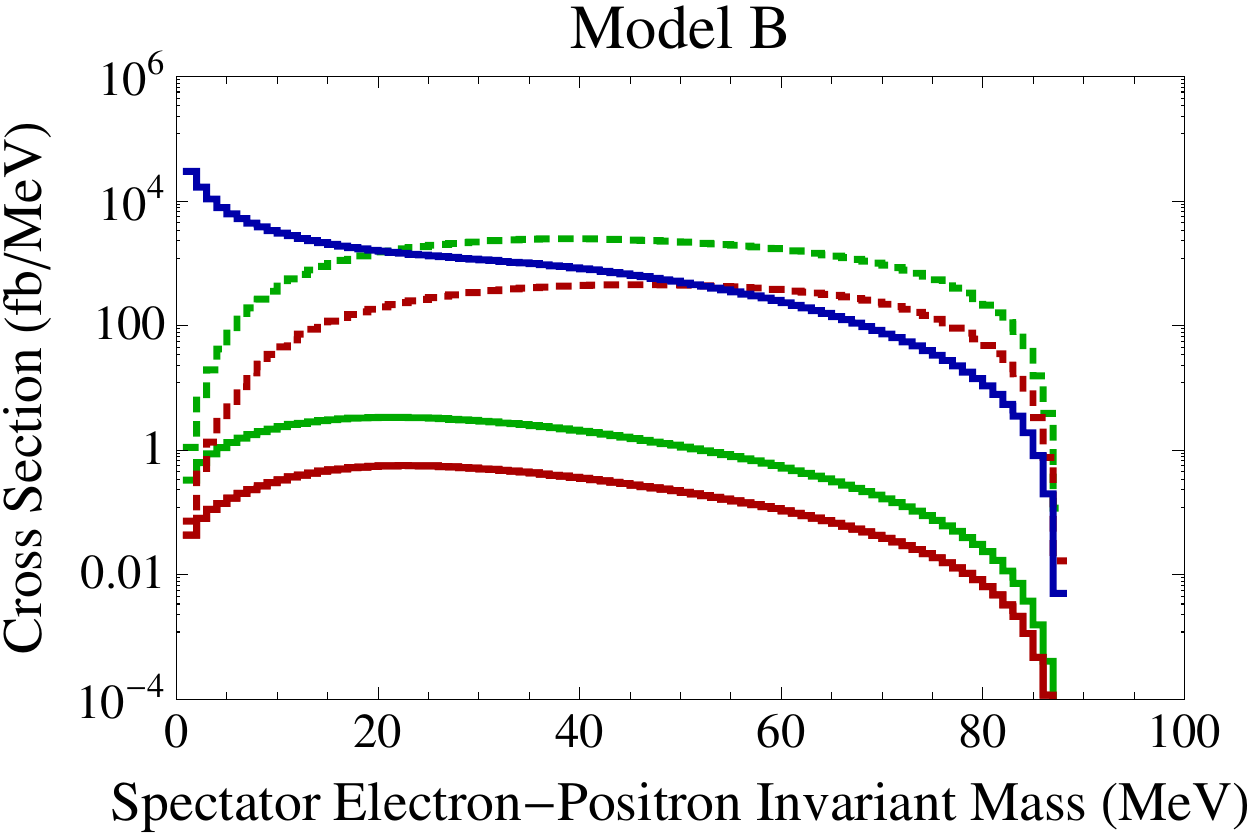}}
~\\
\centerline{\includegraphics[scale=0.45]{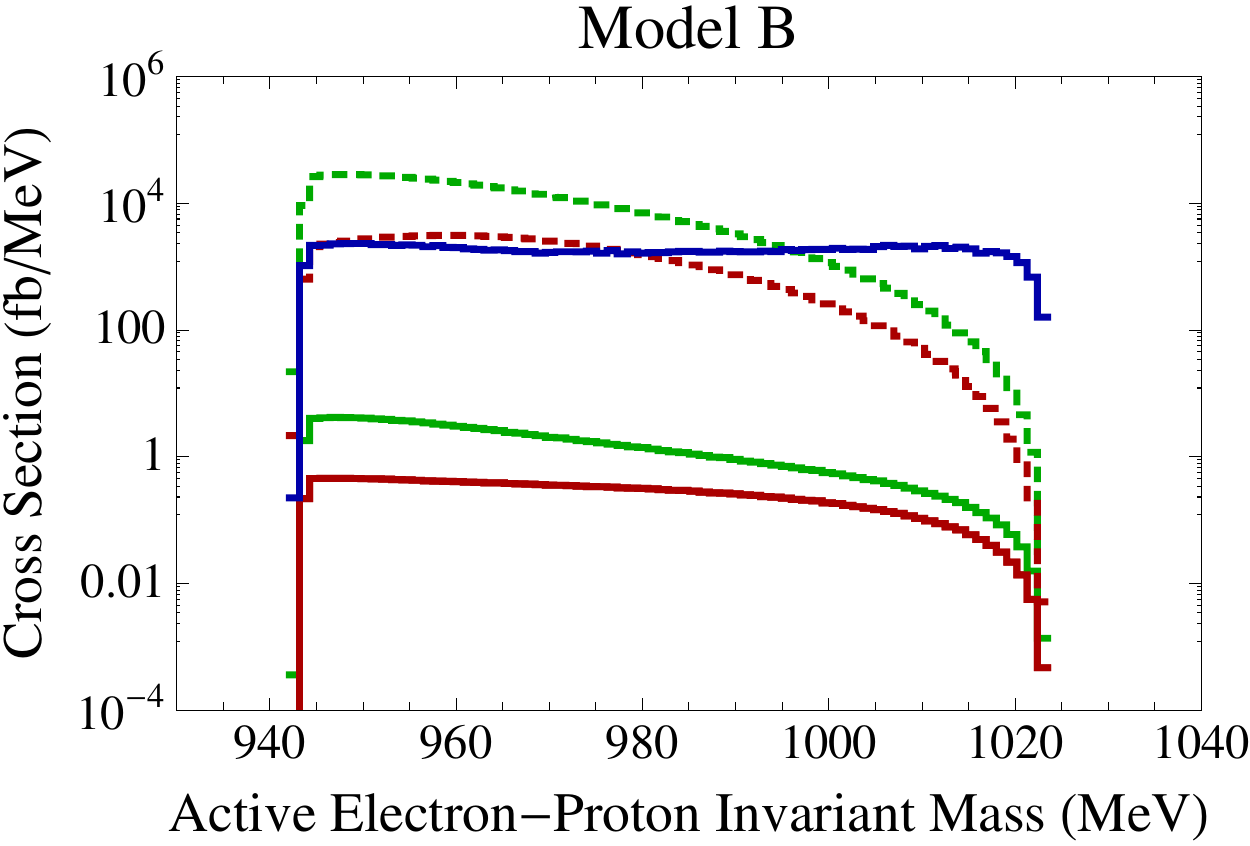} $\qquad$ \includegraphics[scale=0.45]{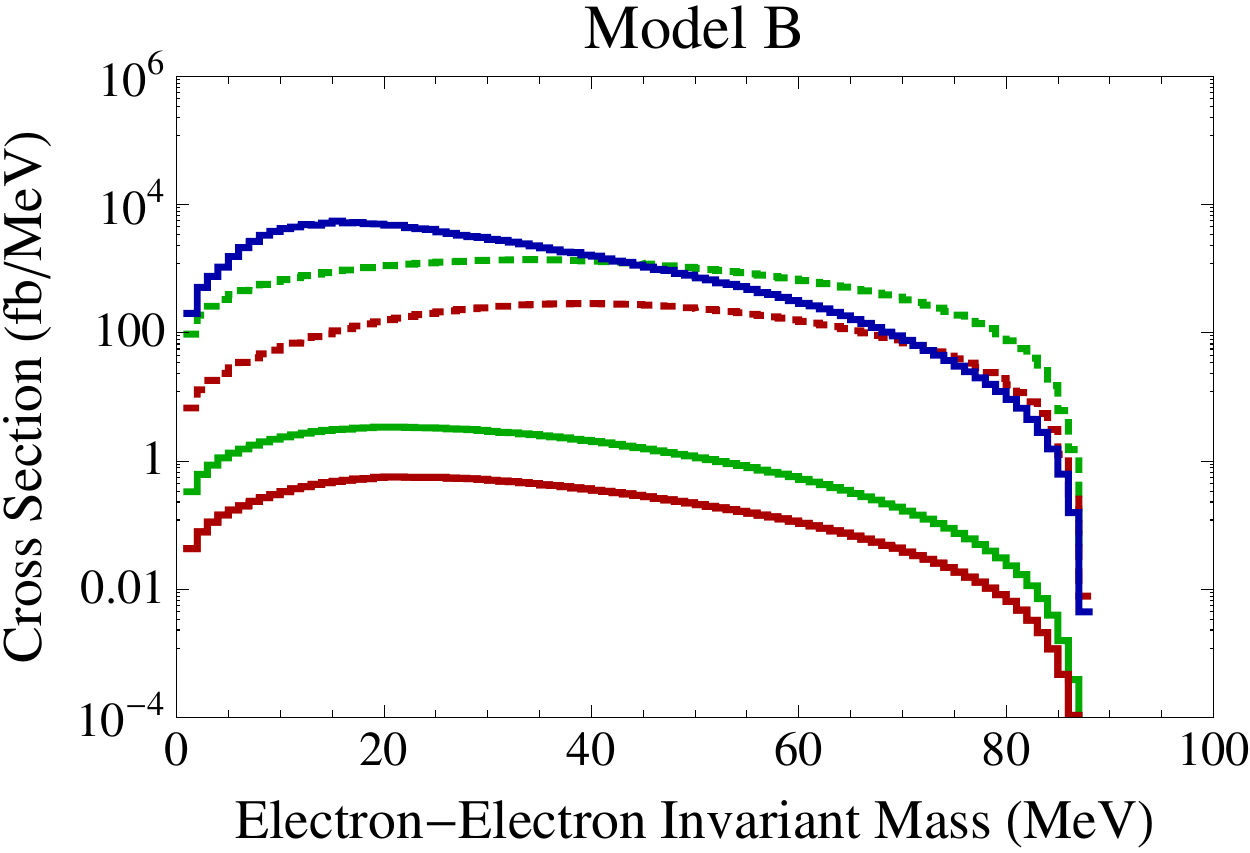}}
~\\
\centerline{\includegraphics[scale=0.45]{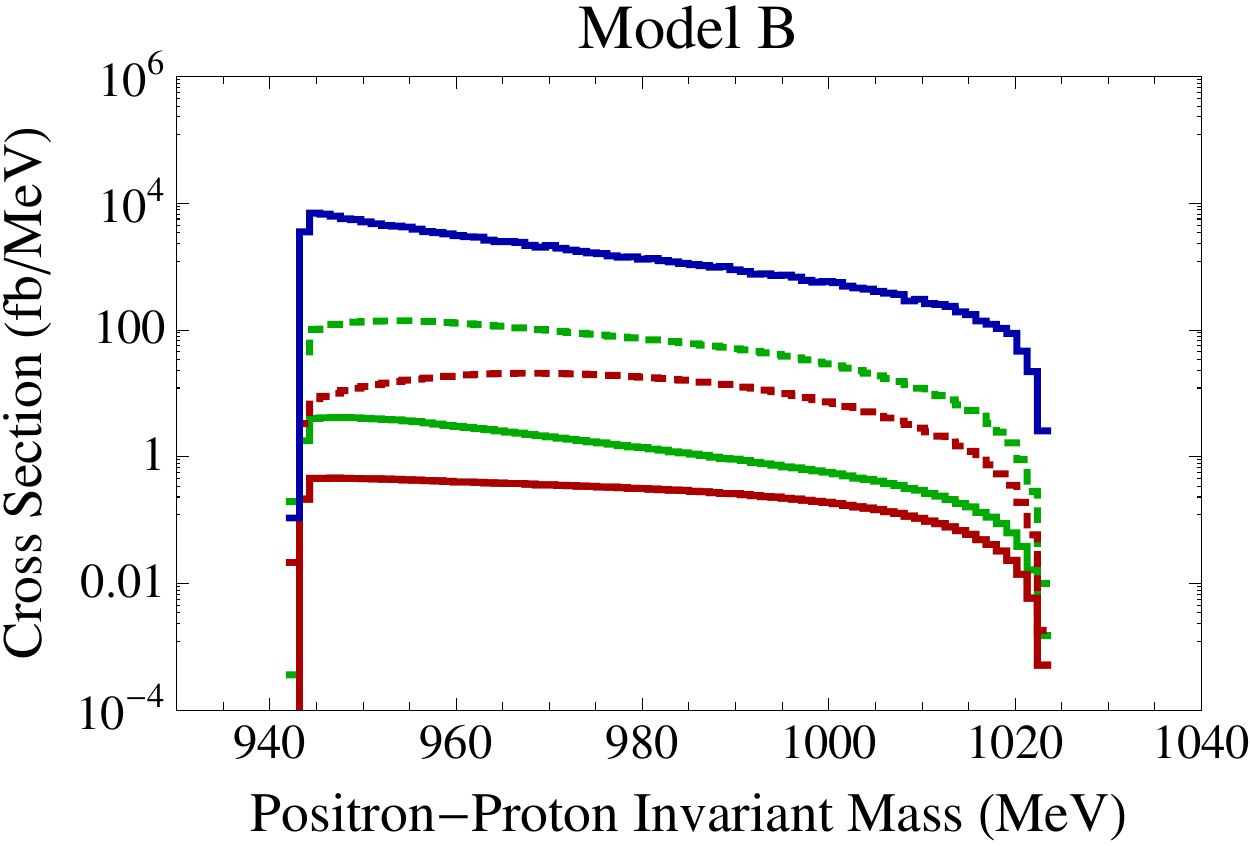} $\qquad$ \includegraphics[scale=0.45]{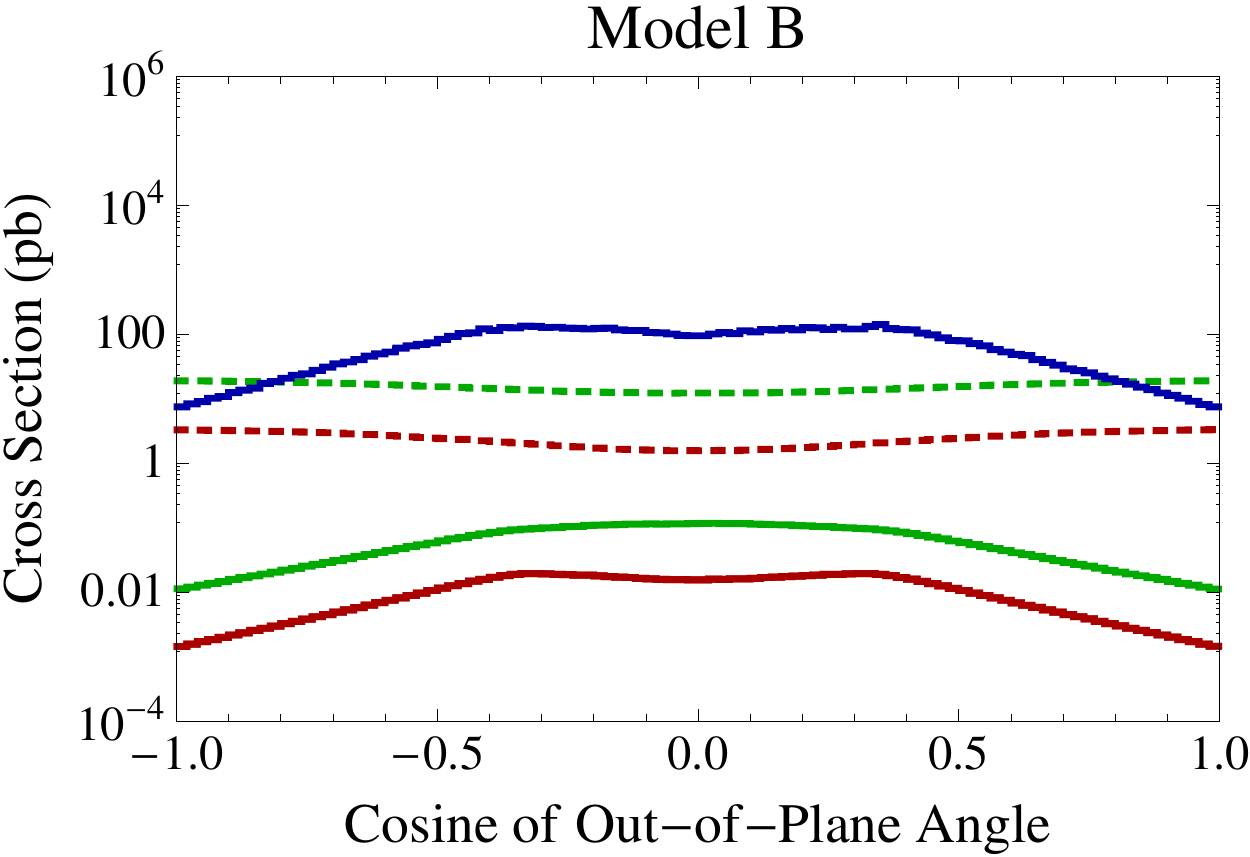}}
~\\
\centerline{\includegraphics[scale=0.45]{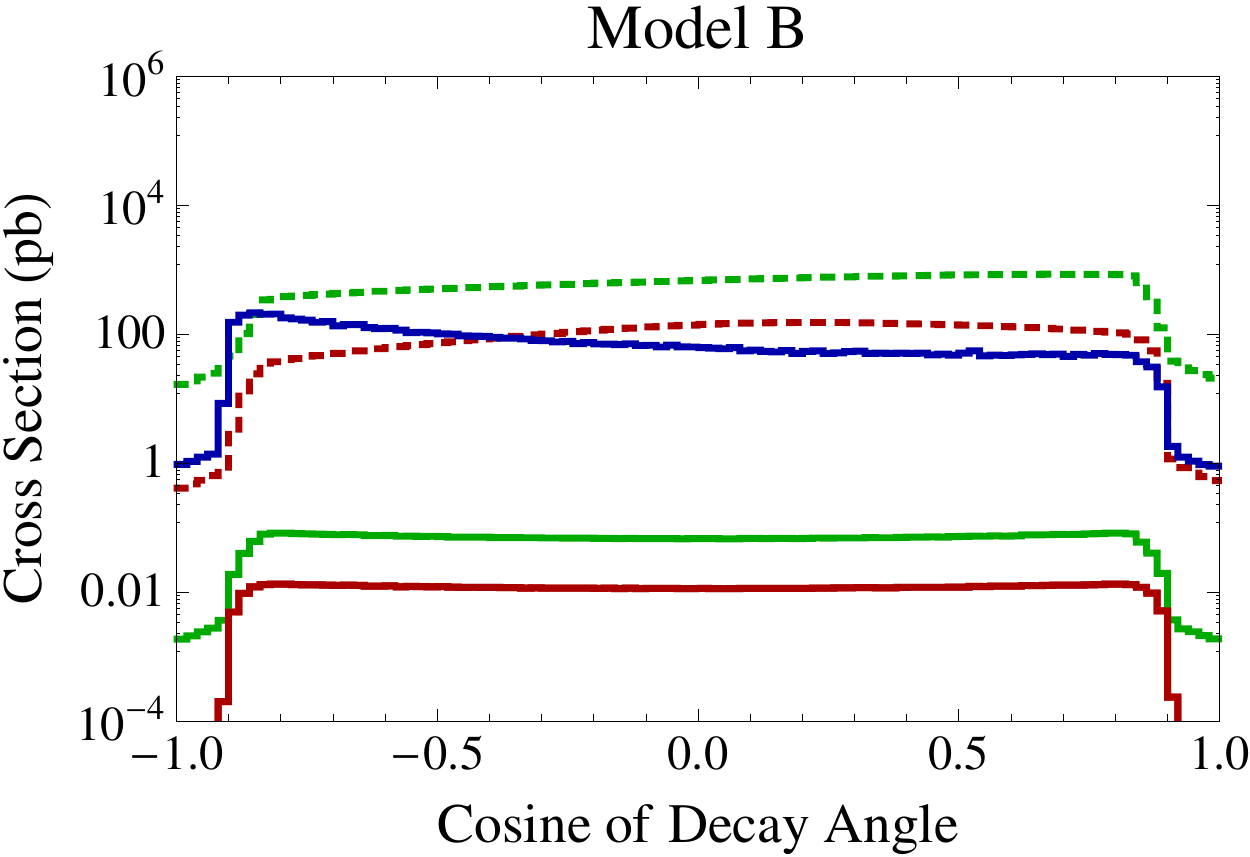} $\qquad$  \includegraphics[scale=0.45]{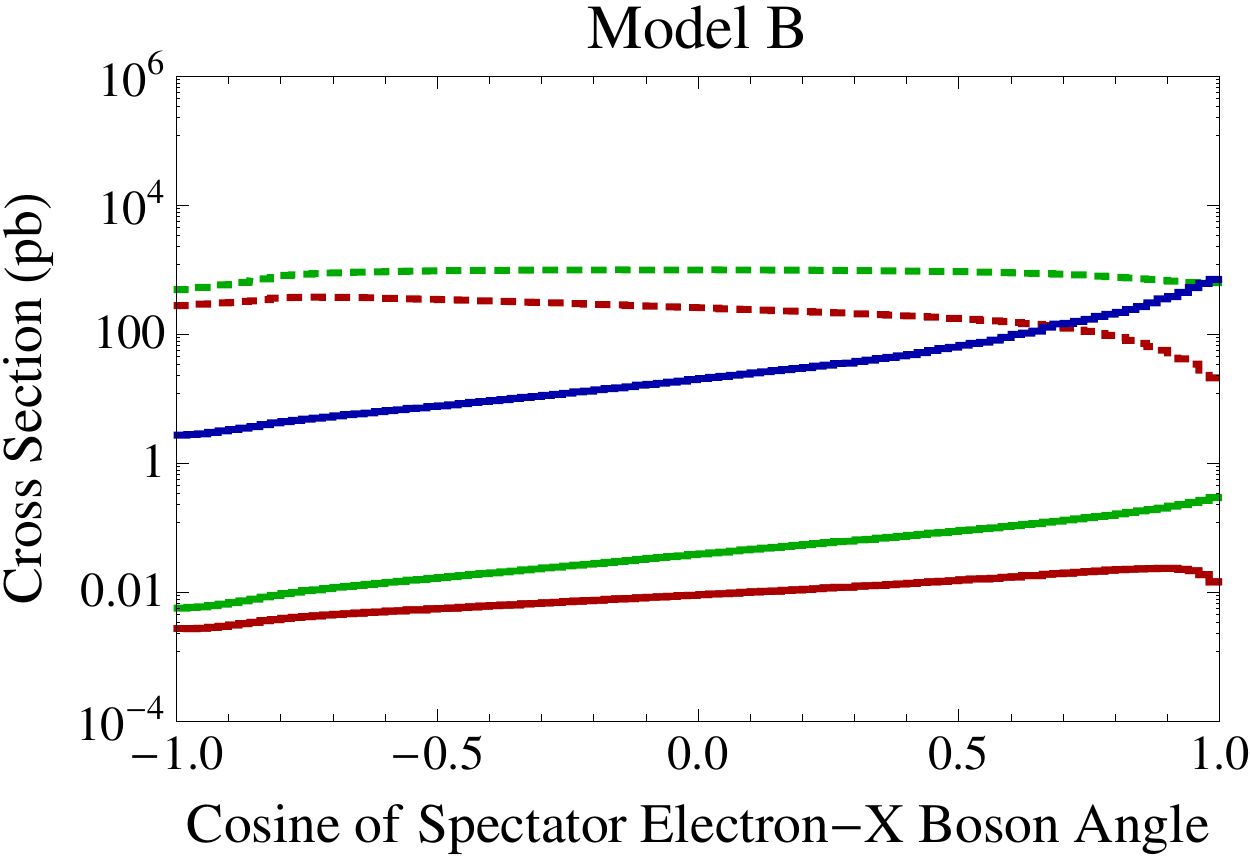}}
~\\
\centerline{\includegraphics[scale=0.55]{figures/KinematicLegend.pdf}}
\caption{Additional kinematic distributions for model B, analogous to \fig{fig:modelAscalar2}.
}
\label{fig:modelBscalar2}
}

The $X$ boson resonance signal could be further enhanced by using the matrix element method of \sec{sec:mereach}, which would be applied at each value of $m_{e^+ e^-}$.  To see how the matrix element method would affect our two benchmark points, we plot some example kinematic distributions in Figs.~\ref{fig:modelAscalar1}, \ref{fig:modelAscalar2}, \ref{fig:modelBscalar1}, and \ref{fig:modelBscalar2}.  Except as indicated, these plots are made using the fiducial detector geometry cuts of $-2 < \eta < 2$, $\mathrm{KE}_p > 0.5 \MeV$, and $\mathrm{KE}_{e^\pm} > 5 \MeV$, and require that at least one electron/positron pair reconstructs $m_X$.  The electron that reconstructs the candidate resonance is the active electron, and the other is the spectator electron.  In addition to the raw background and signal cross sections, we also plot the ideal weighting function from \eq{eq:bestweight}.\footnote{As a sanity check, we verified that the weighting function is approximately equal to the ratio of the signal and background in a given observable bin, appropriately normalized to the phase space volume of that bin.  See \appx{sec:memethod}.}  Large values of the weighting function correspond to regions of phase space that have the highest sensitivity to the $X$ boson.

In \figs{fig:modelAscalar1}{fig:modelBscalar1}, we show the momentum and angular distributions for the four outgoing fermions.  All of the detector geometry cuts are imposed, except that the cut corresponding to the plotted distribution is indicated via shading.

From the momentum distributions, we see that the kinetic energy cuts from \eq{eq;kineticenergycuts} do not cut out much of the most sensitive region, and indeed the cut on proton and spectator electron kinetic energy enhances the signal relative to the background.   For the angular distributions, because the background has a strong peak when an electron scatters in the forward direction, the weighting function is suppressed near $\cos(\theta_{\rm electron}) = 1$.  This explains why in \fig{fig:reachsweepeta}, the reach did not improve much in going from an $|\eta| < 2$ cut to an $|\eta| < 3$ cut.  Away from this peak, the weighting function has roughly flat sensitivity to the electron/positron angle, indicating that large angular acceptance is important for $X$ boson reconstruction.  Because of energy-momentum conservation, the proton can only scatter in the forward direction, and the most sensitive region is in fact in the most forward region.

In \figs{fig:modelAscalar2}{fig:modelBscalar2}, we show the pair-wise invariant mass distributions, as well as three angular distributions.  The out-of-plane angle is the angle the reconstructed $X$ boson moves relative to the plane defined by the incoming electron beam and the spectator electron.  The decay angle is the angle between the $X$ boson momentum and the outgoing positron momentum, as measured in the $X$ boson rest frame.  Also plotted is the angle between the spectator electron and the reconstructed $X$ boson.  

The invariant mass distributions do show some ability to distinguish signal from background.  Especially promising is the invariant mass between the spectator electron and the positron, which peaks at small values for the background.  The angular distributions show less promise, as the ideal weighting functions are relatively flat.  Note that the weighting function for the scalar and vector cases do have different shapes in the angular distributions, which explains why the reach plots in \sec{sec:reach} have different $m_X$ dependence even after correcting for the total signal cross section.  

Once the $X$ boson is discovered, the decay angle could be useful for distinguishing the scalar coupling from the vector case.  This distribution is flat for the scalar signal, but encodes non-trivial angular information in the vector case.  This angular variation is much smaller than the background, though, so a careful analysis would be necessary to extract the nature of the $X$ boson coupling.

\clearpage

\section{Prospects for Displaced Vertices}
\label{sec:displacedpossibility}

We saw in \sec{sec:beamdump} that there is a region of parameter space where the $X$ boson has not yet been ruled out by the beam dump experiments but has a lifetime long enough to leave a displaced vertex in $ep$ collisions.  While the $X$ bosons produced in low energy $ep$ collisions are generically not very boosted, there is enough of a tail in the lifetime distribution that a few $X$ bosons will have large displacements.  Moreover, with a diffuse gas target, one can in principle probe smaller vertex displacements than in a solid target.

Since the QED background does not generically lead to displaced vertices, this suggests that the reach could be extended beyond that presented in \sec{sec:reach}.   Of course, there are instrumental effects that can lead to fake displacement, but we ignore these in this study.  For the signal, the displacement can be straightforwardly calculated by convolving the $X$ boson momentum spectrum with the exponentially falling lifetime curve, properly taking into account boost factors.

\FIGURE[t]{
\centerline{\includegraphics[scale=0.7]{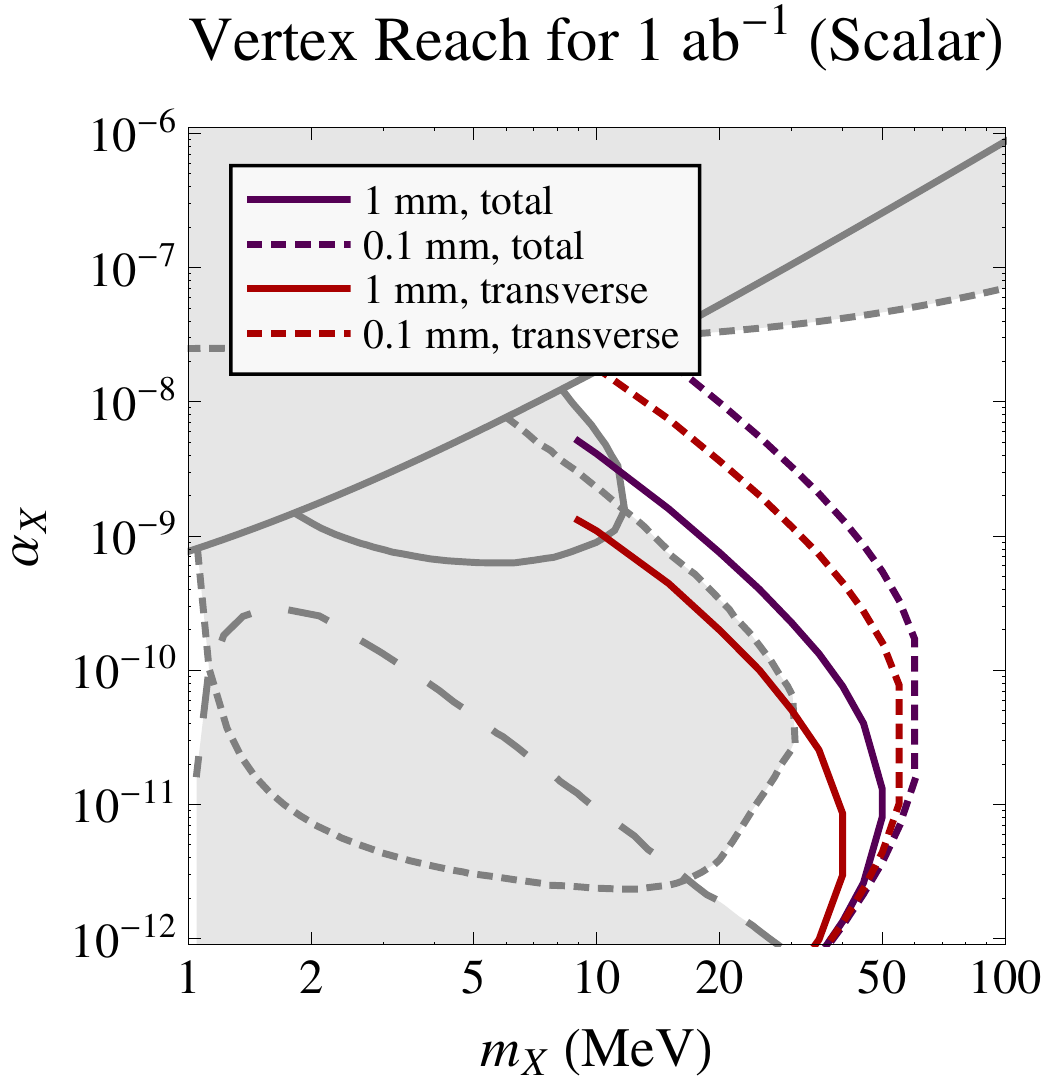} \includegraphics[scale=0.7]{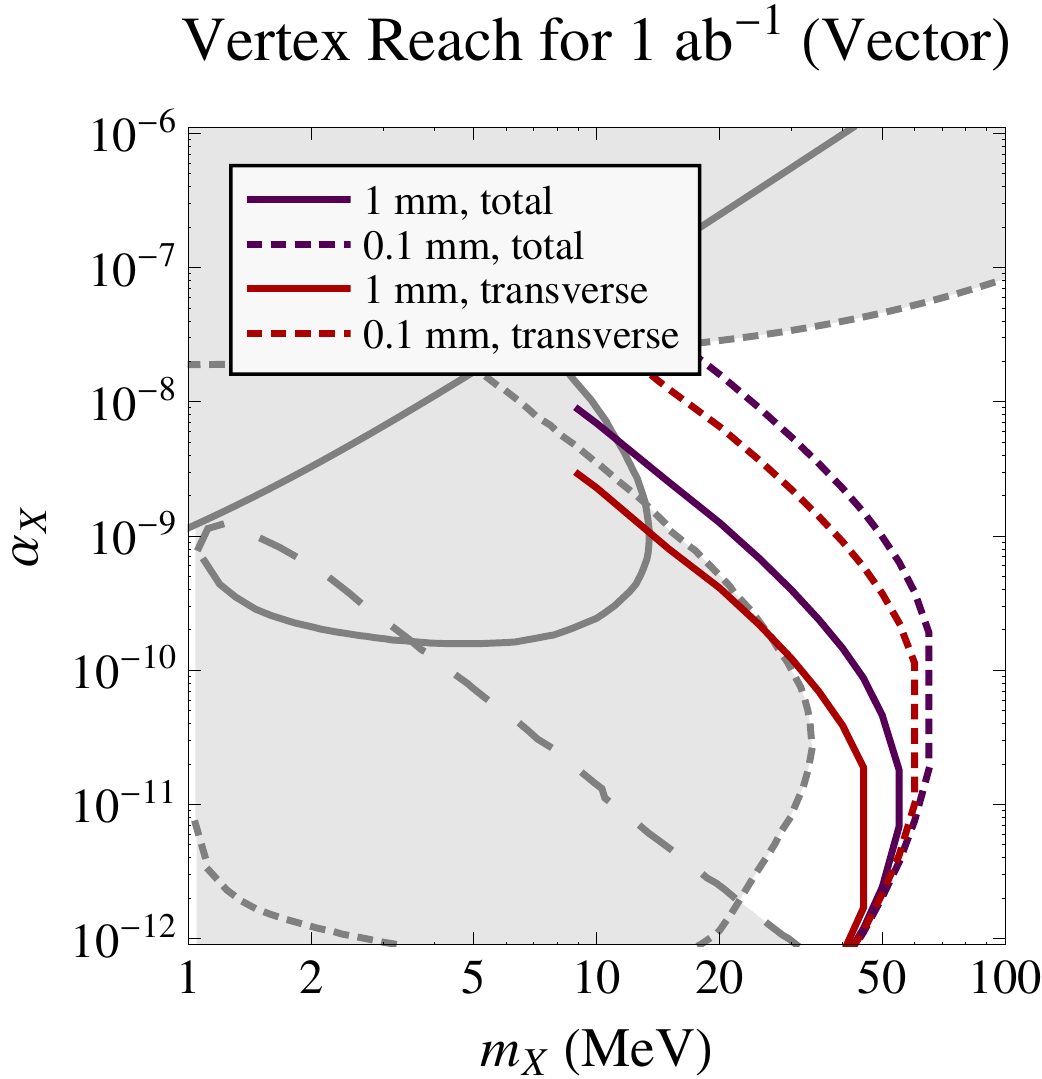}}
\caption{Reach in $ep$ scattering using a displaced vertex search strategy.  The curves correspond to producing 10 signal events with an integrated luminosity of 1 ab$^{-1}$.  The purple curves assume that the total displacement between the interaction vertex and the $X$ boson decay vertex can be measured.  The red curves assume that only the transverse displacement from the beam axis can be measured.  Note that the plotted range for $\alpha_X$ differs from the previous reach plots in this paper, and that an additional bound from the E137 experiment is shown.
}
\label{fig:reachvert}}

With no expected irreducible backgrounds, we estimate the naive reach by considering the region of parameter space where at least 10 displaced vertices would be observed.  The expected reach using two different search strategies is presented in \fig{fig:reachvert}.   The more aggressive strategy attempts to reconstruct the total displacement of the $X$ boson.  That is, the detected recoiling proton and spectator electron define the interaction vertex, and electron-positron pair from the $X$ boson define the decay vertex.  We plot the reach assuming a 1 mm or 0.1 mm total displacement could be observed.  Since there can be multiple scatterings per beam crossing, this method faces a background from uncorrelated scattering events.  

An alternative strategy is to merely reconstruct the transverse displacement of the decay vertex from the beam axis.  Since the $X$ boson momentum is peaked in the forward direction, this noticeably reduces the reach at the same displacement resolution.  Assuming transverse vertex displacement is easier to detect than total displacement, one might imagine that increased resolution might compensate to make this the preferred strategy.  Note that the transverse displacement search faces a possible background from photon conversion within the gas target.

Both displaced vertex strategies probe a different part of the parameter space from the search in \sec{sec:reach}, and therefore can be viewed as complementary to the QED background-limited analysis.  In particular, $X$ bosons with couplings two order of magnitude smaller than those accessible with a direct bump hunt could be seen.  At such small couplings, other prior beam dump experiments come into play beyond those discussed earlier in \sec{sec:beamdump}.  In \fig{fig:reachvert}, the beam dump constraint at the lowest couplings comes from the E137 experiment at SLAC \cite{Bjorken:1988as}, calculated analogously to the constraints in \sec{sec:beamdump}.

\section{Conclusions}
\label{sec:conclude}

Low energy electron-proton scattering is one of the basic processes in the standard model.  It is therefore intriguing that new physics might be discovered in a regime that is thought to be dominated by elastic and quasi-elastic QED processes.   Recent astrophysical anomalies have motivated a new paradigm for dark matter, where heavy dark matter interacts with a light, weakly coupled boson.  New low energy, high intensity scattering experiments are an ideal setting to constrain (or confirm) this exciting scenario.

We have argued an $X$ boson with $\alpha_X \sim 10^{-8}$ and $10 \MeV < m_X < 100 \MeV$ could be discovered in low energy $ep$ scattering with around 1 ab$^{-1}$ of data assuming 1 MeV invariant mass resolution.  Since the search for $X \rightarrow e^+ e^-$ is background limited, it is crucial to have an experiment with good energy resolution and very high statistics.  We believe that the unique combination of high luminosity with full event reconstruction makes this a compelling experimental proposal for the JLab FEL.  This proposal is complementary to the beam dump experiments envisioned in Refs.~\cite{Reece:2009un,Bjorken:2009mm}, which are better suited for smaller value of $\alpha_X$ and larger values of $m_X$.

We have shown that a matrix element method which uses complete kinematic information about the signal and background can increase the sensitivity to the $X$ boson by about a factor of 3.  Though not studied in this paper, a polarized electron beam could be useful in extracting additional matrix element information.  Of course, the most straightforward way to increase the sensitivity of the experiment is to improve the invariant mass resolution beyond our fiducial value of 1 MeV.  Finally, if reconstruction of 1 mm or 0.1 mm displaced vertices are possible, then $ep$ scattering could probe an interesting region of $X$ boson parameter space with smaller couplings.

\acknowledgments{We thank our experimental colleagues at MIT---Peter Fisher, Richard Milner, and Sinh Thong---for extensive discussions on the upcoming JLab FEL proposal.  We also benefitted from experimental advice from Rolf Ent, Ronald Gilman, Yury Kolomensky, Christoph Tschal\"ar, and Bogdan Wojtsekhowski, and theoretical conversations with Rouven Essig, Maxim Pospelov, Philip Schuster, Matthew Strassler, Natalia Toro, Scott Thomas, and Lian-Tao Wang.  G.O. thanks Henry Grishashvili for computing help.  J.T. thanks the Aspen Center for Physics for their hospitality during the completion of this work, and the Miller Institute for Basic Research in Science for funding support.}

\appendix

\section{Finite Mass Calculations}
\label{app:finitemass}

\subsection{Anomalous Magnetic Moment}
\label{app:gminus2detail}

A new light boson, whether it is a scalar or vector, will contribute to the anomalous magnetic moment ($a = \frac{g-2}{2}$) of leptons at the one-loop level.  In the low masses being considered for the $X$ boson, limits on new contributions to the moment are the main indirect constraint and must be calculated accurately.  In the scalar/pseudoscalar case, a convenient choice of parameterization yields \cite{Sinha:1985aw} 
\begin{align}
 \delta a_{s/p} &= \frac{m_\ell^2}{16\pi^2}\int_0^1 dz 
    \frac{\lambda_s^2(1-z)(1-z^2) - \lambda_p^2(1-z)^3}{zm_X^2 + (1-z)^2m_\ell^2}\\
    &\approx \frac{1}{16\pi^2}\frac{m_\ell^2}{m_X^2}
    \left(\lambda_s^2\left(\log\frac{m_\ell^2}{m_X^2} - \frac{7}{6}\right) - \lambda_p^2\left(\log\frac{m_\ell^2}{m_X^2} - \frac{11}{6}\right)\right).
\end{align}
This final form only holds in the limit $m_\ell \ll m_X$, and is thus not appropriate for $a_\mu$.

For the case of a vector coupling, the term proportional to $p^\mu p^\nu$ in the numerator of the propagator drops out of the calculation entirely due to the Ward identity.  However, for axial coupling this is not the case.  At the same time, the extra factor of $1/m^2_X$ this term introduces means it only contributes at $\mathcal{O}(m_e^4/m_X^4)$.  Thus, it can be safely ignored in the electron case, but for constraints from the muon anomalous magnetic moment it must be included for accurate results. The full calculation yields~\cite{Leveille:1977rc}
\begin{align}
  \delta a_{v/a} &= \frac{m_\ell^2}{16\pi^2}\int_0^1 dz 
    \frac{4\lambda_v^2z(1-z)^2 - 4\lambda_a^2\left(z(1-z)(3+z) + 2(1-z)^3\frac{m^2_\ell}{m^2_X}\right)}{zm_X^2 + (1-z)^2m_\ell^2}\\
    &\approx \frac{1}{16\pi^2}\frac{m_\ell^2}{m_X^2}\left(\lambda_v^2\frac{4}{3} - \lambda_a^2\frac{20}{3}\right) ,
\end{align}
where again the approximate form only holds for $m_\ell \ll m_X$. 

\subsection{Lifetime}
\label{app:lifetimedetail}

\FIGURE[t]{
\centerline{\includegraphics[scale=0.7]{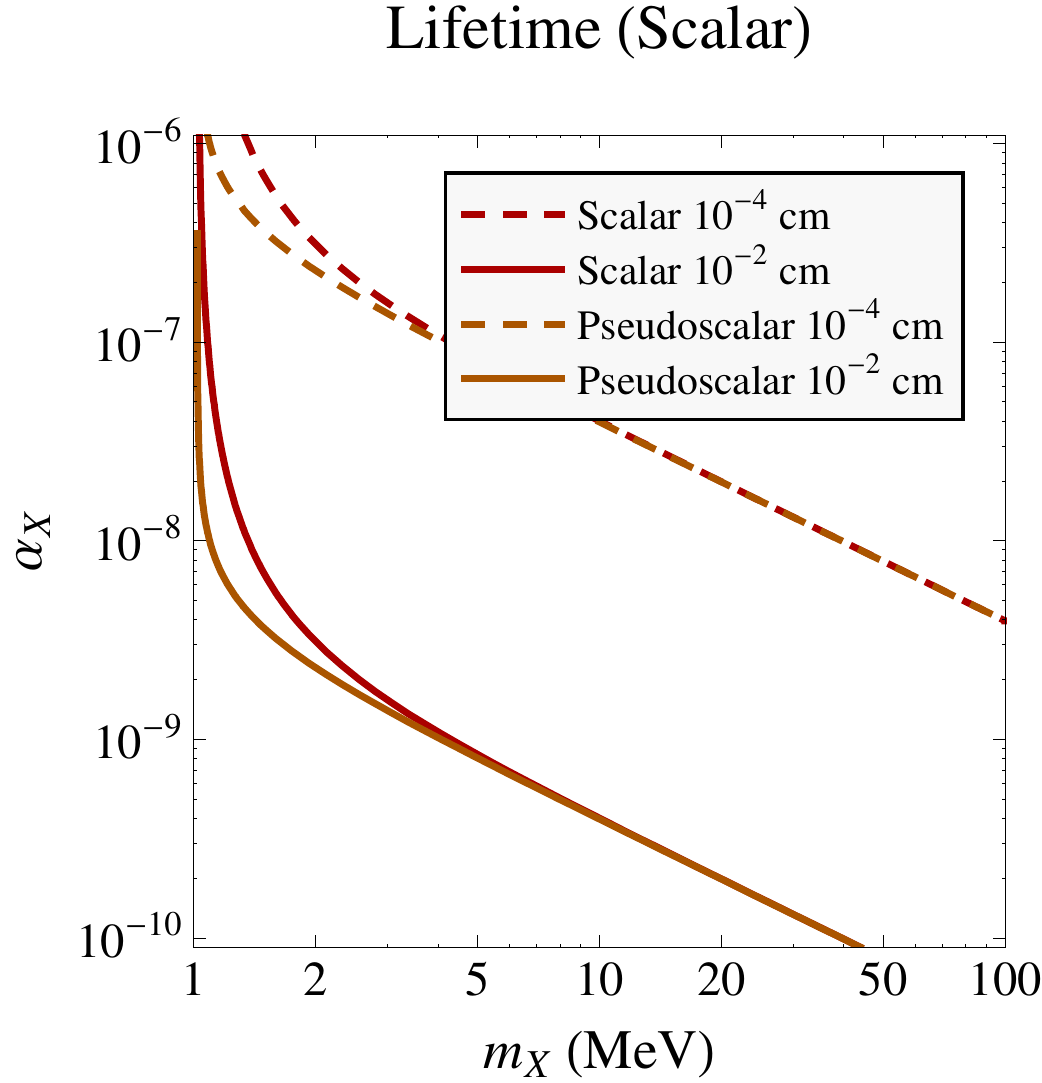} \includegraphics[scale=0.7]{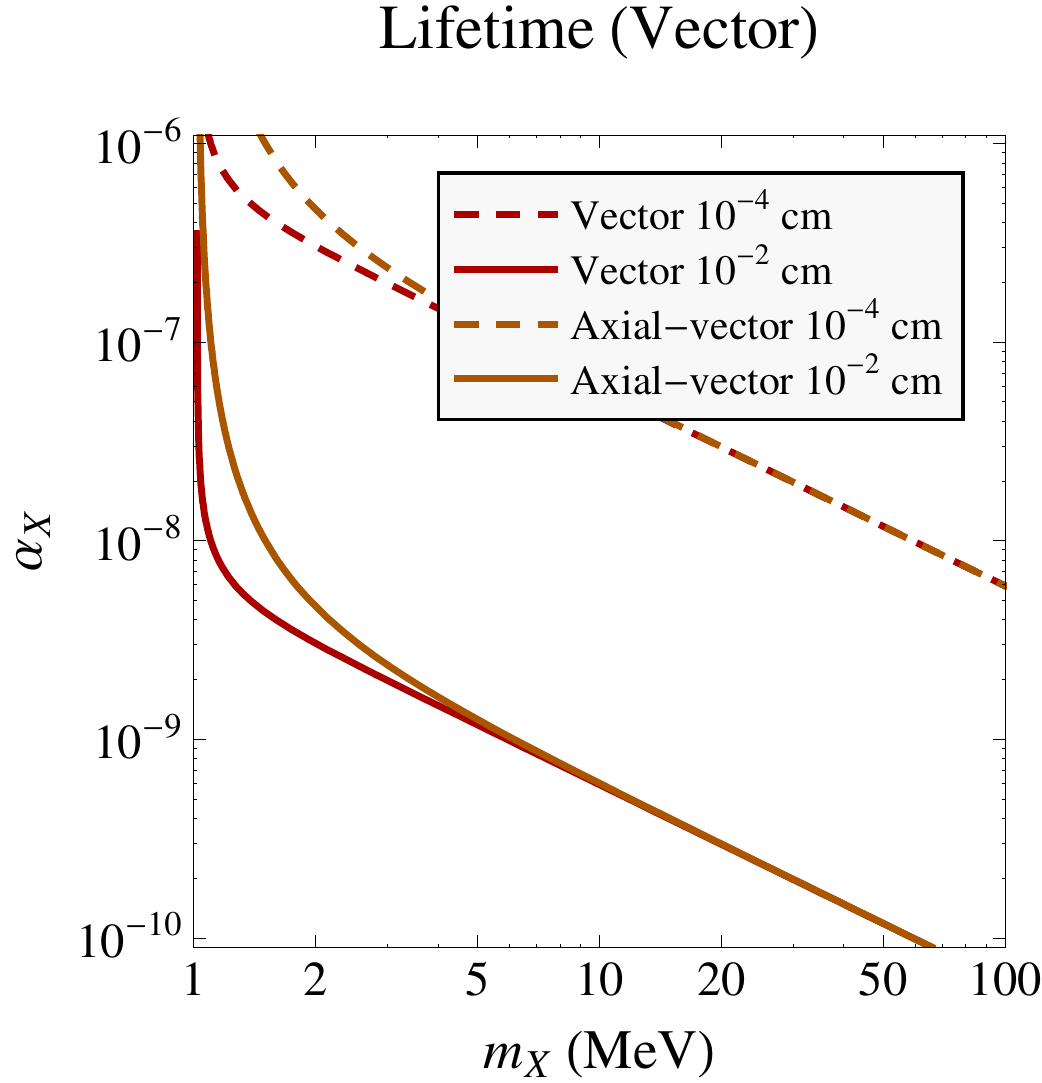}}
\caption{Curves of constant $X$ boson lifetime on the $\alpha_X$ vs. $m_X$ plane.}
\label{fig:xlifetimescalar}
}

If $m_X < 2m_\mu$, then for the couplings envisioned in \sec{sec:theory}, the only allowed decay mode of the $X$ boson is $X \to e^+e^-$.  In more general scenarios, the $X$ boson might decay to neutrinos or photons or other new light fields, but such final states face very different backgrounds and are outside the scope of this study.

The tree-level widths from the couplings described in \sec{sec:theory} are
\begin{align}
  \Gamma_{s/p} &= \frac{m_X}{8\pi}\left((\lambda_s^2 + \lambda_p^2) - \lambda_s^2\frac{4m_e^2}{m_X^2}\right)\sqrt{1-\frac{4m_e^2}{m_X^2}} , \\
  \Gamma_{v/a} &= \frac{m_X}{12\pi}\left((\lambda_v^2 + \lambda_a^2)
    + 2(\lambda_v^2 - 2\lambda_a^2)\frac{m_e^2}{m_X^2}\right)\sqrt{1-\frac{4m_e^2}{m_X^2}}.
\end{align}
In the limit $m_e \ll m_X$, these simplify to $\Gamma_{s/p} = \frac{m_X}{8\pi}(\lambda_s^2 + \lambda_p^2)$ and $\Gamma_{v/a} = \frac{m_X}{12\pi}(\lambda_v^2 + \lambda_a^2)$.  A plot of the $X$ lifetime is shown in \fig{fig:xlifetimescalar} with full $m_e$ dependence, which introduces small differences in the scalar/pseudoscalar and vector/axial-vector cases at masses close to the electron mass.

\section{Signal Calculation}
\label{sec:signal}

In this appendix, we review the narrow width approximation and apply it to $X$ boson production.  The signal for $X$ boson production is $e^{-} p\rightarrow  e^{-} p + X\rightarrow e^{-} p \, e^+ e^-$.  While other diagrams with internal $X$ boson propagators can contribute to the $e^{-} p \, e^{-} e^{+}$ final state, only the diagrams shown in \fig{fig:signal2t3} yield a narrow peak in the $e^+ e^-$ invariant mass distribution near $m_X$.

It is well-known that processes involving a resonance with a small width can be treated by the so-called narrow width approximation (see Ref.~\cite{Uhlemann:2008pm} and references therein).  In the case of $X$ boson production, the idea is to write the square of the full two-to-four matrix element of diagrams in \fig{fig:signal2t3} as
\begin{align}
\vert M(2\rightarrow 4) \vert^2 =\vert \tilde M \vert^2(q, ...)  \times D(q), \qquad D(q)=\frac{1}{(q^2-m_X^2)^2+\Gamma_X^2 m_X^2},
\end{align}
where $q$ is the four-momentum flowing through the $X$ boson propagator.  One then approximates the square of the $X$ boson propagator as
\be
D(q)\approx  \frac{\pi}{m_X \Gamma_X } \delta(q^2-m_X^2).
\ee
Using the cluster decomposition of four-body phase space into a product of three-body and two-body phase space,
\be
\df \Phi_4 = \frac{1}{2\pi}\df \Phi_3 \, \df q^2 \, \df \Phi_2,
\ee
the fully differential cross section takes the form
\begin{align}
\sigma_{\text{signal}} & =\frac{1}{F}\int \df{\Phi_4} \frac{1}{4}\sum_{\text{spins}} \vert M(2\rightarrow 4)\vert^2\nonumber\\
   & = \frac{1}{F}\frac{1}{\Gamma_X}\int \df{\Phi_3}\int \df{\Phi_2}\frac{1}{4}\sum_{\text{spins}} \vert \tilde M(q, ...)\vert^2\frac{1}{2m_X} .\label{signalNW}
\end{align}
Here, $F$ is the incoming flux, which in the case of a fixed target experiment with incoming electron energy $E_e$ equals
\begin{equation}
   F=4 E_e m_p.
\end{equation}

When the $X$ boson has scalar or pseudoscalar couplings, the matrix element of the full two-to-four process takes a convenient factorized form:
\begin{equation}
\label{scalarMEfact}
\vert \tilde M(q, ...)_S\vert^2=\vert M_S(2\rightarrow 3)\vert^2 \, \vert M_S(1\rightarrow 2)\vert^2 .
\end{equation}
For calculating the total cross section, one can perform the $\df\Phi_2$ integral in \eq{signalNW} analytically:
\begin{equation}\label{narrowfin}
  \sigma_{\text{signal scalar}}=\frac{1}{F}\int \df{\Phi_3}\frac{1}{4}\sum_{\text{spins}} \vert M_S(2\rightarrow 3)\vert^2  \left( \frac{\Gamma_{X\rightarrow e^{+}e^{-}}}{\Gamma_X} \right). 
\end{equation}
The factor in parentheses is just the branching fraction of $X$ to the $e^+ e^-$ final state.  Of course, for any real observable, there are always cuts present, at minimum on the detector geometry.  So in practice, one must use the full formula in \eq{signalNW} and multiply the integrand by the desired observable function.

When the $X$ boson has vector or axial-vector couplings, the matrix element does not factorize, but the amplitude does factorize into a contraction between the production of the resonance, the decay of the resonance, and the numerator of the $X$ boson propagator:
\begin{equation}\label{vectorMEfact}
   \vert \tilde M_V(q, ...)\vert^2=\Big\vert M_V^{\mu}(2\rightarrow 3) M_V^{\nu}(1\rightarrow 2) \left( g_{\mu\nu}-\frac{q_{\mu} q_{\nu}}{M_X^2}\right)\Big\vert^2.
\end{equation}
It is worth mentioning that in the vector case, the $q_{\mu} q_{\nu}$ term in the propagator vanishes because of Ward identity:
\begin{equation}
   q_{\nu} M_V^{\nu}(1\rightarrow 2) =0,
\end{equation} 
since the external electron/positron are on-shell.  In the axial-vector case, the Ward identity is no longer true, but because of chiral symmetry, the correction from the $q_{\mu} q_{\nu}$ term is suppressed by powers of $m_e$ and therefore small.  Because of the $X$ boson propagator factor, there are non-trivial angular correlations in the $X$ boson decay.

There is one subtlety in using the narrow width approximation with identical particles in the final state.  Experimentally, we cannot determine which final state electron came from the decay of the resonance and which one is the scattered incoming electron.  Therefore, in principle, we must add to the diagrams in \fig{fig:signal2t3} another two diagrams with the electron legs in the final state interchanged.  However, the effect of adding these diagrams but including a symmetry factor of $1/2$ into the phase space for identical electrons leaves \eq{signalNW} unchanged.  Since \eq{signalNW} is valid for all regions of phase space, the differential distribution in $m_{e^{+}e^{-}}$ will include not just a delta function spike at $m_X$, but also the correct combinatoric background.  For any real experimental observable, this delta function spike will be properly smeared out by the experimental resolution, as long as the experimental resolution is coarser than $\Gamma_X$.

\section{Background Calculation}
\label{sec:background}

The background to $X$ boson production consists of quasi-elastic QED interactions, with example diagrams given in \fig{fig:background}.  Since we are considering incoming beam energies below the pion mass, no QCD interaction are relevant, and the proton remains intact after being struck by the electron.  This simplifies the analysis of the standard model background to the proposed signal dramatically.  Also, since our target is hydrogen gas, we do not have to worry about nuclear excitations.

Given that the background is QED, we can safely consider tree-level diagrams alone, since loops are suppressed by the small electromagnetic coupling.  That said, while the one-loop corrections to the background are much smaller than the tree-level result, they are also expected to be much larger than our signal.  However, one-loop corrections are not expected to dramatically change the shape of the background (and certainly not give a peak at finite $m_{e^+e^-}$), so for the purposes of extracting a signal peak in the invariant mass distribution, the tree-level background result will suffice.

The largest correction we are neglecting comes from the electric form factor of the proton
\be
f_E(q^2) \simeq \left(\frac{m_0^2}{m_0^2 + q^2}\right)^2,
\ee 
which had been included in the study in Ref.~\cite{Heinemeyer:2007sq}.  With $m_0 \sim 700 \MeV$ and $q \lesssim E_e = 100 \MeV$, this will at most yield a 5\% change in the background calculation.  Like the one-loop effects, though, the electromagnetic form factor is not expected to dramatically change the background shape.

There are twelve QED diagrams that contribute to the process $e^-p \rightarrow e^- p \, e^+ e^-$.  Just as in M{\o}ller scattering, one must be mindful that the interchange of identical fermionic legs adds an additional minus sign to the Feynman rules \cite{Peskin:1995ev}.  The background cross section equals
\begin{align}
   &\sigma_{\text{background}} ({\cal{O}})=\frac{1}{F}\int \df{\Phi_4} \frac{1}{2}\frac{1}{4}\sum_{\text{spins}} \vert M_{\text{background}}\vert^2 {\cal{O}}(\Phi_4),\label{backgroundFormula}
\end{align}
where $1/2$ is a symmetry factor from having two identical electrons in the final state, $1/4$ the average over initial polarizations and $\mathcal{O} (\Phi_4)$ is an arbitrary observable.

In particular, for calculating the background in the signal bin for a given value of $m_X$, we use the theta function
\be
{\cal{O}}(\Phi_4) = \theta\left( (m_X - \Delta m/2)^2 < q^2 < (m_X + \Delta m/2)^2 \right),
\ee
where $q^2$ is the invariant mass of an outgoing $e^+e^-$ pair, and $\Delta m$ is the invariant mass resolution.  Because the background is a steeply falling distribution in $q^2$, as shown in \fig{fig:eeinvmass}, this way of calculating the binned background gives slightly more realistic values than
\be
{\cal{O}}(\Phi_4) = 2 m_X \Delta m \, \delta(q^2 - m_X^2),
\ee
though both measurements agree in the small $\Delta m$ limit.

\section{Matrix Element Method}
\label{sec:memethod}

In this appendix, we derive the ideal weighting function to be used in the matrix element method from \sec{sec:mereach}.  To start, consider an unweighted measurement
\be
S = \int \df \Phi \,  S(\Phi), \qquad B = \int \df \Phi \,  B(\Phi).
\ee
The statistical uncertainty in the background $\delta B$ can be determined in terms of the Poisson uncertainty at each point in phase space $\delta B(\Phi)$,
\be
\delta B(\Phi) = \sqrt{B(\Phi)}, \qquad \delta B = \sqrt{\int \df \Phi \,  \left[\delta B(\Phi)\right]^2} = \sqrt{B},
\ee
and we recover the familiar formula that $S/\delta B = S/\sqrt{B}$.\footnote{Strictly speaking, we should really consider statistical uncertainties in both the signal and background, but the background is so much larger than the signal that this is superfluous.}

Now imagine doing a weighted measurement over phase space:
\be
S_{\rm eff} = \int \df \Phi \,  S(\Phi) w(\Phi), \qquad B_{\rm eff} = \int \df \Phi \,  B(\Phi) w(\Phi) .
\ee 
The statistical uncertainty in $B_{\rm eff}$ is
\be
\delta B_{\rm eff} =  \sqrt{\int \df \Phi \,  \left[\delta B(\Phi)\right]^2 w(\Phi)^2} = \sqrt{\int \df \Phi \,  B(\Phi) w(\Phi)^2}.
\ee
Note that the reach in $S_{\rm eff}/\delta B_{\rm eff}$ is independent of the normalization of $w(\Phi)$.

To find the ideal measurement function, we simply need to use a variational method to solve for the condition
\be
\label{eq:weightminimization}
\frac{\mathrm{d}}{\mathrm{d} w(\Phi)} \left( \frac{S_{\rm eff}}{\delta B_{\rm eff}} \right) = 0.
\ee
Using the fact that
\be
\frac{\mathrm{d}}{\mathrm{d} w(\Phi)} S_{\rm eff} = S(\Phi), \qquad \frac{\mathrm{d}}{\mathrm{d} w(\Phi)}  \delta B_{\rm eff} = \frac{B(\Phi) w(\Phi)}{\delta B_{\rm eff}},
\ee
it is straightforward to solve \eq{eq:weightminimization} to find
\be
w_{\rm best}(\Phi) = \frac{ (\delta B_{\rm eff})^2}{S_{\rm eff}} \frac{S(\Phi)}{B(\Phi)} \Rightarrow
\frac{S(\Phi)}{B(\Phi)} ,
\ee
where we have used the fact that the overall normalization of $w_{\rm best}(\Phi)$ is irrelevant for determining the reach.  Plugging $w_{\rm best}(\Phi)$ into $S_{\rm eff}/\delta B_{\rm eff}$, we find
\be
\left[\frac{S_{\rm eff}}{\delta B_{\rm eff}} \right]_{\rm best} = \sqrt{\int \df \Phi \, \frac{S(\Phi)^2}{B(\Phi)}}.
\ee
For $S(\Phi)$ proportional to $B(\Phi)$ (i.e.\ no kinematic shape differences between the signal and background), this formula reverts to the standard $S/\sqrt{B}$.

When we plot the weighting function in Figs.~\ref{fig:modelAscalar1}, \ref{fig:modelAscalar2}, \ref{fig:modelBscalar1}, and \ref{fig:modelBscalar2}, we are actually plotting
\be
\int \df \Phi \, w_{\rm best}(\Phi) \, \mathcal{O}(\Phi),
\ee
where $\mathcal{O}(\Phi)$ is the observable corresponding to a histogram bin.  As a cross check of the weighting function, we checked that for most observables, this function is well-approximated by
\be
\frac{\int \df \Phi \, S(\Phi) \, \mathcal{O}(\Phi)}{\int \df \Phi \, B(\Phi) \,\mathcal{O}(\Phi)}\int \df \Phi \,  \mathcal{O}(\Phi),
\ee
i.e.\ the binned signal over background ratio corrected by a phase space volume factor.

\section{Generalized Couplings}
\label{sec:generalize}

In order to study the $X$ boson phenomenology in a model-independent way, we assumed in \sec{sec:theory} that the $X$ boson only had couplings to electrons and not to protons.  Here, we relax this assumption within the context of several proposed models, to see how the $X$ boson reach using $ep$ scattering is affected.  We find that, for varying reasons, couplings to the proton can be ignored.

\subsection{Kinematic Mixing}
\label{sec:protonpossibility}

In some of the best motivated dark force scenarios, the couplings of the $X$ boson are proportional to the electromagnetic couplings \cite{Pospelov:2007mp,ArkaniHamed:2008qn}. This occurs when the $X$ is a vector boson that kinematically mixes with the photon. That is, the vector $X_\mu$ couples directly to the electromagnetic current, albeit with a suppression factor $\epsilon$,
\be
\mathcal{L} = \epsilon g_\text{em} J_\text{em}^\mu X_\mu.
\ee
This yields $\alpha_X = \epsilon^2 \alpha_\text{EM}$, where in concrete models, typical $\epsilon$ values are $10^{-3}$ to $10^{-4}$.\footnote{In these models, the typical $X$ boson mass is around 1 GeV.  However, models with lighter $X$ bosons closer to 100 MeV are still plausible.}

Adding the $X$ boson coupling to the proton allows for additional diagrams for $X$ boson production. Therefore, there is the potential for noticeable contributions to the signal cross section from the proton coupling, whether through a direct contribution to the cross section or through constructive or destructive interference.  By explicit computation, however, we have checked that the proton coupling is largely irrelevant, and the essential physics can be understood by working out the kinematics of the relevant situations.  

The two differences in producing an $X$ off the photon or electron are the presence of an electron versus a proton propagator and the momentum transfer through the exchanged $t$-channel photon.  If one notes that neither the proton nor the $X$ boson are very relativistic in the energy range under consideration, one can approximately say that $E_p \simeq m_p$, $E_X \simeq m_X$, and $E_e \simeq m_X$ at any point along the electron line.  Then, the fermion propagator can be shown to be $\mathcal{O}(1/m_X)$ in both cases.   However, by seeing how momentum flows through the $t$-channel photon propagator, one can show that in the case of $X$ boson production off the electron line, $|t| \simeq \mathcal{O}(m_X^2)$, while off the proton line, $|t| \simeq \mathcal{O}(m_X m_p)$.  Thus, the cross section for $X$ boson production off the proton line is suppressed by $\mathcal{O}(m_X^2/m_p^2)$.

This suppression can be understood intuitively as arising from the fact that with a proton at rest, enough energy needs to be exchanged in the $t$-channel to actually create an $X$ boson, while with the electron already having sufficient energy, one only needs enough momentum transfer to move the electron into a kinematically valid region for $X$ radiation.  

One might worry that because of interference terms, the suppression would only be $\mathcal{O}(m_X/m_p)$.  However, in addition to the mass dependence, the momentum exchanged through the $t$-channel photon has angular dependence.  This is minimized when the $X$ boson is produced collinearly with the fermion line off which it is produced.  This means that the matrix elements for $X$ production off the electron and off the proton are peaked in entirely different regions of phase space, producing little overlap.   The interference terms end up being suppressed by another 2 orders of magnitude when the diagrams are explicitly calculated and integrated over.  Thus for $m_X \lesssim 100\MeV$, the corrections to the cross sections from including proton couplings are around 1\%, and do not measurably change the reach plots displayed earlier.

\subsection{Axion-Like Coupling}

An alternative framework for the $X$ boson is where the dark sector couples to the standard model through a pseudo Nambu-Goldstone boson, termed an axion portal \cite{Nomura:2008ru}.  In that case, one expects a coupling of $\lambda_p = m_\ell/f_a$ for elementary fermions, where $f_a$ is the axion decay constant.  In such a setup, the constraints from the muon anomalous magnetic moment completely rules out the region of electron couplings that might be probed in our setup, as discussed in \sec{sec:magmoment}. 

However, most axion portal models predict couplings of $\lambda_p \sim m_f/f_a$ for composite particles like the proton.  Thus, with the coupling to the proton $\mathcal{O}(10^3)$ greater than to the electron, one might think it possible to still see a signal from $X$ boson production off the proton, despite the kinematic suppression discussed above.  It turns out, though, that a decay constant small enough for this to be possible has already been ruled out by $K$ decay branching ratios~\cite{Park:2001cv}, which placed a lower bound of $f_a = \mathcal{O}(100\text{ TeV})$ for axion masses $\leq 2\mu$.  This leads to a coupling to the proton of at most $\alpha_X = \mathcal{O}(10^{-10})$ which, combined with the kinematic suppression, would not be detectable with this search.

\bibliography{DF}

\providecommand{\href}[2]{#2}\begingroup\raggedright\begin{thebibliography}{10}

\bibitem{Rubin:1970zz}
V.~C. Rubin and W.~K. Ford, Jr., {\it {Rotation of the Andromeda Nebula from a
  Spectroscopic Survey of Emission Regions}},  {\em Astrophys. J.} {\bf 159}
  (1970) 379--403.

\bibitem{AdelmanMcCarthy:2005se}
{\bf SDSS} Collaboration, J.~K. Adelman-McCarthy {\em et~al.}, {\it {The Fourth
  Data Release of the Sloan Digital Sky Survey}},  {\em Astrophys. J. Suppl.}
  {\bf 162} (2006) 38--48, [\href{http://arxiv.org/abs/astro-ph/0507711}{{\tt
  astro-ph/0507711}}].

\bibitem{Komatsu:2008hk}
{\bf WMAP} Collaboration, E.~Komatsu {\em et~al.}, {\it {Five-Year Wilkinson
  Microwave Anisotropy Probe (WMAP) Observations: Cosmological
  Interpretation}},  {\em Astrophys. J. Suppl.} {\bf 180} (2009) 330--376,
  [\href{http://arxiv.org/abs/0803.0547}{{\tt arXiv:0803.0547}}].

\bibitem{Briel:1997hz}
U.~G. Briel and J.~P. Henry, {\it {An X-ray Temperature Map of Coma}},
  \href{http://arxiv.org/abs/astro-ph/9711237}{{\tt astro-ph/9711237}}.

\bibitem{Finkbeiner:2004us}
D.~P. Finkbeiner, {\it {WMAP microwave emission interpreted as dark matter
  annihilation in the inner Galaxy}},
  \href{http://arxiv.org/abs/astro-ph/0409027}{{\tt astro-ph/0409027}}.

\bibitem{Hooper:2007kb}
D.~Hooper, D.~P. Finkbeiner, and G.~Dobler, {\it {Evidence Of Dark Matter
  Annihilations In The WMAP Haze}},  {\em Phys. Rev.} {\bf D76} (2007) 083012,
  [\href{http://arxiv.org/abs/0705.3655}{{\tt arXiv:0705.3655}}].

\bibitem{Adriani:2008zr}
{\bf PAMELA} Collaboration, O.~Adriani {\em et~al.}, {\it {An anomalous
  positron abundance in cosmic rays with energies 1.5-100 GeV}},  {\em Nature}
  {\bf 458} (2009) 607--609, [\href{http://arxiv.org/abs/0810.4995}{{\tt
  arXiv:0810.4995}}].

\bibitem{Abdo:2009zk}
{\bf The Fermi LAT} Collaboration, A.~A. Abdo {\em et~al.}, {\it {Measurement
  of the Cosmic Ray $e^+ + e^-$ spectrum from 20 GeV to 1 TeV with the Fermi
  Large Area Telescope}},  {\em Phys. Rev. Lett.} {\bf 102} (2009) 181101,
  [\href{http://arxiv.org/abs/0905.0025}{{\tt arXiv:0905.0025}}].

\bibitem{Collaboration:2008aaa}
{\bf H.E.S.S.} Collaboration, F.~Aharonian {\em et~al.}, {\it {The energy
  spectrum of cosmic-ray electrons at TeV energies}},  {\em Phys. Rev. Lett.}
  {\bf 101} (2008) 261104, [\href{http://arxiv.org/abs/0811.3894}{{\tt
  arXiv:0811.3894}}].

\bibitem{Aharonian:2009ah}
{\bf H.E.S.S.} Collaboration, F.~Aharonian {\em et~al.}, {\it {Probing the ATIC
  peak in the cosmic-ray electron spectrum with H.E.S.S}},
  \href{http://arxiv.org/abs/0905.0105}{{\tt arXiv:0905.0105}}.

\bibitem{Weidenspointner:2006nua}
G.~Weidenspointner {\em et~al.}, {\it {The sky distribution of positronium
  annihilation continuum emission measured with SPI/INTEGRAL}},  {\em Astron.
  Astrophys.} {\bf 450} (May, 2006) 1013--1021,
  [\href{http://arxiv.org/abs/astro-ph/0601673}{{\tt astro-ph/0601673}}].

\bibitem{Knodlseder:2003sv}
J.~Knodlseder {\em et~al.}, {\it {Early SPI/INTEGRAL contraints on the
  morphology of the 511 keV line emission in the 4th galactic quadrant}},  {\em
  Astron. Astrophys.} {\bf 411} (2003) L457--L460,
  [\href{http://arxiv.org/abs/astro-ph/0309442}{{\tt astro-ph/0309442}}].

\bibitem{Finkbeiner:2007kk}
D.~P. Finkbeiner and N.~Weiner, {\it {Exciting Dark Matter and the INTEGRAL/SPI
  511 keV signal}},  {\em Phys. Rev.} {\bf D76} (2007) 083519,
  [\href{http://arxiv.org/abs/astro-ph/0702587}{{\tt astro-ph/0702587}}].

\bibitem{Pospelov:2007mp}
M.~Pospelov, A.~Ritz, and M.~B. Voloshin, {\it {Secluded WIMP Dark Matter}},
  {\em Phys. Lett.} {\bf B662} (2008) 53--61,
  [\href{http://arxiv.org/abs/0711.4866}{{\tt arXiv:0711.4866}}].

\bibitem{ArkaniHamed:2008qn}
N.~Arkani-Hamed, D.~P. Finkbeiner, T.~R. Slatyer, and N.~Weiner, {\it {A Theory
  of Dark Matter}},  {\em Phys. Rev.} {\bf D79} (2009) 015014,
  [\href{http://arxiv.org/abs/0810.0713}{{\tt arXiv:0810.0713}}].

\bibitem{Nomura:2008ru}
Y.~Nomura and J.~Thaler, {\it {Dark Matter through the Axion Portal}},
  \href{http://arxiv.org/abs/0810.5397}{{\tt arXiv:0810.5397}}.

\bibitem{Goodsell:2009xc}
M.~Goodsell, J.~Jaeckel, J.~Redondo, and A.~Ringwald, {\it {Naturally Light
  Hidden Photons in LARGE Volume String Compactifications}},
  \href{http://arxiv.org/abs/0909.0515}{{\tt arXiv:0909.0515}}.

\bibitem{Fayet:2007ua}
P.~Fayet, {\it {{U-boson production in $e^+e^-$ annihilations, $\psi$ and
  $\Upsilon$ decays, and light dark matter}}},  {\em Phys. Rev.} {\bf D75}
  (2007) 115017, [\href{http://arxiv.org/abs/hep-ph/0702176}{{\tt
  hep-ph/0702176}}].

\bibitem{Pospelov:2008zw}
M.~Pospelov, {\it {Secluded U(1) below the weak scale}},
  \href{http://arxiv.org/abs/0811.1030}{{\tt arXiv:0811.1030}}.

\bibitem{Kahn:2007ru}
Y.~Kahn, M.~Schmitt, and T.~M.~P. Tait, {\it {Enhanced Rare Pion Decays from a
  Model of MeV Dark Matter}},  {\em Phys. Rev.} {\bf D78} (2008) 115002,
  [\href{http://arxiv.org/abs/0712.0007}{{\tt arXiv:0712.0007}}].

\bibitem{Reece:2009un}
M.~Reece and L.-T. Wang, {\it {Searching for the light dark gauge boson in
  GeV-scale experiments}},  \href{http://arxiv.org/abs/0904.1743}{{\tt
  arXiv:0904.1743}}.

\bibitem{Borodatchenkova:2005ct}
N.~Borodatchenkova, D.~Choudhury, and M.~Drees, {\it {Probing MeV dark matter
  at low-energy e+ e- colliders}},  {\em Phys. Rev. Lett.} {\bf 96} (2006)
  141802, [\href{http://arxiv.org/abs/hep-ph/0510147}{{\tt hep-ph/0510147}}].

\bibitem{Batell:2009yf}
B.~Batell, M.~Pospelov, and A.~Ritz, {\it {Probing a Secluded U(1) at
  B-factories}},  \href{http://arxiv.org/abs/0903.0363}{{\tt arXiv:0903.0363}}.

\bibitem{Essig:2009nc}
R.~Essig, P.~Schuster, and N.~Toro, {\it {Probing Dark Forces and Light Hidden
  Sectors at Low-Energy e+e- Colliders}},
  \href{http://arxiv.org/abs/0903.3941}{{\tt arXiv:0903.3941}}.

\bibitem{Morrissey:2009ur}
D.~E. Morrissey, D.~Poland, and K.~M. Zurek, {\it {Abelian Hidden Sectors at a
  GeV}},  \href{http://arxiv.org/abs/0904.2567}{{\tt arXiv:0904.2567}}.

\bibitem{Yin:2009mc}
P.-f. Yin, J.~Liu, and S.-h. Zhu, {\it {Detecting light leptophilic gauge boson
  at BESIII detector}},  \href{http://arxiv.org/abs/0904.4644}{{\tt
  arXiv:0904.4644}}.

\bibitem{Riordan:1987aw}
E.~M. Riordan {\em et~al.}, {\it {A Search for Short Lived Axions in an
  Electron Beam Dump Experiment}},  {\em Phys. Rev. Lett.} {\bf 59} (1987) 755.

\bibitem{Bross:1989mp}
A.~Bross {\em et~al.}, {\it {A Search for Shortlived Particles Produced in an
  Electron Beam Dump}},  {\em Phys. Rev. Lett.} {\bf 67} (1991) 2942--2945.

\bibitem{Bjorken:2009mm}
J.~D. Bjorken, R.~Essig, P.~Schuster, and N.~Toro, {\it {New Fixed-Target
  Experiments to Search for Dark Gauge Forces}},
  \href{http://arxiv.org/abs/0906.0580}{{\tt arXiv:0906.0580}}.

\bibitem{Heinemeyer:2007sq}
S.~Heinemeyer, Y.~Kahn, M.~Schmitt, and M.~Velasco, {\it {An Experiment to
  Search for Light Dark Matter in Low-Energy ep Scattering}},
  \href{http://arxiv.org/abs/0705.4056}{{\tt arXiv:0705.4056}}.

\bibitem{Boehm:2003hm}
C.~Boehm and P.~Fayet, {\it {Scalar dark matter candidates}},  {\em Nucl.
  Phys.} {\bf B683} (2004) 219--263,
  [\href{http://arxiv.org/abs/hep-ph/0305261}{{\tt hep-ph/0305261}}].

\bibitem{Wojtsekhowski:2009vz}
B.~Wojtsekhowski, {\it {Searching for a U-boson with a positron beam}},  {\em
  AIP Conf. Proc.} {\bf 1160} (2009) 149--154,
  [\href{http://arxiv.org/abs/0906.5265}{{\tt arXiv:0906.5265}}].

\bibitem{Baumgart:2009tn}
M.~Baumgart, C.~Cheung, J.~T. Ruderman, L.-T. Wang, and I.~Yavin, {\it
  {Non-Abelian Dark Sectors and Their Collider Signatures}},  {\em JHEP} {\bf
  04} (2009) 014, [\href{http://arxiv.org/abs/0901.0283}{{\tt
  arXiv:0901.0283}}].

\bibitem{Milner:2009--}
R.~Milner. Private communication, 2009.

\bibitem{Neil:2005jy}
G.~R. Neil {\em et~al.}, {\it {The JLab high power ERL light source}},  {\em
  Nucl. Instrum. Meth.} {\bf A557} (2006) 9--15.

\bibitem{Amsler:2008zzb}
{\bf Particle Data Group} Collaboration, C.~Amsler {\em et~al.}, {\it {Review
  of particle physics}},  {\em Phys. Lett.} {\bf B667} (2008) 1.

\bibitem{Sinha:1985aw}
R.~Sinha, {\it {Anomalous Magnetic Moment of Electron and Constraint on
  Composite Bosons of Weak Interaction}},  {\em Phys. Rev.} {\bf D34} (1986)
  1509.

\bibitem{Leveille:1977rc}
J.~P. Leveille, {\it {The Second Order Weak Correction to (g-2) of the Muon in
  Arbitrary Gauge Models}},  {\em Nucl. Phys.} {\bf B137} (1978) 63.

\bibitem{:2009cp}
{\bf The BABAR} Collaboration, B.~Aubert {\em et~al.}, {\it {Search for Dimuon
  Decays of a Light Scalar in Radiative Transitions $\Upsilon(3S) \to \gamma
  A_0$}},  \href{http://arxiv.org/abs/0902.2176}{{\tt arXiv:0902.2176}}.

\bibitem{Aubert:2009pw}
{\bf The BABAR} Collaboration, B.~Aubert {\em et~al.}, {\it {Search for a
  Narrow Resonance in $e^+e^-$ to Four Lepton Final States}},
  \href{http://arxiv.org/abs/0908.2821}{{\tt arXiv:0908.2821}}.

\bibitem{Bergsma:1985qz}
{\bf CHARM} Collaboration, F.~Bergsma {\em et~al.}, {\it {Search for Axion Like
  Particle Production in 400-GeV Proton-Copper Interactions}},  {\em Phys.
  Lett.} {\bf B157} (1985) 458.

\bibitem{Batell:2009di}
B.~Batell, M.~Pospelov, and A.~Ritz, {\it {Exploring Portals to a Hidden Sector
  Through Fixed Targets}},  \href{http://arxiv.org/abs/0906.5614}{{\tt
  arXiv:0906.5614}}.

\bibitem{Turner:1987by}
M.~S. Turner, {\it {Axions from SN 1987a}},  {\em Phys. Rev. Lett.} {\bf 60}
  (1988) 1797.

\bibitem{Abouzaid:2006kk}
{\bf KTeV} Collaboration, E.~Abouzaid {\em et~al.}, {\it {Measurement of the
  rare decay $\pi^0 \to e^+ e^-$}},  {\em Phys. Rev.} {\bf D75} (2007) 012004,
  [\href{http://arxiv.org/abs/hep-ex/0610072}{{\tt hep-ex/0610072}}].

\bibitem{Boos:2009un}
E.~Boos {\em et~al.}, {\it {CompHEP 4.5 Status Report}},
  \href{http://arxiv.org/abs/0901.4757}{{\tt arXiv:0901.4757}}.

\bibitem{Alwall:2007st}
A.~Johan {\em et~al.}, {\it {MadGraph/MadEvent v4: The New Web Generation}},
  {\em JHEP} {\bf 09} (2007) 028, [\href{http://arxiv.org/abs/0706.2334}{{\tt
  arXiv:0706.2334}}].

\bibitem{Abulencia:2005uq}
{\bf CDF} Collaboration, A.~Abulencia {\em et~al.}, {\it {Top quark mass
  measurement from dilepton events at CDF II}},  {\em Phys. Rev. Lett.} {\bf
  96} (2006) 152002, [\href{http://arxiv.org/abs/hep-ex/0512070}{{\tt
  hep-ex/0512070}}].

\bibitem{Abazov:2004cs}
{\bf D0} Collaboration, V.~M. Abazov {\em et~al.}, {\it {A precision
  measurement of the mass of the top quark}},  {\em Nature} {\bf 429} (2004)
  638--642, [\href{http://arxiv.org/abs/hep-ex/0406031}{{\tt hep-ex/0406031}}].

\bibitem{Abazov:2006bd}
{\bf D0} Collaboration, V.~M. Abazov {\em et~al.}, {\it {Measurement of the top
  quark mass in the lepton + jets final state with the matrix element method}},
   {\em Phys. Rev.} {\bf D74} (2006) 092005,
  [\href{http://arxiv.org/abs/hep-ex/0609053}{{\tt hep-ex/0609053}}].

\bibitem{Kolomensky:2009--}
Y.~G. Kolomensky. Private communication, 2009.

\bibitem{Bjorken:1988as}
J.~D. Bjorken {\em et~al.}, {\it {Search for Neutral Metastable Penetrating
  Particles Produced in the SLAC Beam Dump}},  {\em Phys. Rev.} {\bf D38}
  (1988) 3375.

\bibitem{Uhlemann:2008pm}
C.~F. Uhlemann and N.~Kauer, {\it {Narrow-width approximation accuracy}},  {\em
  Nucl. Phys.} {\bf B814} (2009) 195--211,
  [\href{http://arxiv.org/abs/0807.4112}{{\tt arXiv:0807.4112}}].

\bibitem{Peskin:1995ev}
M.~E. Peskin and D.~V. Schroeder, {\it {An Introduction to quantum field
  theory}}, . Reading, USA: Addison-Wesley (1995) 842 p.

\bibitem{Park:2001cv}
{\bf HyperCP} Collaboration, H.~K. Park {\em et~al.}, {\it {Observation of the
  Decay $K^- \to \pi^- \mu^+ \mu^-$ and Measurements of the Branching Ratios
  for $K^\pm \to \pi^\pm \mu^+ \mu^-$}},  {\em Phys. Rev. Lett.} {\bf 88}
  (2002) 111801, [\href{http://arxiv.org/abs/hep-ex/0110033}{{\tt
  hep-ex/0110033}}].

\end{thebibliography}\endgroup

\end{document}